\documentclass[12pt]{article}	
\usepackage{amsmath,amssymb,bm,graphicx}
\usepackage{here}
\usepackage{color} 	

\newcommand{\Dslash}{\not \!\! D}
\newcommand{\Aslash}{\not \!\! A}

\newcommand{\kslash}{\not \! k}

\newcommand{\delslash}{\not \! \partial}

\makeatletter

\@addtoreset{figure}{section}
\makeatother

\begin{document}

\begin{flushright}
\end{flushright}

\vskip 0.5 truecm

\begin{center}
{\Large{\bf  Berry's phase and chiral anomalies
}}
\end{center}
\vskip .5 truecm
\centerline{\bf Kazuo Fujikawa~$^1$
 {\rm and} Koichiro Umetsu~$^2$ }
\vskip .4 truecm
\centerline {\it $^1$~Interdisciplinary Theoretical and Mathematical Sciences Program (iTHEMS), 
}
\centerline {\it   RIKEN, Wako 351-0198, 
Japan}
\vskip 0.4 truecm
\centerline {\it $^2$~Laboratory of Physics, College of Science and Technology, and Junior College, }
\centerline{\it Funabashi Campus, Nihon University, Funabashi, Chiba 274-8501, Japan}
\vskip 0.5 truecm

\makeatletter
\@addtoreset{equation}{section}
\def\theequation{\thesection.\arabic{equation}}
\makeatother

\begin{abstract}

 The  basic materials of Berry's phase and chiral anomalies are presented to appreciate the  phenomena related to those notions. 
 As for Berry's phase,  a general survey of the subject   
 is presented using both Lagrangian and Hamiltonian formalisms. 
 The canonical Hamiltonian formalism of the Born-Oppenheimer approximation, when applied to the anomalous Hall effect, can incorporate the gauge symmetry of Berry's connection but unable to incorporate 
 the electromagnetic vector potential simultaneously.   
 Transformed to the Lagrangian formalism with a time-derivative term allowed, the Born-Oppenheimer approximation  can incorporate the electromagnetic vector potential simultaneously with Berry's connection, but the consistent  canonical property is lost and thus becomes classical. The Lagrangian formalism can thus incorporate both gauge symmetries simultaneously but spoils the basic quantum symmetries, and thus results in classical anomalous Poisson brackets and the classical Nernst effect as in the conventional formalism.

As for chiral anomalies, we present basic materials by the path integral formulation with an emphasis on fermions on the lattice. A chiral fermion defined by $\gamma_{5}$ on the lattice does not contain the chiral anomaly for the non-vanishing lattice spacing $a\neq0$.   
 The idea of a spectral flow on the lattice does not lead to an anomaly for each species doubler separately but rather to a pair production in a general sense.
We also mention that a specific construction called the Ginsparg-Wilson fermion, which is free of species doublers, may practically be useful. 
We discuss 
the representative applications of Berry's phase and chiral anomalies in nuclear physics and related fields to illustrate the use of  these two basic notion.

\end{abstract}

\vspace{1cm}

\clearpage

\tableofcontents

\clearpage
\section{Introductory remarks}
The notion of Berry's phase~\cite{Berry-review, Higgins, Berry, Simon} is widely used in physics nowadays, but at the same time the terminology of Berry's phase does not appear to be sharply defined in some applications.  In the present review, we use  the terminology of Berry's phase in a narrow sense; we are going to use Berry's phase as the phenomena related to the adiabatic theorem in quantum mechanics \cite{Born, Kato}, in particular, the topological properties related to the level crossing phenomena  in quantum physics. We thus distinguish the Aharonov-Bohm effect \cite{Aharonov-Bohm}, for example, from Berry's phase despite the known similarity between them \cite{Berry}.

The notions of  topology, level crossing and the adiabatic theorem also appear in a specific Hamiltonian formulation of chiral gauge anomaly by Nelson and Alvarez-Gaume \cite{Nelson}. On the basis of their formulation, one of the present authors (KF) made the following observation in 2006 \cite{Fujikawa-2006}:  The chiral anomaly is characterized by the inevitable failure of the adiabatic theorem, namely, the inevitable occurrence of level crossing (or state change) forced by the topology of the external gauge field even in the ideal adiabatic process.  In contrast, the precise topology of Berry's phase of Dirac monopole-type is self-generated only when the precise adiabatic condition is satisfied. The topology of Berry's phase does not cause the level crossing by breaking the adiabatic theorem. From this point of view, the basic ideas of these two phenomena are very distinct. It was also suggested in  \cite{Fujikawa-2006} that ``similar phenomena'' may better be distinguished from the ``same phenomena'' in connection with Berry's phase and chiral anomalies \cite{Bertlmann}.

Nevertheless,  the interesting phenomena of Berry's phase which are related to chiral anomalies  have been widely discussed. It has been argued in ingenious ways that Berry's phase gives rise to the phenomena    closely related to chiral anomalies \cite{Stone, Jackiw-IJMP} and in some special case the chiral anomaly itself \cite{Son-Yamamoto, Stephanov}. One of the purposes of this review article is to present the basic formulation of Berry's phase together with an account of  chiral anomalies by placing an emphasis on fermions on the lattice that are useful to assess these interesting proposals. The basic view of the present review is that these two basic ideas are different by the very basic construction of these two notions, although superficially one recognizes interesting similarities.  The present review is thus different from the conventional review articles in that we mainly present the basic materials instead of the analyses of extensive applications.

In Section 2, we briefly summarized the standard derivation of Berry's phase and its essential aspects using the second quantized path integral.   The topological properties in the adiabatic limit was also noted.
In Section 3, we present a systematic way to derive the commonly used effective action of the anomalous Hall effect starting with the canonical quantum system 
on the basis of the adiabatic condition 
\begin{eqnarray}
|\epsilon_{n\pm1}-\epsilon_{n}|\gg 2\pi\hbar/T.
\end{eqnarray}
Here  $2\pi\hbar/T$ is the typical energy scale of the slower system $X_{k}$ and $\epsilon_{n}$ is the $n$-th energy level of the fast system. The crucial approximation one makes in this derivation is that one discards many terms linear in the time derivative, which are regarded to be small in the adiabatic approximation.  But these terms are crucially important to determine the equal-time commutation relations in either the conventional canonical quantization starting with a given action or other methods of quantization such as the Bjorken-Johnson-Low (BJL) prescription. We emphasize the BJL method in this review mainly because the importance of the terms linear in time-derivative become very explicit \footnote{ The basic idea of the Bjorken-Johnson-Low (BJL) prescription, which is more general than the canonical quantization in the sense that it works for the system with quantum anomalies also, is that the equal-time commutation relation of two operators $A(t)$ and 
 $B(t)$ is determined by the analysis of the short-time
 limit of the time-ordered product $T A(t_{1})B(t_{2})$ with $t_{1}\rightarrow t_{2}$. This means that 
 the infinitely large frequencies of the Fourier transform of $T A(t_{1})B(t_{2})$ determine the equal-time commutation relation. See Appendix A for more details.}.  

When one examines the same anomalous Hall effect by a canonical Hamiltonian formalism in the Born-Oppenheimer approximation with a point-like Berry's connection, one recognizes that no anomalous commutation relations appear without the electromagnetic vector potential. In fact, one shows that the electromagnetic vector potential is not incorporated in a canonical Hamiltonian formalism of the adiabatic approximation.
We thus adopt a view that the quantum mechanical anomalous Hall effect is described by the system without the electromagnetic vector potential, and a non-canonical classical system is induced if one adds the electromagnetic vector potential and thus results in the classical Nernst effect. This picture is adopted throughout the present review.

In Section 4, we review the path integral formulation of quantum anomalies with an  emphasis on the basic properties of fermions defined on the lattice with the finite lattice spacing $a\neq 0$. The notable feature of fermions on the lattice is the so-called species doubling \cite{Wilson-1974, Karsten-Smit, Nielsen-Ninomiya}. If one attempts to define a single species of fermion on the lattice using a chiral ($\gamma_{5}$) invariant Lagrangian, one encounters 16 species of fermions all together in the case of Lorentz invariant 4-dimensional case.  The species doublers are defined independently of chiral anomalies. We emphasize that chiral anomaly for each species doubler separately is ill-defined (or vanish)
in the actual evaluation of lattice gauge theory for $a\neq0$, since each species doubler, which is defined in a part of the Brillouin zone in momentum space, is not a local field in space-time. In some sense, the notions of the species doubler and the chiral anomaly for each species doubler separately are complementary; if one defines the species doubler explicitly, it implies the spectrum smoothly connected over the doublers, which in turn implies that the definition of anomaly for each species doubler separately becomes ill-defined. Rather, the pair production in a general sense, as is explained in further detail later, is a natural notion valid for species doublers  connected by a smooth spectrum. Also, the effective chiral fermion appears in the lattice of condensed matter physics, for which no fundamental chiral symmetry in the Hamiltonian level is specified; thus  the basic fermions appearing in the (conventional) lattice gauge theory can be completely different from those effective chiral fermions in the condensed matter physics when it comes to the fundamental properties related to chiral properties. One needs to exercise due care. 

We also mention the Ginsparg-Wilson fermion \cite{Ginsparg-Wilson} which avoids the species doublers on the lattice; it has an  interesting property with regard to chiral anomalies. We illustrate how to define a massless Dirac fermion on the lattice which may practically be useful.
Simply stated, one can define a single massless Dirac fermion on the lattice without species doublers for $a\neq0$ but with a modified definition of the chiral operator $\Gamma_{5}$.  
 
In Section 5, we discuss some representative topics in nuclear physics and related fields, as applications of the basic materials explained in this review. As concrete subjects, we mention  the general connections of the interesting phenomenon called ``chiral magnetic effect'' widely discussed in nuclear physics with the general subject of chiral anomalies and the associated sphaleron effect. We also discuss the possible connection of the chiral anomaly with the classical anomalous Hall effect, which is often discussed in nuclear physics; we show that the chiral anomaly thus defined differs from the conventional chiral anomaly one is familiar with.

In the applications of Berry's phase and chiral anomalies in a wider area of physics, physically interesting phenomena often appear  in the settings which are not defined in an ideal manner. Namely, some other important notions and ideas enter the problem. In such cases, we comment on only those aspects which are understood using the notions of chiral anomaly and Berry's phase.  

In Appendix A, a minimum explanation of the BJL method is given. In Appendix B, a detailed account of what is the monopole appearing in Berry's phase is given based on an exactly solvable model. The subject discussed in Appendix B does not have a much direct connection with the presentation in this review, except for the conceptual contexts discussed in subsections 3.2.3 and 3.3.  But, as is emphasized later, the subject has a deep meaning as to  Berry's phase beyond phenomenological applications such as the anomalous Hall effect.

\newpage
\section{Elements of geometric phases }

In this article, we use mainly the second quantized path integral approach to geometric phases.
 For the more conventional approaches to the subject, the readers are referred to  review articles and textbooks in \cite{ Holstein, Mead, Sakurai, A-Bohm}.

\subsection{Path integral and second quantization}
The basic idea of the adiabatic theorem is well-known \cite{Born} and the precise proof is given~\cite{Kato}. In the physical applications, we treat the slow variables as literally slowly moving particles~\cite{Higgins, Berry} instead of the ideal infinitely slow movement~\cite{Simon}. In this subsection we present a  convenient definition of adiabatic geometric phases  using the path integral formulation. (We use the term ``geometric phase'' and the term ``Berry's phase'' interchangeably, except for the non-adiabatic geometric phase.) 
We start with the generic hermitian Hamiltonian 
\begin{eqnarray}\label{second quantization0}
\hat{H}=\hat{H}(\hat{\vec{p}},\hat{\vec{x}},X(t))
\end{eqnarray}
for a single particle theory in a set of slowly varying background C-number
variables $X(t)=(X_{1}(t),X_{2}(t),...)$. For example, in the case of a simplified molecular system, the variables $x_{k}$ stand for the coordinates of the electrons and the background variables $X_{k}(t)$ stand  for the coordinates of the atomic nuclei.
The path integral for this theory for the time interval
$0\leq t\leq T$ in the second quantized 
formulation is given by 
\begin{eqnarray}\label{second quantization1}
Z&=&\int{\cal D}\psi^{\dagger}{\cal D}\psi
\exp\Bigg\{ \frac{i}{\hbar}\int_{0}^{T}dtd^{3}x \Bigg[
\psi^{\dagger}(t,\vec{x})i\hbar\frac{\partial}{\partial t}
\psi(t,\vec{x})\nonumber\\
&&-\psi^{\dagger}(t,\vec{x})
\hat{H}\left( \frac{\hbar}{i}\frac{\partial}{\partial\vec{x}},
\vec{x},X(t)\right)\psi(t,\vec{x} ) \Bigg] \Bigg\}
\end{eqnarray}
which reproduces the quantum mechanical (operator) equation
\begin{eqnarray}
\langle i\hbar\frac{\partial}{\partial t}
\psi(t,\vec{x}) -
\hat{H}\left( \frac{\hbar}{i}\frac{\partial}{\partial\vec{x}},
\vec{x},X(t) \right)\psi(t,\vec{x})\rangle = 0.
\end{eqnarray}
We then define a complete set of instantaneous eigenfunctions
\begin{eqnarray}\label{basis set}
&&\hat{H}\left( \frac{\hbar}{i}\frac{\partial}{\partial\vec{x}},
\vec{x},X(t) \right)v_{n}(\vec{x},X(t))
=E_{n}(X(t))v_{n}(\vec{x},X(t)), \nonumber\\
&&\int d^{3}x v^{\dagger}_{n}(\vec{x},X(t))v_{m}(\vec{x},X(t))
=\delta_{n,m},
\end{eqnarray}
and expand the classical field $\psi(t,\vec{x})$ in the path integral which is a Grassmann number for a fermion, for example,
as
\begin{eqnarray}\label{field expansion}
\psi(t,\vec{x})=\sum_{n}b_{n}(t)v_{n}(\vec{x},X(t)).
\end{eqnarray}
We then have
${\cal D}\psi^{\dagger}{\cal D}\psi
=\prod_{n}{\cal D}b_{n}^{\star}
{\cal D}b_{n}$
and the path integral in the second quantized formulation is 
written as~\cite{deguchi}
\begin{eqnarray}\label{second quantization2}
Z&=&\int \prod_{n}{\cal D}b_{n}^{\star}{\cal D}b_{n}
\exp \Bigg\{ \frac{i}{\hbar}\int_{0}^{T}dt \Bigg[
\sum_{n}b_{n}^{\star}(t)i\hbar\frac{\partial}{\partial t}
b_{n}(t)\nonumber\\
&&+\sum_{n,m}b_{n}^{\star}(t)
\langle n|i\hbar\frac{\partial}{\partial t}|m\rangle
b_{m}(t)
-\sum_{n}b_{n}^{\star}(t) E_{n}(X(t))b_{n}(t) \Bigg] \Bigg\}
\end{eqnarray}
where the second term in the action, which is defined by 
\begin{eqnarray}\label{geometric phase}
\langle n|i\hbar\frac{\partial}{\partial t}|m\rangle
&\equiv& \int d^{3}x v^{\dagger}_{n}(\vec{x},X(t))
i\hbar\frac{\partial}{\partial t}v_{m}(\vec{x},X(t))\nonumber\\
&=&\langle n|i\hbar\frac{\partial}{\partial X_{k}(t)}|m\rangle\frac{dX_{k}(t)}{dt}
\end{eqnarray}
stands for what
	commonly referred to as {\em Berry's phase}  \cite{ Higgins, Berry, Simon} and its 
off-diagonal generalization.  
 We take the time $T$ 
as a period of the  variables, $X_{k}(T)=X_{k}(0)$, in the analysis
of geometric phases for the reasons stated below. The above formula \eqref{second quantization2} itself is exact (in the framework of the so-called snap shot picture).   
 Intuitively, the adiabatic 
process means that $T$ is much larger than the typical
time scale implied by $\hbar/\Delta E_{n}(X(t))$ with, for example, $\Delta E_{n}(X(t))= E_{n+1}(X(t))-E_{n}(X(t))$ for each $n$, or equivalently, 
\begin{eqnarray}
\Delta E_{n}(X(t))\gg \hbar/T
\end{eqnarray}
The adiabatic theorem is defined by $T\rightarrow\infty$ \cite{Simon, Born, Kato}, but a more practical definition as above for finite $T$ is commonly used in  applications.

Translated into the operator formulation,
we thus obtain the effective Hamiltonian (depending on Bose or 
Fermi statistics)
\begin{eqnarray}\label{effective Hamiltonian1}
\hat{H}_{eff}(t)&=&\sum_{n}\hat{b}_{n}^{\dagger}(t)
E_{n}(X(t))\hat{b}_{n}(t)
-\sum_{n,m}\hat{b}_{n}^{\dagger}(t)
\langle n|i\hbar\frac{\partial}{\partial t}|m\rangle
\hat{b}_{m}(t)
\end{eqnarray}
with $[\hat{b}_{n}(t), \hat{b}^{\dagger}_{m}(t)]_{\mp}
=\delta_{n,m}$.
All the information about geometric phases  is included in 
this effective Hamiltonian and in this sense geometric phases are {\em dynamical}.
See also Berry~\cite{Berry} for a related observation. 
When one defines the Schr\"{o}dinger picture 
$\hat{{\cal H}}_{eff}(t)$ by replacing all $\hat{b}_{n}(t)$ by
$\hat{b}_{n}(0)$ in the above $\hat{H}_{eff}(t)$,
the second quantization formula for the evolution operator 
gives rise to the result of the first quantization in the form~\cite{deguchi} 
\begin{eqnarray}\label{Schroedinger picture}
\langle m|T^{\star}\exp\left\{ -\frac{i}{\hbar}\int_{0}^{T}
\hat{{\cal H}}_{eff}(t)
dt \right\} |n\rangle 
=
\langle m(T)|T^{\star}\exp \left\{-\frac{i}{\hbar}\int_{0}^{T}
\hat{H}(\hat{\vec{p}}, \hat{\vec{x}},  
X(t))dt \right\}|n(0)\rangle \nonumber\\
\end{eqnarray}
where $T^{\star}$ stands for the time ordering operation.
The state vectors in the second quantization  on the left-hand 
side are defined by $
|n\rangle=\hat{b}_{n}^{\dagger}(0)|0\rangle$,
and the state vectors on the right-hand side  stand for the 
first quantized states defined by
$\langle\vec{x}|n(t)\rangle=v_{n}(\vec{x},X(t))$.
Both-hand sides of the above equality \eqref{Schroedinger picture} are exact, but the 
difference is that the geometric phases, both  diagonal and 
off-diagonal, are explicit in the second quantized formulation 
on the left-hand side. In other words, if one evaluates the right-hand side exactly by converting it to the path integral (in the first quantization formalism), for example, all the effects of Berry's phases are included in principle but not seen.  {\em The adiabatic Berry's phase becomes visible only in the adiabatic approximation}.

The exact probability amplitude which satisfies Schr\"{o}dinger 
equation with the initial condition $\psi_{n}(\vec{x},0; X(0))=v_{n}(\vec{x};X(0))$
is given by
\begin{eqnarray}\label{physical amplitude0}
\psi_{n}(\vec{x},t; X(t))=
\langle 0|\hat{\psi}(t,\vec{x})\hat{b}^{\dagger}_{n}(0)|0\rangle
\end{eqnarray}
since $i\hbar\partial_{t}\hat{\psi}=\hat{H}\hat{\psi}$ in the 
present problem. To be explicit, we have 
\begin{eqnarray}\label{physical amplitude1}
\psi_{n}(\vec{x},t; X(t))
=\sum_{m} v_{m}(\vec{x};X(t))
\langle m|T^{\star}\exp\Bigg\{ -\frac{i}{\hbar}\int_{0}^{t}
\hat{{\cal H}}_{eff}(t)dt \Bigg\}|n\rangle
\end{eqnarray}
by noting that \eqref{Schroedinger picture}  is given by 
$\langle0| \hat{b}_{m}(t)\hat{b}^{\dagger}_{n}(0)|0\rangle$.
This formula is also derived by noting the conventional formula
\begin{eqnarray}\label{physical amplitude2}
&&\psi_{n}(\vec{x},t; X(t))\nonumber\\
&&=\langle \vec{x}|T^{\star}\exp \left\{-\frac{i}{\hbar}\int_{0}^{t}
\hat{H}(\hat{\vec{p}}, \hat{\vec{x}},  
X(t))dt \right\}|n(0)\rangle\nonumber\\ 
&&=\sum_{m} v_{m}(\vec{x};X(t))
\langle m(t)|T^{\star}\exp \left\{-\frac{i}{\hbar}\int_{0}^{t}
\hat{H}(\hat{\vec{p}}, \hat{\vec{x}},  
X(t))dt \right\}|n(0)\rangle
\end{eqnarray}
and the relation \eqref{Schroedinger picture}.

In the limit of the slowly varying background variables, namely, if the variation of $X_{k}(t)$ is slow at each moment and $T\rightarrow {\rm large}$, one may assume the dominance of
 diagonal elements in  \eqref{effective Hamiltonian1} and \eqref{physical amplitude1} \cite{Born}, and one thus has \footnote{Intuitively, all the time derivative terms are small in the adiabatic limit $T\rightarrow {\rm large}$ and thus the diagonal terms coming together with the energy eigenvalues $E_{n}$ give the dominant contributions in the linear order of small quantities.}
\begin{eqnarray}\label{adiabatic amplitude1}
&&\psi_{n}(\vec{x},T; X(T))\nonumber\\
&&\simeq v_{n}(\vec{x};X(T))
\exp\left\{ -\frac{i}{\hbar}\int_{0}^{T}\left[ E_{n}(X(t))
-\langle n|i\hbar\frac{\partial}{\partial t}|n\rangle \right]dt \right\}.
\end{eqnarray}
 The last term in the exponential is the common expression of adiabatic Berry's phase, which is written using the diagonal term in  \eqref{geometric phase} in the form 
 \begin{eqnarray}\label{2-Berry's phase}
\int_{0}^{T} dt\langle n|i\hbar\frac{\partial}{\partial t}|n\rangle
&=&\oint{\cal A}_{k}(X) \frac{dX_{k}(t)}{dt}dt
\end{eqnarray}
for the periodic motion $X_{k}(0)=X_{k}(T)$ with 
\begin{eqnarray}
\langle n|i\hbar\frac{\partial}{\partial X_{k}(t)}|n\rangle
={\cal A}_{k}(X) 
\end{eqnarray}
but this expression by itself does not explicitly show that the Berry's phase is topological, although the time dependence of the right-hand side of \eqref{2-Berry's phase} formally disappears.

\subsection{Hidden local gauge symmetry}

The path integral formula \eqref{second quantization2} is based on the expansion \eqref{field expansion} and 
the starting second-quantized path integral \eqref{second quantization1} depends only on 
the field
variable $\psi(t,\vec{x})$, not on  $\{ b_{n}(t)\}$
and $\{v_{n}(\vec{x},X(t))\}$ separately. This fact shows that 
our formulation contains an exact hidden local gauge symmetry
which keeps the field variable $\psi(t,\vec{x})$ invariant \cite{fujikawa} 
\begin{eqnarray}\label{hidden local}
&&v_{n}(\vec{x},X(t))\rightarrow v^{\prime}_{n}(t; \vec{x},X(t)) \equiv
e^{i\alpha_{n}(t)}v_{n}(\vec{x},X(t)),\nonumber\\
&&b_{n}(t) \rightarrow b^{\prime}_{n}(t)=
e^{-i\alpha_{n}(t)}b_{n}(t), \ \ \ \ n=1,2,3,...,
\end{eqnarray}
where the gauge parameter $\alpha_{n}(t)$ is a general 
function of $t$. 
This gauge symmetry (or substitution rule) states the fact 
that the choice of coordinates in the functional space is 
arbitrary and this symmetry by itself does not give any conservation law. This symmetry is exact under a 
rather mild condition that the basis set \eqref{basis set} is not singular, 
namely, it is exact not only for the adiabatic case but also for
 the non-adiabatic case. Consequently, physical observables 
should always respect this symmetry. Also, by using this local 
gauge freedom, one can choose the phase convention of the basis 
set $\{v_{n}(t,\vec{x},X(t))\}$ at one's will such that the 
analysis of geometric phases becomes simplest.

Our next observation is that $\psi_{n}(\vec{x},t; X(t))$ in \eqref{physical amplitude0}
transforms under the hidden local gauge symmetry \eqref{hidden local} as
\begin{eqnarray} \label{ray representation}
\psi^{\prime}_{n}(\vec{x},t; X(t))=e^{i\alpha_{n}(0)}
\psi_{n}(\vec{x},t; X(t))
\end{eqnarray}
{\em independently} of the value of $t$. 
This transformation is derived using the exact 
representation \eqref{physical amplitude0}, and it implies that 
$\psi_{n}(\vec{x},t; X(t))$ is a physical object since 
$\psi_{n}(\vec{x},t; X(t))$  stays in the {\em same 
ray}~\cite{dirac, streater} 
under an arbitrary hidden local gauge transformation.
This transformation is explicitly checked for the formula of the adiabatic
approximation \eqref{adiabatic amplitude1} also.

The product
$\psi_{n}(\vec{x},0; X(0))^{\dagger}\psi_{n}(\vec{x},T; X(T))$
is thus manifestly independent of the choice of the (time dependent)  phase 
convention of the basis set $\{v_{n}(t,\vec{x},X(t))\}$. 
For the adiabatic formula \eqref{adiabatic amplitude1}, the gauge invariant quantity
 is given by
\begin{eqnarray}\label{gauge invariant quantity}
&&\psi_{n}(\vec{x},0; X(0))^{\dagger}\psi_{n}(\vec{x},T; X(T))\nonumber\\
&&=v_{n}(0,\vec{x}; X(0))^{\dagger}v_{n}(T,\vec{x};X(T))
\nonumber\\
&&\times\exp\left\{-\frac{i}{\hbar}\int_{0}^{T} \left[{\cal E}_{n}(X(t))
-\langle n|i\hbar\frac{\partial}{\partial t}|n\rangle \right]dt \right\}.
\end{eqnarray}
We then observe that by choosing the hidden gauge such that 
$v_{n}(T,\vec{x};X(T))=v_{n}(0,\vec{x}; X(0))$, the prefactor 
$v_{n}(0,\vec{x}; X(0))^{\dagger}v_{n}(T,\vec{x};X(T))$ becomes 
real and positive. Note that we are assuming the {\em periodic
evolution of the external parameters}, $X(T)=X(0)$ \footnote{If the variation of $X(t)$ is not periodic and thus $v_{n}(T,\vec{x};X(T))$ is not cyclic (i.e., periodic up to a phase),  the factor $v_{n}(0,\vec{x}; X(0))^{\dagger}v_{n}(T,\vec{x};X(T))$ is not chosen to be real and positive for all $\vec{x}$ by a suitable choice of the hidden gauge symmetry in general. The geometric phase is not uniquely defined in such a case.}. Then the 
phase factor in \eqref{gauge invariant quantity} defines a physical quantity uniquely. 
After this gauge 
fixing, the phase in \eqref{gauge invariant quantity} is still invariant under residual 
gauge transformations satisfying the periodic boundary condition
$\alpha_{n}(0)=\alpha_{n}(T)$,
in particular, for the gauge of the form 
\begin{eqnarray}
\alpha_{n}(X(t))
\end{eqnarray}
which gives the common gauge transformation of Berry's 
phase \cite{Berry}. 

The hidden local gauge symmetry is also used \cite{fujikawa} to describe the so-called ``non-adiabatic Aharonov-Anandan phase'' instead of  using the notion of parallel transport \cite{Aharonov}.


\subsection{Two-band model of Berry's phase}

To simplify the analysis of topological properties, we now assume that 
the level crossing takes place only between 
the two levels. This simplification is 
expected to be approximately  valid to analyze Berry's phase  
in the  neighborhood of the specific level crossing for a multi-level system. 

We start with a complete set of generic eigenfunctions of the Hamiltonian\\
 $\hat{H}(\frac{\hbar}{i}\frac{\partial}{\partial\vec{x}},\vec{x},X(0))$ defined at $t=0$, instead of the instantaneous eigenfunctions of the time dependent Hamiltonian discussed so far,
\begin{eqnarray}\label{generic eigenfunctions}
&&\hat{H}\left( \frac{\hbar}{i}\frac{\partial}{\partial\vec{x}},
\vec{x},X(0)\right)u_{n}(\vec{x},X(0))
=\lambda_{n}u_{n}(\vec{x},X(0)), \nonumber\\
&&\int d^{3}xu_{n}^{\star}(\vec{x},X(0))u_{m}(\vec{x},X(0))=
\delta_{nm},
\end{eqnarray}
and expand 
\begin{eqnarray}
\psi(t,\vec{x})=\sum_{n}a_{n}(t)u_{n}(\vec{x},X(0)).
\end{eqnarray}
We then have 
${\cal D}\psi^{\star}{\cal D}\psi=\prod_{n}{\cal D}a_{n}^{\star}
{\cal D}a_{n}$ 
and the path integral \eqref{second quantization1} is written as 
\begin{eqnarray}\label{second quantization3}
Z&=&\int \prod_{n}{\cal D}a_{n}^{\star}
{\cal D}a_{n}
\exp \Bigg\{ \frac{i}{\hbar}\int_{0}^{T}dt \Bigg[
\sum_{n}a_{n}^{\star}(t)i\hbar\frac{\partial}{\partial t}
a_{n}(t)\nonumber\\
&&-\sum_{n,m}a_{n}^{\star}(t)E_{nm}(X(t))a_{m}(t) \Bigg] \Bigg\}
\end{eqnarray}
where 
\begin{eqnarray}\label{eigenvalues}
E_{nm}(X(t))=\int d^{3}x u_{n}^{\star}(\vec{x},X(0))
\hat{H}\left( \frac{\hbar}{i}\frac{\partial}{\partial\vec{x}},
\vec{x},X(t) \right)u_{m}(\vec{x},X(0)).
\end{eqnarray}

Unlike the path integral \eqref{second quantization2}, which formally diagonalizes the Hamiltonian at each moment, the Hamiltonian is not diagonalized in the  present path integral \eqref{eigenvalues}.
We thus diagonalize the energy eigenvalues in a more detailed manner to define Berry's phase more explicitly. 
The effective Hamiltonian to be analyzed 
in the path integral \eqref{second quantization3} is then defined  by the $2\times 2$ hermitian
matrix $ h(X(t))=\left(E_{nm}(X(t))\right)$ in the two-level truncation.
If one assumes that the level crossing takes place at the 
origin of the parameter space $X(t)=0$, one needs to analyze
the matrix
\begin{eqnarray}\label{effective Hamitonian1}
h(X(t)) &=&  \left(E_{nm}(X(t))\right)\nonumber\\
&=& \left(E_{nm}(0)\right) + 
\left(\frac{\partial}{\partial X_{k}}E_{nm}(X)|_{X=0}\right)
X_{k}(t) + ....
\end{eqnarray}
 for sufficiently small $(X_{1}(1),X_{2}(1), ... )$ but we consider all orders in $X(t)$ beyond the linear approximation. By a time 
independent unitary transformation, which does not induce 
an extra geometric term, the first term in \eqref{effective Hamitonian1} is diagonalized; we thus assume that the first term is already diagonal for simplicity in the following.

In the present approximation, essentially the four dimensional 
sub-space of the parameter space is relevant, and after a 
suitable re-definition of the parameters by taking
combinations of  $X_{k}(t)$, we write the matrix \eqref{effective Hamitonian1} as~\cite{Berry}
\begin{eqnarray}\label{effective Hamitonian2}
h(X(t))
&=&\left(\begin{array}{cc}
            E(0)+p_{0}(t)&0\\
            0&E(0)+p_{0}(t)
            \end{array}\right)
        -\mu \sigma^{l}p_{l}(t)\nonumber\\
\end{eqnarray}
where $\sigma^{l}$ stands for the Pauli matrices, and $\mu$ is a 
suitable coupling constant; we assume the level crossing at $X_{k}=0$ and thus the diagonal energy eigenvalues are degenerate. The choice of the signature of $\mu$ is arbitrary. We choose the form of the effective Hamiltonian \eqref{effective Hamitonian2} in analogy with a dipole moment placed in a magnetic field in Appendix B. This parametrization in 
terms of the variables $p_{l}$ is in fact valid beyond the linear 
approximation and thus our analysis is generic for a two-band problem.  But the two-band approximation itself is expected to 
be accurate only near the level crossing point.

\subsubsection{Weyl-type fermion}
 
 The essence of this model is described by an effective  Hamiltonian \footnote{Berry's phase in a realistic two-band model in condensed matter physics has been analyzed in great detail by Nagaosa \cite{Nagaosa}. His result agrees with our analysis of an idealized two-band model in the precise adiabatic limit. }, which is generally called a Weyl fermion in the Brillouin zone,
\begin{eqnarray}\label{level crossing Hamiltonian}
H=-\mu \vec{p}(t)\cdot \vec{\sigma}
\end{eqnarray}
where $\vec{\sigma}$ stands for the pseudo spin that describes the upper and lower crossing bands with the slowly varying  angular variables $\vec{p}(t)$ ; the naive band-crossing takes place at $\vec{p}(t)=0$. 
We are going to explicitly analyze the behavior of  Berry's phase at the two limiting cases, at adiabatic (slowly varying angular freedom) limit and non-adiabatic (fast varying angular freedom)  limit.

We shall follow the procedure of Stone~\cite{Stone} who initiated the general formulation of Berry's phase of a Weyl-type  fermion.
We thus start with  the Schr\"{o}dinger equation with \eqref{level crossing Hamiltonian}
\begin{eqnarray}\label{Schroedinger equation}
i\hbar \partial_{t}\psi(t)=H\psi(t)
\end{eqnarray}
 or the Lagrangian (in the spirit of the second quantization) given by
\begin{eqnarray}\label{Lagrangian}
L=\psi^{\dagger}(t)[i\hbar \partial_{t}-H]\psi(t)
\end{eqnarray}
where the two-component spinor $\psi(t)$ specifies the movement of upper and lower levels which appear in the band-crossing problem. In the present context, the {\em fast variables} are given by the spin freedom $\psi(t)$ characterized by the energy scale $\mu|\vec{p}(t)|$, and the slower variables are given by the angular freedom of $\vec{p}(t)$. For simplicity, we assume the magnitude $|\vec{p}(t)|$ to be time independent.
We then perform
a time-dependent unitary transformation
\begin{eqnarray}\label{unitary1}
\psi(t)= U(\vec{p}(t))\psi^{\prime}(t),\ \ 
\psi^{\dagger}(t)={\psi^{\prime}}^{\dagger}(t) 
U^{\dagger}(\vec{p}(t))
\end{eqnarray}
with
\begin{eqnarray}
U(\vec{p}(t))^{\dagger}\mu\vec{p}(t)\cdot\vec{\sigma}
U(\vec{p}(t))
=\mu|\vec{p}|\sigma_{3}.
\end{eqnarray} 
This unitary transformation is explicitly given by a $2\times2$ matrix
$U(\vec{p}(t))=\left(
             v_{+}(\vec{p})\  v_{-}(\vec{p}) \right)$,
where
\begin{eqnarray}
v_{+}(\vec{p})=\left(\begin{array}{c}
            \cos\frac{\theta}{2}e^{-i\varphi}\\
            \sin\frac{\theta}{2}
            \end{array}\right), \ \ \ 
v_{-}(\vec{p})=\left(\begin{array}{c}
            \sin\frac{\theta}{2}e^{-i\varphi}\\
            -\cos\frac{\theta}{2}
            \end{array}\right)
\end{eqnarray}
which correspond to the  use of {\em instantaneous eigenfunctions} of the operator $\mu\vec {p}(t)\cdot\vec{\sigma}$, namely, $\mu\vec {p}(t)\cdot\vec{\sigma}v_{\pm}(\vec{p})=\pm \mu|\vec{p}|v_{\pm}(\vec{p})$, where
\begin{eqnarray}
\vec {p}(t)=|\vec{p}|(\sin\theta\cos\varphi, \sin\theta\sin\varphi, \cos\theta)
\end{eqnarray}
 with time dependent $\theta(t)$ and $\varphi(t)$.

Based on this transformation, 
 the equivalence of two 
Lagrangians is derived, namely, $L$ in \eqref{Lagrangian}
and 
\begin{eqnarray}
L^{\prime}=
{\psi^{\prime}}^{\dagger}[i\hbar\partial_{t}
+\mu|\vec{p}|\sigma_{3}+U(\vec{p}(t))^{\dagger}
i\hbar\partial_{t}U(\vec{p}(t))]\psi^{\prime}.
\end{eqnarray}
 The starting Hamiltonian \eqref{level crossing Hamiltonian} is thus replaced by (using the argument of path integral \cite{Stone})
\begin{eqnarray}\label{geometric-phase}
H^{\prime}(t)&=&
-\mu|\vec{p}|\sigma_{3}- U(\vec{p}(t))^{\dagger}
i\hbar\partial_{t}U(\vec{p}(t))\nonumber\\
&=& -\mu|\vec{p}|\sigma_{3} -\hbar\left(\begin{array}{cc}
\frac{(1+\cos\theta)\dot{\varphi}}{2}&\frac{\dot{\varphi}\sin\theta+i\dot{\theta}}{2}\\
            \frac{\dot{\varphi}\sin\theta-i\dot{\theta}}{2}&
\frac{(1-\cos\theta)\dot{\varphi}}{2}
            \end{array}\right).
\end{eqnarray}
In the next section (see \eqref{fundamental condition}), we give a general criterion of the adiabatic condition for a multi-level system
\begin{eqnarray}\label{adiabatic condition-2}
|E_{n\pm1}(\overline{P}) -E_{n}(\overline{P})|\gg 2\pi\hbar/T
\end{eqnarray}
where $T$ stands for the time scale of the slower system (in the present two-level system, we adopt $\dot{\varphi} \sim 2\pi/T$). In the present case, one may choose  
 $2\mu|\vec{p}|= |E_{n\pm1}(\overline{P}) -E_{n}(\overline{P})|$ in the adiabatic condition.
In the adiabatic limit \eqref{adiabatic condition-2}
\begin{eqnarray}\label{adiabatic condition in band-crossing}
2\mu|\vec{p}| \gg 2\pi\hbar/T,
\end{eqnarray}
one thus has 
\begin{eqnarray}\label{adiabatic Stone phase}
H^{\prime}_{ad} \simeq  -\mu|\vec{p}|\sigma_{3} -\hbar\left(\begin{array}{cc}
\frac{(1+\cos\theta)\dot{\varphi}}{2}&0\\
            0&
\frac{(1-\cos\theta)\dot{\varphi}}{2}
            \end{array}\right).
\end{eqnarray}
If $T$ is sufficiently large $2\mu|\vec{p}| \gg 2\pi\hbar/T$,
one may neglect the off-diagonal parts in \eqref{geometric-phase} and retain only the diagonal components as in \eqref{adiabatic Stone phase}; this is the common assumption of the {\em adiabatic} (slow motion) limit \cite{Born, Kato}.

Stone then finds
 that the adiabatic Berry's phase for the $++$ component which has the lower energy~\cite{Stone}
\begin{eqnarray}\label{adiabatic Stone phase2}
&&\exp\left[ -i/\hbar \oint {H^{\prime}}^{(++)}_{ad}dt \right] \nonumber\\
&=&\exp \left[ i\mu/\hbar \int_{0}^{T}|\vec{p}|dt+i \int_{0}^{T} \frac{(1+\cos\theta)}{2}d\varphi \right],
\end{eqnarray}
namely, the monopole-like flux \footnote{This monopole-like term has been suggested as an analogue of the Wess-Zumino term by Stone \cite{Stone}, but this term is specific to each energy level separately, while chiral anomaly is more globally defined and valid for all the levels uniformly as was emphasized in \cite{Fujikawa-2006}.}
\begin{eqnarray}
\Omega_{mono}=\hbar\int_{0}^{T} \frac{(1+\cos\theta)}{2}d\varphi.
\end{eqnarray}
The monopole-like potential 
\begin{eqnarray}\label{adiabatic Stone phase3}
{\cal A}_{\varphi}=\hbar\frac{(1+\cos\theta)}{2|\vec{p}|\sin\theta} =\frac{e_{M}}{4\pi|\vec{p}|\sin\theta}(1+\cos\theta)
\end{eqnarray}
with ${\cal A}_{\theta}={\cal A}_{p}=0$ and $e_{M}=2\pi\hbar$ and regular in the {\em lower hemisphere}, 
is generated by
$H=-\mu \vec{p}(t)\cdot \vec{\sigma}$  in \eqref{level crossing Hamiltonian} which describes the two-level crossing. 
Berry's phase is also written in a common form
\begin{eqnarray}
\int_{0}^{T}{\cal A}_{k}(p)\dot{p}_{k}dt=\int_{0}^{T}v^{\dagger}_{+}(p)i\hbar\partial_{t}v_{+}(p)dt
\end{eqnarray}
in the adiabatic limit.
One can also write it for the upper hemisphere
\begin{eqnarray}\label{adiabatic Stone phase4}
{\cal A}_{\varphi}=-\frac{e_{M}}{4\pi|\vec{p}|\sin\theta}(1-\cos\theta)
\end{eqnarray}
after a gauge transformation of Wu and Yang \cite{Wu-Yang}.
 We note that  the magnetic charge of the monopole-like potential \eqref{adiabatic Stone phase3} is given by 
\begin{eqnarray}\label{monopole charge}
e_{M}=2\pi\hbar
\end{eqnarray}
which shows that Berry's phase in the present context is a quantum effect.

To the opposite to the adiabatic limit, one  may consider the limit where 
 $\hbar$ times the frequency of the angular freedom of $\vec{p}(t)$, estimated by $2\pi\hbar/T$, is much larger than the magnitude of the level crossing interaction (i.e., level spacing) $2\mu|\vec{p}|$ or if the particle approaches the monopole position $|\vec{p}|\rightarrow 0$ for any finite $T$ in the relation \eqref{geometric-phase}; this is defined as the {\em non-adiabatic limit}.  Then
\begin{eqnarray}\label{non-adiabatic condition}
2\mu|\vec{p}| \ll 2\pi\hbar/T,
\end{eqnarray}
and the term with $\dot\varphi\sim 2\pi\hbar/T$, which is linear in the time derivative, dominates the energy splitting  $\mu|\vec{p}|\sigma_{3}$ term. 

To see the implications of the non-adiabatic condition \eqref{non-adiabatic condition} explicitly, it is convenient to perform a further unitary transformation of the 
fermionic  variable
\begin{eqnarray}
\psi^{\prime}(t)= U(\theta(t))\psi^{\prime\prime}(t), \ \ \  
{\psi^{\prime}(t)}^{\dagger}=
{\psi^{\prime\prime}}^{\dagger}(t) 
U^{\dagger}(\theta(t))
\end{eqnarray}
with~\cite{deguchi}
\begin{eqnarray}\label{unitary2}
U(\theta(t))=\left(\begin{array}{cc}
            \cos\frac{\theta}{2}&-\sin\frac{\theta}{2}\\
            \sin\frac{\theta}{2}&\cos\frac{\theta}{2}
            \end{array}\right)
\end{eqnarray}
in addition to \eqref{unitary1}, which diagonalizes the dominant Berry's phase term.
The Hamiltonian \eqref{geometric-phase} then becomes
\begin{eqnarray}\label{exact non-adiabatic}
H^{\prime\prime}(t)
&=&
-\mu|\vec{p}| U(\theta(t))^{\dagger}
\sigma_{3}U(\theta(t))\nonumber\\
&&+ (U(\vec{p}(t))U(\theta(t)))^{\dagger}
\frac{\hbar}{i}\partial_{t}(U(\vec{p}(t))U(\theta(t)))
\nonumber\\
&=&
-\mu|\vec{p}| \left(\begin{array}{cc}
            \cos\theta&-\sin\theta\\
            -\sin\theta&-\cos\theta
            \end{array}\right)
-\hbar\left(\begin{array}{cc}
            \dot{\varphi}&0\\
            0&0
            \end{array}\right).
\end{eqnarray}
Note that the first term is bounded by $\mu|\vec{p}|$ and the second term is dominant for the non-adiabatic case $2\mu|\vec{p}| \ll 2\pi\hbar/T$. We emphasize that both \eqref{geometric-phase} and \eqref{exact non-adiabatic} are {\em exact} expressions (in the framework of instantaneous solutions).  

The Hamiltonian (in the non-adiabatic limit) by ignoring small off-diagonal terms then becomes
\begin{eqnarray}\label{non-adiabatic Stone phase}
H^{\prime\prime}_{\rm nonad} \simeq
-\mu|\vec{p}| \left(\begin{array}{cc}
            \cos\theta&0\\
            0&-\cos\theta
            \end{array}\right)
-\hbar\left(\begin{array}{cc}
            \dot{\varphi}&0\\
            0&0
            \end{array}\right).
\end{eqnarray}
The topological Berry's phase given by the second term thus either vanishes or becomes trivial in the non-adiabatic limit 
\begin{eqnarray}\label{trivial phase2}
\exp\left\{ \frac{i}{\hbar}\oint {\cal A}_{\varphi}|\vec{p}|\sin\theta \dot{\varphi}dt \right\}=\exp\left\{i\oint \dot{\varphi}dt \right\}=\exp\{2\pi i\}=1
\end{eqnarray}
independently of $\theta$ for the very rapid movement of the angular freedom of $\vec{p}(t)$, namely, $T\rightarrow 0$ with fixed $\mu|\vec{p}|$, or very close to the monopole position, namely, $\mu|\vec{p}|\rightarrow 0$ with fixed $T$.
Note that the trivial phase \eqref{trivial phase2} is equivalent to the vanishing Berry's curvature, or corresponds to the pure gauge term of a monopole in the sense of Wu and Yang.

It is crucial to examine the adiabatic limit of the fast system, namely, the Weyl term, in the two-level system to identify the adiabatic Berry's phase by discarding  terms linear in the time derivatives; one thus finds the conventional Berry's phase as in \eqref{adiabatic Stone phase}. On the other hand, the commutation relations of slower variables are determined with the BJL prescription 
in the non-adiabatic limit of the slower variables (see Appendix A), including all the terms linear in the time derivatives  
of Berry's phase such as \eqref{exact non-adiabatic}. 
 This is also the case with the canonical prescription of commutation relations by using  the symplectic forms including  all the terms linear in the time derivatives. The crucial fact to be noted  is that the adiabatic Berry's phase and the commutation relations of the slower system are determined by the very different {\em two aspects} of the action.

In the two-level crossing model of a Weyl fermion, it is possible to characterize these two different aspects as a characterization of the generic Berry's phase.
 We have shown that the monopole-like object associated with generic Berry's phase appearing in the two-level crossing is parameterized by 
\begin{eqnarray}\label{parameter eta}
\eta=\frac{\mu|\vec{p}|T}{\pi\hbar}=\frac{\mu|\vec{p}|}{\pi\hbar/T} = \frac{{\rm static\ energy}}{{\rm kinetic\ energy}},
\end{eqnarray}
namely, the {\em functional form} of Berry's phase depends on $\eta$ in a symbolic notation~\footnote{This general parametrization of Berry's phase \eqref{Berry connection} is shown to be valid for the case of an exactly solvable model of Berry's phase of a two-level crossing problem, which is explained in Appendix B; there one replaces $\vec{p}(t)\rightarrow \vec{B}(t)$.},
\begin{eqnarray}\label{Berry connection}
{\cal A}_{k}(\vec{p}; \eta)\dot{p}_{k} = {\cal A}_{k}(\vec{p}; \mu|\vec{p}|T/\pi\hbar)\dot{p}_{k} = {\cal A}_{k}(\vec{p}; 2\mu|\vec{p}|/\hbar\omega)\dot{p}_{k}.
\end{eqnarray}
The parameter $T=2\pi/\omega$ stands for the period of a closed motion of the angular freedom of $\vec{p}(t)$ to define a non-trivial Berry's phase uniquely, as already explained. 

The value $\eta=\infty$ corresponds to the adiabatic limit where the conventional Berry's phase is well-defined, and the value of the parameter $\eta$ 
\begin{eqnarray}
\eta \simeq 1, \ \ {\rm namely,} \ \ \mu|\vec{p}| \simeq \frac{\pi\hbar}{T}
\end{eqnarray}
is the domain where the adiabatic approximation starts to break down; one can see this fact in \eqref{geometric-phase} where all the terms there become approximately equal in magnitude and the notion of adiabatic approximation is not applicable. The value $\eta\rightarrow 0$ corresponds to the non-adiabatic limit where Berry's phase becomes trivial and no monopole appears.

\subsubsection{Some technical remarks}
Berry's phase arises from the adiabatic theorem of  quantum mechanics which is controlled by the canonical quantum commutation relations. Berry's phase does not control the quantum mechanical properties. In this viewpoint, the conventional treatment of Berry's phase emphasizes too much  the quantum mechanical properties implied by  Berry's phase at a specific parameter domain, i.e., adiabatic domain. In general, the time derivative terms such as $\dot{\varphi}$ in \eqref{geometric-phase} determine the topological properties, and at the same time the dynamical canonical properties are determined by the large frequency limit of those parameters, as in the BJL prescription.
In this sense, the totality of the domains covered by changing the parameters $\eta$ in \eqref{parameter eta}, as is analyzed in the previous subsection, is crucial to analyze the reliable quantum and topological properties of Berry's phase. This aspect is emphasized in Appendix B in the analysis of magnetic monopoles and the Longuet-Higgins phase change rule. A careful analysis of the exactly solvable model of Berry's phase, which assumes a Dirac monopole form  in the adiabatic limit, shows that Berry's phase becomes trivial in the non-adiabatic parameter domain as is shown in Appendix B; this helps maintain the consistency of Berry's phase with the principle of quantum mechanics in the form of canonical commutation relations \cite{Deguchi-Fujikawa-AP}. 

As a specific example in the adiabatic limit $\eta\rightarrow\infty$, Berry's phase \eqref{Berry connection} gives the monopole-like
phase as in \eqref{adiabatic Stone phase3}, while in the non-adiabatic limit $\eta\rightarrow 0$, Berry's phase becomes trivial
(i.e., curvature vanishes) as in \eqref{trivial phase2}.  This latter property shows that Berry's phase has no singularity at the origin $|\vec{p}|\rightarrow 0$ with fixed $T$, although the common assumption of a  genuine Dirac monopole-form (of Berry's phase) would suggest such a singularity.  Rather, one can confirm that the monopole curvature vanishes there.  See also subsection 3.3 for an implication of the Weyl-fermion model.

\section{Berry's phase and the principle of quantum mechanics}

\subsection{Anomalous Hall effect and anomalous commutation relations}
 
The applications  of Berry's phase to the analyses of the anomalous Hall effect \cite{Karplus, Fang} and the spin Hall effect \cite{Hirsch} appears to have opened a new research area \cite{Nagaosa-review, Niu-review}, namely, the non-commutative geometry $[x^{k}(t),x^{l}(t)] = O(\hbar^{2})$ and a dynamical system which is not quantized in a conventional manner.  
 The main purpose of this section is to clarify the mathematical aspects related to the anomalous commutation relations induced by  
Berry's phase.

The effective equations of motion, which 
incorporate Berry's phase  near the level crossing point and the electromagnetic vector potential $eA_{k}(x)$, are customarily adopted as \cite{Niu}
\begin{eqnarray}\label{semi-classical equation}
\dot{x}_{k}=-\Omega_{kl}(\vec{p})\dot{p}_{l} +\frac{\partial \epsilon_{n}(\vec{p})}{\partial p_{k}}, \ \
\dot{p}_{k}=-eF_{kl}(\vec{x})\dot{x}_{l} + e\frac{\partial}{\partial x_{k}}\phi(\vec{x})
\end{eqnarray}
by adding the adiabatic Berry's phase $\Omega_{kl}\dot{p}_{l}$ to the equations of motion as an extra induced term. Here $\epsilon_{n}(\vec{p})$ essentially  stands for the $n$-th energy level in the band structure. These effective equations that incorporate the adiabatic Berry's phase, which originally  appears as a phase of the wave function of the $n$-th level and is of order $\hbar$, are used to analyze the intra-band phenomena.   The magnetic flux $\Omega_{kl}(\vec{p})$ of Berry's phase, which is  assumed to be a {\em genuine Dirac monopole form}~\cite{Dirac} for the moment independently of other parameters of the model, and the electromagnetic tensor $F_{kl}$ are defined by  
\begin{eqnarray}
\Omega_{kl}=\frac{\partial}{\partial p_{k}}{\cal A}_{l}-\frac{\partial}{\partial p_{l}}{\cal A}_{k}, \ \  \ F_{kl}=\frac{\partial}{\partial x_{k}}A_{l}-\frac{\partial}{\partial x_{l}} A_{k},
\end{eqnarray}
respectively. The monopole is located at the origin of momentum space, i.e., at the level crossing point in the adiabatic level-crossing problem, but it is now assumed to be a genuine particle in the momentum space.  Here we defined $p_{l}=\hbar k_{l}$ to write all the equations in terms of $p_{l}$ 
compared to the notation in \cite{Niu}, to keep track of the $\hbar$ factor in a transparent way.

Duval et al.~\cite{Duval,Duval2} have shown that the equations of motion  
\eqref{semi-classical equation} are derived from the  effective action 
\begin{eqnarray}\label{action-1}
S=\int dt\left[ p_{k}\dot{x}_{k} - eA_{k}(\vec{x})\dot{x}_{k} + {\cal A}_{k}(\vec{p})\dot{p}_{k}-\epsilon_{n}(\vec{p})+e\phi(\vec{x})\right],
\end{eqnarray}  
 {\em without using any commutation relations}; a simplified version of this action was used in \cite{Niu}.  They have shown that the action \eqref{action-1} with Berry's phase included belongs to a {\em non-canonical system} \cite{Niu}, namely, not quantized in a conventional manner. It is surprising that the supposed to be canonical system, \eqref{action-1}, obtained  by considering the adiabatic limit of the conventional canonical system as is shown in the next subsection,  is converted to a non-canonical system. Without asking the reason why a non-canonical system is obtained for the moment, it is natural to regard the basic equation \eqref{semi-classical equation} as classical and study its implications.

We are going to study the implications of the classical action \eqref{action-1}.  Duval et al. inverted a symplectic matrix defined by the action in the extended phase space formalism \cite{Faddeev-Jackiw}. They then derived the anomalous Poisson brackets induced by the genuine monopole curvature $\Omega_{kl}=\epsilon^{klm}\Omega_{m}$,
\begin{eqnarray}\label{Poisson bracket} 
&&\{x_{k},x_{l}\}=\frac{\epsilon^{klm}\Omega_{m}}{1+e\vec{B}\cdot\vec{\Omega}} , \ \ \ \{p_{k},x_{l}\}=-\frac{\delta_{kl}+e\Omega_{k}B_{l}}{1+e\vec{B}\cdot\vec{\Omega}},\nonumber\\
&&\{p_{k},p_{l}\}=- \frac{\epsilon^{klm}eB_{m}}{1+e\vec{B}\cdot\vec{\Omega}},
\end{eqnarray}
where the factors containing $\Omega_{k}$ are anomalous \cite{Duval,Duval2}.
  
This action \eqref{action-1} was later analyzed using the Bjorken-Johnson-Low (BJL) prescription~\cite{BJL}, which is summarized in Appendix A,  and a path integral formalism to confirm the above derivation. 
The path integral analysis used in~\cite{fujikawa-prd2018} is based on the approximate quadratic expansion of the action \eqref{action-1} around a classical solution in the phase space $(\vec{x}_{(0)},\vec{p}_{(0)})$ by replacing $(\vec{x},\vec{p})\rightarrow (\vec{x}_{(0)},\vec{p}_{(0)})+(\vec{x},\vec{p})$,
\begin{align}\label{action2}
S=\int dt \Big[ p_{k}\dot{x}_{k} -\frac{e}{2}F_{lk}(\vec{x}_{(0)})x_{l}\dot{x}_{k} + \frac{1}{2}\Omega_{lk}(\vec{p}_{(0)})p_{l}\dot{p}_{k}
 -\frac{\vec{p}^{2}}{2m}+\frac{e}{2}\partial_{k}\partial_{l}\phi(\vec{x}_{(0)})x_{k}x_{l} \Big].
\end{align}
To be definite we choose the kinetic energy term $\epsilon_{n}=\vec{p}^{2}/2m$. The last two terms in the action are not important to define the commutation relations as is explained in Appendix A and we choose convenient ones for actual calculations.
The commutators are given by the path integral and the BJL prescription  (on the understanding that $kl$ matrix element on the right-hand sides is taken) \footnote{The results \eqref{commutator-1} agree with Poisson brackets in \eqref{Poisson bracket} (to the accuracy of the approximation), if one uses an identity $(\frac{1}{1-e\Omega F})_{kl}=(\delta_{kl}+eB_{k}\Omega_{l})/(1+e(B_{m}\Omega_{m}))$ which is valid when one defines $F_{kl}=\epsilon^{klm}B_{m}$ and $\Omega_{kl}=\epsilon^{klm}\Omega_{m}$. Similarly, $(\frac{1}{1-eF\Omega})_{kl}=(\delta_{kl}+e\Omega_{k}B_{l})/(1+e(B_{m}\Omega_{m}))$.}
\begin{eqnarray}\label{commutator-1}
&&[x_{k},x_{l}]=i\hbar\frac{1}{1-e\Omega F}\Omega, \ \ \ [p_{k},x_{l}]=-i\hbar\frac{1}{1-eF\Omega},\nonumber\\
&&[p_{k},p_{l}]=-i\hbar eF\frac{1}{1-e\Omega F}.
\end{eqnarray}
These commutation relations are expected to be generic for the effective action \eqref{action-1} in the spirit of the background field method. 
These commutation relations \eqref{commutator-1}  
are consistent with the quantum interpretation of Poisson brackets \eqref{Poisson bracket} given by the modified canonical formalism.

We are later going to show that the action \eqref{action-1} is derived from the conventional (canonical) path integral formula by the adiabatic approximation, namely, by discarding the most terms linear in the time derivative, which are supposed to be small, in the action.  Thus the action \eqref{action-1} is an adiabatic action and not suitable for the use of canonical quantization in general. The exact commutation relations are determined by considering all the terms linear  in time derivative (for example, the symplectic forms consisting of terms linear in the time derivative in the canonical treatment), which may not be  important in the analysis of the adiabatic approximation and thus may be neglected in the derivation of the action \eqref{action-1}.

Our analysis of this issue given below 
 shall  mainly follow  the recent article \cite{Fujikawa-Umetsu-2022}.

\subsection{ Consistency of  Berry's phase with the principle of quantum mechanics}
\subsubsection{Derivation of the action of anomalous Hall effect}
We use the path integral formulation of Berry's phase discussed in Section 2. We adopt the Hamiltonian of the form $H=H_{0} + H_{1}$ assuming that the slower particle is charged with $q=-e$ ($e>0$) and the fast particle is neutral for simplicity,
\begin{eqnarray}\label{starting Hamiltonian2}
H_{0}(X, P+ eA(X))&=&\frac{1}{2M}(P_{k} + eA_{k}(X))^{2} -e\phi(X),\nonumber\\
H_{1}(x, p; P+ eA(X))&=&H_{1}(x_{k}, p_{k} ; P_{k} + eA_{k}(X))
\end{eqnarray}
with the canonical quantization of fast variables
 \begin{eqnarray}\label{standard commutator-0}
 [p_{k}, x_{l}]=\frac{\hbar}{i}\delta_{kl}, \ \ [p_{k}, p_{l}]=0, \ \ [x_{k}, x_{l}]=0
 \end{eqnarray}
 and  the canonical quantization of slower variables
 \begin{eqnarray}\label{standard commutator-2}
 [P_{k}, X_{l}]=\frac{\hbar}{i}\delta_{kl}, \ \ [P_{k}, P_{l}]=0, \ \ [X_{k}, X_{l}]=0
 \end{eqnarray}
by treating $H_{0}(X, P + eA(X))$ as the slower system; from now on, we use the capital characters $X_{k}$ and $P_{k}$ for slower variables without stated otherwise; as for lower case letters, they are used for slower variables also when we refer to equations appearing in the past works.  The use of covariant derivative, $P_{k} + eA_{k}(X)$,  ensures the electromagnetic gauge invariance. We discuss the case of a single fast particle and a single slow particle, for simplicity, although our use of the second quantization can cover a slightly more general case~\footnote{The general proof of the adiabatic theorem by T. Kato \cite{Kato} includes the case of degenerate states. We can incorporate the case with the degeneracy of $n$-states of the fast system using an internal $U(n)$ symmetry for a single particle. See also \cite{Wilczek}.}.
We start with the fundamental path integral
\begin{align}\label{starting path integral}
&Z
=\int {\cal D}\overline{P}_{k}{\cal D}X_{k}\exp \left\{ \frac{i}{\hbar}\int_{0}^{T} dt \left[(\overline{P}_{k}(t)-eA_{k}(X(t)))\dot{X}_{k}(t)
- H_{0}(X(t), \overline{P}(t)) \right] \right\}\notag\\
&\times\int {\cal D}\psi^{\star}{\cal D}\psi\exp \left\{\frac{i}{\hbar}\int_{0}^{T} dt\int d^{3}x \left[\psi(t,\vec{x})^{\star}i\hbar \partial_{t}\psi(t,\vec{x}) - \psi(t,\vec{x})^{\star}H_{1}\left(\frac{\hbar}{i}\vec{\nabla}, \vec{x}; \overline{P}(t) \right)\psi(t,\vec{x}) \right] \right\} 
\end{align}
with 
\begin{eqnarray}
\overline{P}_{k}(t)=P_{k}(t) + eA_{k}(X(t))
\end{eqnarray}
where we used the invariance of the path integral measure
${\cal D}P_{k}{\cal D}X_{k}={\cal D}\overline{P}_{k}{\cal D}X_{k}$.  
We use the second quantization scheme for the fast system, which is convenient to analyze the level crossing and Berry's phase as was demonstrated in Section 2.
It is confirmed that the BJL prescription (or canonical consideration) reproduces the canonical commutation relations  \eqref{standard commutator-2} and 
\begin{eqnarray}
[\psi(t,\vec{x}), \psi^{\dagger}(t,\vec{y})]=\delta(\vec{x}-\vec{y})
\end{eqnarray}
in the path integral formula of \eqref{starting path integral}.
We expand the field variable $\psi(t,\vec{x})$ into a complete set of orthonormal bases $\{\phi_{k}(\vec{x}; \overline{P}(t))\}$ defined by 
\begin{eqnarray}\label{instantaneous mode expansion}
&&H_{1}\left(\frac{\hbar}{i}\vec{\nabla}, \vec{x}; \overline{P}(t) \right)\phi_{n}(\vec{x}; \overline{P}(t))
=E_{n}(\overline{P}(t))\phi_{n}(\vec{x}; \overline{P}(t)),\nonumber\\
&&\psi(t,\vec{x})=\sum_{n}a_{n}(t)\phi_{n}(\vec{x}; \overline{P}(t)).
\end{eqnarray}
 We then have 
\begin{align}
Z
 =&\int {\cal D}\overline{P}_{k}{\cal D}X_{k}\exp \left\{\frac{i}{\hbar}\int_{0}^{T} dt \left[(\overline{P}_{k}(t)-eA_{k}(X(t)))\dot{X}_{k}(t)
- H_{0}(X(t), \overline{P}(t)) \right] \right\}\nonumber\\
&\times\int \Pi_{n}{\cal D}a^{\star}_{n}{\cal D}a_{n}\exp \left\{ \frac{i}{\hbar}\int_{0}^{T} dt\sum_{n} \left[ a_{n}^{\star}(t)i\hbar \partial_{t}a_{n}(t)- E_{n}(\overline{P}(t)) a_{n}^{\star}(t)a_{n}(t) \right] \right\}\nonumber\\
&\hspace{3cm} \times \exp \left\{\frac{i}{\hbar}\int_{0}^{T} dt \sum_{n,l}\langle n,t|i\hbar\partial_{t}|l,t\rangle a_{n}^{\star}(t)a_{l}(t) \right\}\nonumber\\
\simeq&\int {\cal D}\overline{P}_{k}{\cal D}X_{k}\exp\left\{ \frac{i}{\hbar}\int_{0}^{T} dt \left[ (\overline{P}_{k}-eA_{k}(X))\dot{X}_{k}
- H_{0}(X, \overline{P}) \right] \right\}\nonumber\\
&\times\int \Pi_{n}{\cal D}a^{\star}_{n}{\cal D}a_{n}\exp \left\{\frac{i}{\hbar}\int_{0}^{T} dt\sum_{n} \left[ a_{n}^{\star}(t)i\hbar \partial_{t}a_{n}(t)- \left(E_{n}(\overline{P}) -{\cal A}^{(n)}_{k}(\overline{P})\dot{\overline{P}^{k}}\right)a_{n}^{\star}(t)a_{n}(t) \right] \right\}\label{final result of second quantization}
\end{align} 
where we defined
\begin{eqnarray}\label{diagonal approximation}
\langle n,t|i\hbar\partial_{t}|l,t\rangle &=& \int d^{3}x\phi^{\dagger}_{n}(\vec{x}; \overline{P}(t))i\hbar\partial_{t}\phi_{l}(\vec{x}; \overline{P}(t)),\nonumber\\
{\cal A}^{(n)}_{k}(\overline{P})\dot{\overline{P}^{k}}(t) &\equiv& \langle n,t|i\hbar\partial_{t}|n,t\rangle\nonumber\\
&=& \int d^{3}x\phi^{\dagger}_{n}(\vec{x}; \overline{P}(t))i\hbar\frac{\partial}{\partial \overline{P}^{k}(t)}\phi_{n}(\vec{x}; \overline{P}(t))\dot{\overline{P}^{k}}(t).
\end{eqnarray} 
The diagonal element ${\cal A}^{(n)}_{k}(\overline{P})\dot{\overline{P}^{k}}$ is commonly called Berry's phase. The assumption of the diagonal dominance, which was used in the last step of the above path integral formula, corresponds to the {\em adiabatic approximation} and it is valid only for 
\begin{eqnarray}\label{fundamental condition}
 |E_{n\pm1}(\overline{P}) -E_{n}(\overline{P})| \gg \hbar \frac{2\pi}{T_{s}}
\end{eqnarray}
where $T_{s}$ stands for the typical time scale of the slower system; usually $T_{s}$ is estimated by the period of the  slowly varying variable $\overline{P}_{k}(t)$, $\overline{P}_{k}(0)=\overline{P}_{k}(T_{s})$. We understand that $\hbar \frac{2\pi}{T_{s}}$ stands for the typical energy scale contained in $\overline{P}_{k}(t)$. (We use $T_{s}$ for the period of the slower dynamical system to distinguish it from the T-ordering operation whenever necessary.) It is convenient to choose the upper-bound of the time integral in the path integral \eqref{final result of second quantization} to agree with this period $T_{s}$. This condition means that the  typical energy of the slower system is much smaller than the level  spacing of the fast system.
If this adiabaticity condition is not satisfied, one needs to retain the off-diagonal transition elements $\langle n,t|i\hbar\partial_{t}|l,t\rangle$ in \eqref{final result of second quantization}, and the path integral is reduced to the starting expression \eqref{starting path integral}  after summing over $n$ and $l$.  This shows that one would recover the starting universal canonical commutation relations for the slower variables \eqref{standard commutator-2} if one evaluates the fast system exactly without using the adiabatic approximation. 
Similarly, no anomalous Poisson brackets for slower variables are generated without the adiabatic approximation in the fast variables.

The slower system standing on the specific n-th level 
\begin{eqnarray}
a_{n}^{\dagger}(0)|0\rangle
\end{eqnarray}
 of the fast system, namely, the fast system being constrained to the n-th level,  is described by the path integral derived from \eqref{final result of second quantization}~\footnote{The conversion of the path integral to the time-evolution operator for the variables $\{a_{n}, a_{n}^{\dagger}\}$ is 
\begin{eqnarray}
&&\int \Pi_{n}{\cal D}a^{\star}_{n}{\cal D}a_{n}\exp\Bigg\{\frac{i}{\hbar}\int_{0}^{T} dt \Bigg[\sum_{n} a_{n}(t)^{\star}i\hbar \partial_{t}a_{n}(t)\nonumber\\
&&\hspace{4cm} - \left(E_{n}(\overline{P}(t)) -{\cal A}^{(n)}_{k}(\overline{P}(t))\dot{\overline{P}^{k}}(t)\right)a_{n}^{\star}(t)a_{n}(t) \Bigg] \Bigg\}\nonumber\\
&&\rightarrow \exp \left\{\sum_{n}\frac{-i}{\hbar}\int_{0}^{T}dt 
\left(E_{n}(\overline{P}(t)) -{\cal A}^{(n)}_{k}(\overline{P}(t))\dot{\overline{P}^{k}}(t)\right)a^{\dagger}_{n}(0)a_{n}(0)
\right\}\nonumber
\end{eqnarray}
by noting $a^{\dagger}_{n}(t)a_{n}(t)=a^{\dagger}_{n}(0)a_{n}(0)$.}
\begin{align}\label{final path integral formula}
Z_{n}
 =&\int {\cal D}\overline{P}_{k}{\cal D}X_{k}\exp\left\{ \frac{i}{\hbar}\int_{0}^{T} dt \left[(\overline{P}_{k}-eA_{k}(X))\dot{X}_{k}
- H_{0}(X, \overline{P})- (E_{n}(\overline{P}) -{\cal A}^{(n)}_{k}(\overline{P})\dot{\overline{P}_{k}}) \right] \right\}
\end{align} 
with $H_{0}(X, \overline{P})=\frac{1}{2M}\overline{P}_{k}^{2} -e\phi(X)$. This formula is valid under the crucial constraint \eqref{fundamental condition}.
The Lagrangian appearing in this path integral
\begin{eqnarray}\label{effective Lagrangian-n}
L_{n}=(\overline{P}_{k}-eA_{k}(X))\dot{X}_{k} +{\cal A}^{(n)}_{k}(\overline{P})\dot{\overline{P}_{k}}
- (\frac{1}{2M}\overline{P}_{k}^{2}+E_{n}(\overline{P})) + e\phi(X)
\end{eqnarray}
precisely  agrees with the common Lagrangian in condensed matter physics \eqref{action-1} if one identifies
\begin{eqnarray}
\overline{P}_{k}\rightarrow p_{k}, \ \ X_{k}\rightarrow x_{k},\ \ {\rm and}\ \ \epsilon_{n}(p)=\frac{1}{2M}\overline{P}_{k}^{2} + E_{n}(\overline{P}), 
 \end{eqnarray}
 and if one uses the adiabatic Berry's phase ${\cal A}^{(n)}_{k}$ associated with the specific $n$-th level of the fast system.  Our path integral formulation thus naturally reproduces the common formulas \eqref{action-1} in the precise adiabatic limit but with the adiabatic Berry's phase  ${\cal A}^{(n)}_{k}$ instead of a genuine Dirac monopole. The action appearing in \eqref{final path integral formula} is invariant under a gauge transformation of Berry's phase 
\begin{eqnarray}\label{gauge transformation of Berry's phase}
{\cal A}^{(n)}_{k}(\overline{P}) \rightarrow {\cal A}^{(n)}_{k}(\overline{P}) + \frac{\partial}{\partial \overline{P}_{k}} \omega^{(n)}(\overline{P})
\end{eqnarray}
with $\omega^{(n)}(\overline{P})$ for each ${\cal A}^{(n)}_{k}(\overline{P})$ if one chooses the periodic boundary condition $\overline{P}_{k}(0)=\overline{P}_{k}(T)$. This gauge invariance is a direct consequence of the adiabatic (diagonal) approximation in \eqref{diagonal approximation}, which is based on the crucial condition \eqref{fundamental condition}. This gauge invariance is thus a measure of the validity of the adiabatic approximation.

\subsubsection{Equations of motion without using commutation relations}
It has been shown that the slower system standing on the specific $n$-th level  $a_{n}^{\dagger}(0)|0\rangle$ of the fast system, namely, the slower system being constrained to the infinitesimal neighborhood of the $n$-th level of the fast system,  is (approximately) described by the path integral \eqref{final path integral formula} with the Lagrangian \eqref{effective Lagrangian-n}.
 It is important that this
path integral is an approximation in the following two senses: Firstly, it is an approximation since all the states other than $n$ and their mixings with the state $n$ have been neglected. Secondly, to justify the above truncation, the valid energy  domain of the path integral formula (for slower variables) is limited to the  neighborhood of the specific state $n$. If one goes outside this energy domain (and to the full dynamical domain), the above path integral formula with the given $L_{n}$ is not accurate. 

Besides these limitations, the Lagrangian \eqref{effective Lagrangian-n} is known that it does not satisfy the canonical properties \cite{Niu, Duval}  for $eA_{k}\neq0$.  
This implies that the separate  path integral \eqref{final path integral formula} needs to be treated with due care compared to the well-defined formula \eqref{final result of second quantization} before the adiabatic approximation.  
It may,  however,  be allowed to assume that the classical equations of motion, as an 
stationary condition $\delta S\left(\overline{P}, X\right)= 0$  in the validity domain of the action \eqref{fundamental condition}, is valid 
\begin{eqnarray}\label{semi-classical equation-2}
\dot{X}_{k}=-\Omega^{(n)}_{kl}(\vec{\overline{P}})\dot{\overline{P}}_{l} +\frac{\partial \epsilon_{n}(\vec{\overline{P}})}{\partial \overline{P}_{k}}, \ \
\dot{\overline{P}}_{k}=-eF_{kl}(\vec{X})\dot{X}_{l} + e\frac{\partial}{\partial X_{k}}\phi(\vec{X})
\end{eqnarray}
which gives a  meaning to the classical equations \eqref{semi-classical equation}. This set of semi-classical equations, which are independent of commutation relations, are assumed usually to be  valid in the {\em adiabatic limit}. 

For $\Omega^{(n)}_{kl}(\vec{\overline{P}})\dot{\overline{P}}_{l}=0$, this set of equations \eqref{semi-classical equation-2} satisfy the canonical conditions and thus can be promoted to the quantum equations.  It is also possible to realize the canonical system if one sets $eA_{k}(X)=0$, since the action is then written as (using $X_{k}=\overline{X}_{k}+ {\cal A}^{(n)}_{k}(\overline{P}))$
\begin{eqnarray}
S&=& \int dt \{\overline{P}_{k}\dot{X}_{k} +{\cal A}^{(n)}_{k}(\overline{P})\dot{\overline{P}_{k}}
- (\frac{1}{2M}\overline{P}_{k}^{2}+E_{n}(\overline{P})) + e\phi(X)\}\nonumber\\
&=& \int dt \{\overline{P}_{k}\dot{\overline{X}}_{k} 
- (\frac{1}{2M}\overline{P}_{k}^{2}+E_{n}(\overline{P})) + e\phi(\overline{X}_{k}+ {\cal A}^{(n)}_{k}(\overline{P}))\}
\end{eqnarray} 
and the last system defines a canonical Hamiltonian system with 
\begin{eqnarray}\label{gauge covariant Hamiltonian}
H&=& \int dt \{ (\frac{1}{2M}\overline{P}_{k}^{2}+E_{n}(\overline{P})) - e\phi(\overline{X}_{k}+ {\cal A}^{(n)}_{k}(\overline{P}))\},\nonumber\\
{[}\overline{P}_{k}, \overline{X}_{l}] &=&-i\hbar\delta_{kl}, \ [\overline{X}_{k}, \overline{X}_{l}]=0, \ [\overline{P}_{k}, \overline{P}_{l}]=0
\end{eqnarray} 
for a point-like monopole ${\cal A}^{(n)}_{k}(\overline{P})$ approximation.
One can thus describe the quantum mechanical equations for the anomalous Hall effect without the electromagnetic vector potential in the present approximation; the treatment of the vector-potential $eA_{k}(X)$ separately from the scalar potential $e\phi(X)$ is not very satisfactory, but it may physically be allowed. 

In our later analysis of the Born-Oppenheimer approximation, we show the {\em incompatibility} of Berry's connection, which is defined in a manner similar to the present Berry's phase, and the electromagnetic vector potential in the canonical Hamiltonian formalism. This may provide a clear physical explanation why the dynamical system \eqref{action-1} with non-vanishing electromagnetic potential $eA_{k}(X)\neq 0$ gives rise to a non-canonical classical system.

\subsubsection{Canonical commutation relations} 

The fundamental canonical commutation relations are always valid in the precise treatment without adiabatic approximations. This observation is natural and consistent with our basic path integral formulation \eqref{final result of second quantization}; if one sums over all the levels of the fast system, one comes back to the starting fundamental formula \eqref{starting path integral} for which one obviously recovers the basic canonical commutation relations for the slower variables.  The basic canonical commutation relations are valid exactly, which covers all the allowed energy range of the fast and slower systems. 
 It is important how one incorporates the above general observations into the understanding of the slower system
\begin{align}\label{effective Lagrangian-2}
L_{n}=\left(\overline{P}_{k}(t)-eA_{k}(X(t)) \right)\dot{X}_{k}(t) +{\cal A}^{(n)}_{k}(\overline{P})\dot{\overline{P}_{k}}
- \left( \frac{1}{2M}\overline{P}_{k}^{2}+E_{n}(\overline{P}) \right) + e\phi(X)
\end{align}
which  agrees with the common Lagrangian in condensed matter physics \eqref{action-1} after suitable re-interpretation. 

When one discusses the commutation relations, one may consider two main options: 

(i) One may presume the existence of a master Lagrangian which is valid for any energy domain, to be explicit, one may consider the Lagrangian appearing in \eqref{starting path integral} which satisfies the ordinary canonical commutation relations. In each adiabatic domain of the slower dynamical system, this master Lagrangian is represented by the effective Lagrangian \eqref{effective Lagrangian-2}
which  agrees with the common Lagrangian of the slower system in condensed matter physics \eqref{action-1} for each energy level of the fast system.
 The Lagrangian \eqref{effective Lagrangian-2} is defined in the adiabatic limit of movement by discarding most of the terms linear in the time derivative. Thus it is not used to quantize the slower variables by the conventional canonical analysis that is performed using the symplectic forms or the BJL prescription, which includes all the discarded 
 time-derivative terms.

Namely, a natural extension of the above Lagrangian to the non-adiabatic domain, where one defines the commutation relations by the conventional canonical analysis or the BJL procedure in a reliable way,  is to go back to the starting well-defined path integral formula \eqref{starting path integral} which gives rise to the normal canonical commutation relations 
\begin{eqnarray}\label{standard commutator-3}
 [P_{k}, X_{l}]=\frac{\hbar}{i}\delta_{kl}, \ \ [P_{k}, P_{l}]=0, \ \ [X_{k}, X_{l}]=0
 \end{eqnarray}
with 
$\overline{P}_{k}(t)=P_{k}(t) + e A_{k}(X(t))$.
The appearance of the adiabatic Berry's phase is a secondary effect of an adiabatic approximation in quantum mechanics, and thus Berry's phase would not modify the principle of quantum mechanics. 
This appears to be  logically consistent. The procedure described here is based on an assumption that the two aspects, namely, the adiabatic aspects of the anomalous Hall effect and the aspects of commutation relations are (approximately) described by the two very distinct aspects of the same theory.  In this option, the path integral \eqref{final path integral formula} is a formal object that may be used to derive the adiabatic equations of motion of the anomalous Hall effect \eqref{semi-classical equation} {\em approximately} by a classical action principle, which is done without using commutation relations. No anomalous commutation relations nor Nernst effect are induced in this interpretation.  The formula \eqref{semi-classical equation} is regarded as a {\em useful but  approximate semi-classical formula} valid only in the standard adiabatic condition \eqref{fundamental condition} \footnote{  In this view point, our understanding of the adiabatic  Berry's phase in the anomalous Hall effect  is analogous to that of Schwinger's anomalous magnetic moment in QED, which modifies the low-energy effective equations of motion of spin by an order $\hbar$ correction but does not modify the canonical commutation relations of the electron field.  }.

(ii) An alternative view  may be to regard that the adiabatic limit which generates adiabatic Berry's connection approximately is always consistent with the principle of quantum mechanics, as is indicated by \eqref{gauge covariant Hamiltonian}. As a specific reason  for the appearance of non-canonical property (i.e., not quantized in a conventional manner)  of the action \eqref{action-1} or \eqref{effective Lagrangian-2}, we attribute it to the truncation of the well-defined quantum system \eqref{starting path integral} to sub-systems in which  the fundamental gauge symmetry of the electromagnetic vector potential $eA_{k}$ and the newly introduced  gauge symmetry of the adiabatic Berry's phase \eqref{gauge transformation of Berry's phase} become incompatible. If one avoids the appearance of the electromagnetic gauge potential $eA_{k}(X)$, one can maintain the consistency of the adiabatic Berry's phase with the principle of quantum  mechanics (even for a point-like Berry's phase assumed). A detailed  account why the algebraic incompatibility of the electromagnetic vector potential $eA_{k}(X)$ with the gauge invariance of Berry's connection appears in  the canonical formulation of the Born-Oppenheimer approximation shall be given in subsection 3.5 later. In the present review, we mainly adopt this second view since it clarifies the interesting fact why the non-canonical system appears in the conventional approach of \eqref{action-1}.  In the present review,  we do not address the basic issue  how the deformation of Berry's phase itself, which is induced when one moves away from the precise adiabatic domain, influences the commutation relations; this issue is fundamental in the option (i) above. 

In this second view, the conventional treatment of anomalous Hall effect  \cite{Niu, Duval} is regarded to be  based on an {\em intentional application} of the additional electromagnetic vector potential $eA_{k}$ to the canonical system with adiabatic Berry's phase. The resulting system then inevitably becomes non-canonical and classical.
The anomalous commutation relations and the Nernst effect are thus the consequences of the additional electromagnetic vector potential which breaks the canonical system by a clear physical mechanism shown by the Born-Oppenheimer approximation (see subsection 3.5 later). This broken canonical system may be naturally called ``classical'', since the non-canonical behavior is anticipated and intended.  
This would correspond to the scheme usually adopted in the literature and we illustrated in subsection 3.1. 
We then find anomalous Poisson brackets \eqref{Poisson bracket} or (approximate) anomalous commutation relations \eqref{commutator-1}.


\subsection{Exact treatment of the two-level model} 
The role of the two-level system 
is to provide Berry' phase term in \eqref{action-1} \cite{Nagaosa}. We here comment on this simple system since it is convenient to clarify our basic view points discussed in the preceding subsection.
 We first rewrite the exact formula \eqref{geometric-phase} (in the snapshot approximation) as 
\begin{eqnarray}\label{geometric-phase2}
H^{\prime}(t)
= -\mu|\vec{p}|\sigma_{3} -\hbar\left(\begin{array}{cc}
\frac{(1+\cos\theta)}{2}&\frac{\sin\theta}{2}\\
            \frac{\sin\theta}{2}&
\frac{(1-\cos\theta)}{2}
            \end{array}\right)\dot{\varphi}
            -\hbar\left(\begin{array}{cc}
0&\frac{i}{2}\\
\frac{-i}{2}&0
            \end{array}\right)\dot{\theta}.
\end{eqnarray}
This separation of the fast variables in the first term, and the slower variables in the second and the third terms, is necessary to discuss the adiabatic approximation.
Eq.\eqref{geometric-phase2} shows that $\dot{\theta}$ has no non-trivial conjugate variable in the phase space. It vanishes when one integrates $\int dt H^{\prime}(t)$ to define the action. The other slow variable $\dot{\varphi}$ appears to have  non-trivial conjugate variables, which are Berry's phases and their off-diagonal partners, and thus could contribute to the modification of commutation relations.  But one can confirm that the matrix multiplying $\dot{\varphi}$ has a vanishing determinant and a unit trace, 
\begin{eqnarray}
{\rm det}\left(\begin{array}{cc}
\frac{(1+\cos\theta)}{2}&\frac{\sin\theta}{2}\\
            \frac{\sin\theta}{2}&
\frac{(1-\cos\theta)}{2}
            \end{array}\right)=0, \ \ \ 
            {\rm Tr}\left(\begin{array}{cc}
\frac{(1+\cos\theta)}{2}&\frac{\sin\theta}{2}\\
            \frac{\sin\theta}{2}&
\frac{(1-\cos\theta)}{2}
            \end{array}\right)=1
\end{eqnarray}
which imply the eigenvalues $1$ and $0$ {\em independently} of $\theta$, and thus dynamically trivial. In fact, the canonically equivalent exact \eqref{exact non-adiabatic},
\begin{eqnarray}\label{exact non-adiabatic2}
H^{\prime\prime}(t)
=
-\mu|\vec{p}| \left(\begin{array}{cc}
            \cos\theta&-\sin\theta\\
            -\sin\theta&-\cos\theta
            \end{array}\right)
-\hbar\left(\begin{array}{cc}
            1&0\\
            0&0
            \end{array}\right)\dot{\varphi}
\end{eqnarray}
shows that $\dot{\varphi}$, which multiplies would-be Berry's phases and their off-diagonal partners,  does not have any non-trivial conjugate variable, and thus does not contribute to the modification of commutation relations. The first term in  \eqref{exact non-adiabatic2} is essentially the same as the starting expression \eqref{level crossing Hamiltonian} if one considers the azimuthally symmetric configurations. 

Thus would-be Berry's phases and their off-diagonal partners in a precise  (snapshot) treatment of a two-level system generate no extra non-trivial derivative couplings in the action, and thus causes no renewed quantization and thus no anomalous commutation relations.
This fact agrees with the observation made in connection with \eqref{final result of second quantization} in the preceding subsection
\begin{align}\label{final result of second quantization-11}
Z
 =&\int {\cal D}\overline{P}_{k}{\cal D}X_{k}\exp \left\{ \frac{i}{\hbar}\int_{0}^{T} dt \left[(\overline{P}_{k}(t)-eA_{k}(X(t)))\dot{X}_{k}(t)
- H_{0}(X, \overline{P}) \right] \right\}\nonumber\\
&\times \int \Pi_{n}{\cal D}a^{\star}_{n}{\cal D}a_{n}\exp \left\{\frac{i}{\hbar}\int_{0}^{T} dt\sum_{n} \left[ a_{n}^{\star}(t)i\hbar \partial_{t}a_{n}(t)- E_{n}(\overline{P}) a_{n}^{\star}(t)a_{n}(t) \right] \right\}\nonumber\\
&\hspace{3cm} \times \exp \left\{\frac{i}{\hbar}\int_{0}^{T} dt \sum_{n,l}\langle n,t|i\hbar\partial_{t}|l,t\rangle a_{n}^{\star}(t)a_{l}(t) \right\}
\end{align} 
in which, the second quantized part except for the time derivative terms $a_{n}^{\star}(t)i\hbar \partial_{t}a_{n}(t)$ corresponds to the present  \eqref{geometric-phase2} or equivalently \eqref{exact non-adiabatic2}. If one sums over $n$ and $l$ in \eqref{final result of second quantization-11},
the path integral is reduced to the original path integral \eqref{starting path integral}, in which the canonical commutation relations of slower variables $
[X^{k}(t), P^{l}(t)]= i\hbar \delta_{kl}, \ \ [X^{k}(t), X^{l}(t)]= 0,\ \
[P^{k}(t), P^{l}(t)]= 0$
with $P_{k}(t) = \overline{P}_{k}(t)-eA_{k}X(t)$
are recovered by the canonical analysis or the BJL prescription.

The anomalous commutation relations of slower variables do not appear if no  adiabatic approximations are made in the sector of fast variables. 
It is important that we are not talking about the cancellation of a monopole and an anti-monopole, as is suggested in the adiabatic formula \eqref{adiabatic Stone phase2}, but rather no anomalous commutation relations appear from the beginning if one uses the exact formula \eqref{geometric-phase2} or the equivalent \eqref{exact non-adiabatic2}. 

The common argument for Berry's phase corresponds to the assumption that the
adiabatic approximation such as \eqref{adiabatic condition in band-crossing} and  \eqref{adiabatic Stone phase}, which we replicate here, 
\begin{eqnarray}\label{adiabatic Stone phase2}
H^{\prime}_{ad} \simeq  \left(\begin{array}{cc}
-\mu|\vec{p}| - \hbar\frac{(1+\cos\theta)\dot{\varphi}}{2}&0\\
            0&
\mu|\vec{p}| -\hbar\frac{(1-\cos\theta)\dot{\varphi}}{2}
            \end{array}\right)
\end{eqnarray}
is approximately valid for the fixed values of the slower angular variables.  Namely, off-diagonal time-derivative terms are assumed to be neglected. The  Hamiltonian in the adiabatic approximation describes the monopole configurations. This picture of the exact original commutation relations as  in \eqref{exact non-adiabatic2} and an approximate adiabatic formula with Berry's phase \eqref{adiabatic Stone phase2} is consistent with the option (i) in subsection 3.2.3. 

On the contrary,  one may adopt a picture that (point-like) Berry's phase in \eqref{adiabatic Stone phase2} is absorbed into the kinetic term to form a covariant  derivative and thus does not spoil the canonical structure of the slower system as in \eqref{gauge covariant Hamiltonian}, in the absence of the electromagnetic vector potential.  If one does not ask how Berry's phase arises, this scheme then realizes the option (ii) in subsection 3.2.3.  The drawback of this view is that Berry's phase is assumed to be given by the (time independent) monopole factor in \eqref{adiabatic Stone phase2}
regardless the time dependence of the canonical variable $\dot{\varphi}$ multiplying it. The (time) dependence of Berry's phase on the slower variable $\dot{\varphi}$ is important (see also Appendix B) but it is neglected in the common applications in \cite{Niu, Duval}.

\subsection{Born-Oppenheimer approximation and  covariant derivatives}

The analysis of a manageable  model  of  the Born-Oppenheimer approximation shall be given in this subsection to illustrate that the Born-Oppenheimer  approximation  which operates within the scheme of the canonical Hamiltonian formalism, does not deform the principle of quantum mechanics in the slower system. 
The common use of gauge invariant  auxiliary variables $X_{k}+{\cal A}_{k}(P)$ to describe the anomalous Hall effect, which formally satisfy the anomalous commutation relations,  actually does not deform the canonical commutation relations \cite{Blount}. Also, the Born-Oppenheimer approximation, which operates in the canonical Hamiltonian formalism, cannot incorporate the electromagnetic vector potential $eA_{k}(X)$, namely, it is shown that it cannot satisfy the gauge symmetries of Berry's connection in terms of the covariant derivative  $X_{k}+{\cal A}_{k}(P)$ and in terms of  the covariant electromagnetic vector potential $P_{k}+eA_{k}(X)$,  simultaneously. This is consistent with the well-known fact \cite{Niu, Duval, Duval2} that the inevitable failure of the canonical formulation of Berry's phase in the presence of the vector potential $eA_{k}(X)$ in \eqref{action-1}. 

 In the formulation of the Born-Oppenheimer approximation, one starts with the time-independent master Schr\"{o}dinger equation (using a simplified model described by $x^{k}$ and $X^{k}$, as an example),
 \begin{eqnarray}\label{BO-1}
 [H_{0}(X, P) + H_{1}(x, p; P)]\Psi(x, P)=E\Psi(x, P)
  \end{eqnarray}
 which implies that we adopt the representation of $x_{k}$ and $P_{k}$ diagonal \footnote{For this technical reason, we use $P_{k}$ instead of $\overline{P}_{k}= P_{k}+eA_{k}(X)$ since we want to have $P_{k}$ diagonal.}.
 The canonical quantization of the fast variables is 
 \begin{eqnarray}
 [p_{k}, x_{l}]=\frac{\hbar}{i}\delta_{kl}, \ \ [p_{k}, p_{l}]=0, \ \ [x_{k}, x_{l}]=0
 \end{eqnarray}
 and the canonical quantization of the slower variables is given by
 \begin{eqnarray}\label{BO-standard commutator-2}
 [P_{k}, X_{l}]=\frac{\hbar}{i}\delta_{kl}, \ \ [P_{k}, P_{l}]=0, \ \ [X_{k}, X_{l}]=0.
 \end{eqnarray}
 The present analysis goes through for any finite $N$ number of fast coordinates $x_{k}$ without any significant modification. We however consider a single freedom for each of the fast and slower systems, for simplicity. 
  
As a first step, we  confirm that the typical quantum mechanical solutions of $H_{0}(X,P)$ generate slow motions compared to the expected motion in $H_{1}$; we thus treat $P_{k}$ as slower variables.  We then expand the total wave function
 \begin{eqnarray}\label{BO-2}
 \Psi(x, P)=\sum_{n}\varphi_{n}(P)\phi_{n}(x, P)
 \end{eqnarray}
 by solving the equation
 \begin{eqnarray}\label{BO-3}
 H_{1}(x, p; P)\phi_{n}(x, P)=E_{n}(P)\phi_{n}(x, P)
 \end{eqnarray}
with $P_{k}$ treated as background variables; the states $\{\phi_{n}(x, P) \}$ are assumed to form a complete orthonormal basis set of the fast system. By inserting \eqref{BO-2} into the equation \eqref{BO-1} and multiplying by $\phi^{\star}_{l}(x,P)$ and  integrating over $x_{k}$,  one obtains 
\begin{eqnarray}
\sum_{n} \left\{ \int d^{3}x \phi_{l}(x, P)^{\star}H_{0}(X, P)\phi_{n}(x, P)+ E_{n}(P)\delta_{l,n} \right\} \varphi_{n}(P) =E\varphi_{l}(P).
\end{eqnarray}
We tentatively adopt in this subsection for simplicity
\begin{eqnarray}\label{Hamiltonian of slow variables-2}
H_{0}=\frac{1}{2M}P_{k}^{2} +\frac{M\omega_{0}^{2}}{2}X_{k}^{2}
\end{eqnarray}
 which makes the analysis transparent without extra technical complications. We later give a non-trivial example.
Using the completeness relation $\sum_{l^{\prime}}\phi^{\star}_{l^{\prime}}(x,P)\phi_{l^{\prime}}(y,P)=\delta^{3}(x^{k}-y^{k})$, we have
\begin{eqnarray}\label{exact Born-Oppenheimer}
&&\sum_{n}\sum_{l^{\prime}}\Bigg\{ \frac{M\omega_{0}^{2}}{2}\left(\delta^{ll^{\prime}}\frac{-\hbar}{i}\nabla_{k}+{\cal A}^{ll^{\prime}}_{k}(P) \right) \left(\delta^{l^{\prime}n}\frac{-\hbar}{i}\nabla_{k}+{\cal A}^{l^{\prime}n}_{k}(P)\right) + \frac{1}{2M}P_{k}^{2}\delta_{ln}\nonumber\\
&&\hspace{1.5cm} +E_{n}(P)\delta_{l,n} \Bigg\}\varphi_{n}(P)
=E\varphi_{l}(P)
\end{eqnarray}
where $X_{k}=\frac{-\hbar}{i}\frac{\partial}{\partial P_{k}}= \frac{-\hbar}{i}\nabla_{k}$ and 
\begin{eqnarray}
{\cal A}^{ll^{\prime}}_{k}(P)=\int d^{3}x  \phi_{l}(x,P)^{\star}\frac{-\hbar}{i}\frac{\partial}{\partial P_{k}}\phi_{l^{\prime}}(x,P).
\end{eqnarray}
 If one assumes the diagonal dominance ({\em adiabatic approximation}), namely, if one assumes that the slower variables $P_{k}$ do not cause a sizeable mixing of fast systems described by $\phi_{l}(x,P)$, or assuming that 
 the properties of the slower system $\varphi_{l}(P)$ are well described by ignoring the effects of all the states $\phi_{l^{\prime}}(x,P)$ with $l^{\prime}\neq l$ in \eqref{exact Born-Oppenheimer}, 
 one obtains 
 \begin{eqnarray}\label{time development of slow variables}
 \left\{ \frac{M\omega_{0}^{2}}{2}(X_{k}+{\cal A}^{(l)}_{k}(P))(X_{k}+{\cal A}^{(l)}_{k}(P)) +\frac{1}{2M}P_{k}^{2} +E_{l}(P) \right\} \varphi_{l}(P)
=E\varphi_{l}(P)
 \end{eqnarray}
where we defined~\footnote{To analyze the deformation of $ {\cal A}^{(l)}_{k}(P)$ off the precise adiabatic domain, one would need to incorporate the off-diagonal elements in \eqref{Berry connection in BO}. }
 \begin{eqnarray}\label{Berry connection in BO}
 {\cal A}^{(l)}_{k}(P)\equiv {\cal A}^{ll}_{k}(P) = \int d^{3}x  \phi_{l}(x,P)^{\star}\frac{-\hbar}{i}\frac{\partial}{\partial P_{k}}\phi_{l}(x,P),
 \end{eqnarray}
  which is often called Berry's {\em connection} for the specific level $l$; ${\cal A}^{(l)}_{k}(P)$, which is assumed to be a point-like, is an order 
  $\hbar$ quantity. The above is the standard formula of the Born-Oppenheimer approximation, namely, an {\em approximate formula} although useful one.  (See also the condition  \eqref{fundamental condition} to define Berry's phase in an adiabatic approximation).
 
 It is thus obvious that the slower variables $X^{k}$ are quantized in the standard manner {\em only once} as in 
\eqref{BO-standard commutator-2},
and the noncommutative geometry,  $[X^{k}(t), X^{l}(t)] \neq 0$, is not induced by Berry's connection.
The system of the $l$-th subsector \eqref{time development of slow variables} has an interesting gauge symmetry defined by the simultaneous transformations 
\begin{eqnarray}\label{state specific gauge symmetry}
&&{\cal A}^{(l)}_{k}(P)\rightarrow{\cal A}^{(l)}_{k}(P) +\hbar \partial_{k}\alpha^{(l)}(P),\nonumber\\
&&\varphi_{l}(P)\rightarrow e^{i \alpha^{(l)}(P)}\varphi_{l}(P)
\end{eqnarray}
 which keeps  $\Psi_{l}(x,P)=\varphi_{l}(P)\phi_{l}(x, P)$ of each $l$-th subsector invariant; the gauge variation of ${\cal A}^{(l)}_{k}(P)$, which is  induced by the phase change of $\phi_{l}(x,P)$, keeps $\Psi_{l}(x, P)$ invariant if 
one compensates it by the phase change  of $\varphi_{l}(P)$. This gauge symmetry is thus an indicator of the independence of the specific $\Psi_{l}(x,P)$ from other sectors with $l^{\prime} \neq l$, if one chooses $\alpha^{(l)}(P)$ different for each different $l$, and it is a measure of the validity of the adiabatic approximation.  This gauge symmetry is manifest in the equation \eqref{time development of slow variables} if one recalls that the combination \cite{Fujikawa-Umetsu-2022}
\begin{eqnarray}\label{covariant coordinate}
X^{(l)}_{k} = X_{k}+{\cal A}^{(l)}_{k}(P)
\end{eqnarray}
defines a {\em covariant derivative} with respect to \eqref{state specific gauge symmetry}~\cite{Blount, Nagaosa}, 
which satisfies 
\begin{eqnarray}
[X^{(l)}_{k}, X^{(l)}_{m}]=-\frac{\hbar}{i}[\partial_{k} {\cal A}^{(l)}_{m}(P) -\partial_{m} {\cal A}^{(l)}_{k}(P) ]
\end{eqnarray}
although $[X_{k}, X_{m}]=0$.
In this derivation, we used  the momentum representation where the momentum is diagonal,
\begin{eqnarray}
X_{k}=-\frac{\hbar}{i}\frac{\partial}{\partial P_{k}}
\end{eqnarray}
which satisfies the standard canonical commutation relations \eqref{BO-standard commutator-2}.

We now discuss the corresponding equations of motion of slower variables to compare the final formula with the one given by Berry's phase. 
The effective Hamiltonian for the slower system constructed on the $l$-th level of the fast system is given by \eqref{time development of slow variables}
\begin{eqnarray}\label{BO effective Hamiltonian}
H_{l}(P)=\frac{1}{2M}P_{k}^{2} +\frac{M\omega_{0}^{2}}{2}\left(X_{k}+{\cal A}^{(l)}_{k}(P) \right) \left(X_{k}+{\cal A}^{(l)}_{k}(P) \right) + E_{l}(P)
\end{eqnarray}
with $E_{l}(P)$ arising from the $l$-th level of the fast system.
We examine the quantum mechanical equations of motion of slower variables generated by the effective Hamiltonian constrained to the $l$-th level of the fast system using the canonical commutation relations (in the Heisenberg picture)
\begin{eqnarray}\label{BO-4}
\dot{P}_{m}&=&\frac{i}{\hbar}[H_{l}, P_{m}]=-M\omega_{0}^{2}(X_{m}+{\cal A}^{(l)}_{m}(P)), \nonumber\\
\dot{X}_{m}&=&\frac{i}{\hbar}[H_{l}, X_{m}]\nonumber\\
&=& \frac{M\omega_{0}^{2}}{2}\left[ (X_{k}+{\cal A}^{(l)}_{k}(P))\frac{\partial}{\partial P_{m}}{\cal A}^{(l)}_{k}(P) +\frac{\partial}{\partial P_{m}}{\cal A}^{(l)}_{k}(P)(X_{k}+{\cal A}^{(l)}_{k}(P)) \right]
\nonumber\\
&&+\frac{\partial}{\partial P_{m}} \left[ \frac{P^{2}_{k}}{2M} + E(P) \right]\nonumber\\
&=&-\frac{1}{2} \left[\dot{P}_{k}\frac{\partial}{\partial P_{m}}{\cal A}^{(l)}_{k}(P)+ \frac{\partial}{\partial P_{m}}{\cal A}^{(l)}_{k}(P)\dot{P}_{k} \right]+\frac{\partial}{\partial P_{m}} \left[ \frac{P^{2}_{k}}{2M}+ E_{l}(P) \right]\nonumber\\
&=&- \frac{\partial}{\partial P_{m}}{\cal A}^{(l)}_{k}(P)\dot{P}_{k}+\frac{\partial}{\partial P_{m}} \left[ \frac{P^{2}_{k}}{2M}+ E_{l}(P) \right]
\end{eqnarray}
where the last expression is valid when one ignores the possible operator ordering problem \footnote{The operator ordering appears when one equates the first two terms in the third line to the first term in the last line in the second equation of \eqref{BO-4}. This ordering is ignored if one is interested in the quantity only in the order $\hbar$ since ${\cal A}^{(l)}_{k}(P)$ is of order of $\hbar$. }.
If one uses the {\em auxiliary variables} $X^{(l)}_{m}\equiv X_{m}+{\cal A}^{(l)}_{m}(P)$ in \eqref{covariant coordinate} specific to the $l$-th level ,
 one  has an equivalent set of equations of motion (again by ignoring the possible operator ordering problem) \cite{Blount}
\begin{eqnarray}\label{BO-5}
\dot{P}_{m}&=&-M\omega_{0}^{2}X^{(l)}_{m}, \nonumber\\
\dot{X}^{(l)}_{m}&=& - \Omega^{(l)}_{mk}(P)\dot{P}_{k}
  +\partial_{m}\left[ \frac{P^{2}_{k}}{2M}+ E_{l}(P) \right]
\end{eqnarray}
with
\begin{eqnarray}\label{BO magnetic flux}
\Omega^{(l)}_{mk}(P)=\frac{\partial}{\partial P_{m}}{\cal A}^{(l)}_{k}(P) - \frac{\partial}{\partial P_{k}}{\cal A}^{(l)}_{m}(P).
\end{eqnarray}
If one uses the electromagnetic scalar potential $e\phi(X)$ instead of the harmonic potential, \eqref{BO-5} becomes 
\begin{eqnarray}\label{BO-6}
\dot{P}_{m}&=&e\frac{\partial}{\partial X^{(l)}_{m}}\phi(X^{(l)}), \nonumber\\
\dot{X}^{(l)}_{m}&=& - \Omega^{(l)}_{mk}(P)\dot{P}_{k}
  +\partial_{m}\left[ \frac{P^{2}_{k}}{2M}+ E_{l}(P) \right]
\end{eqnarray}
The set of equations \eqref{BO-6} correspond to the common {\em quantum equations} of the anomalous Hall effect in condensed matter physics \eqref{semi-classical equation} \cite{Blount}, although without $eA_{k}$. 

One may notice that the auxiliary variables $X^{(l)}_{m}$ \eqref{covariant coordinate} give rise to the non-commutative geometry
\begin{eqnarray}\label{canonical commutator in l-th level}
[P_{k}, P_{l}]=0, \ \ [P_{k}, X^{(l)}_{m}]=\frac{\hbar}{i}\delta_{kl}, \ \ 
[X^{(l)}_{k}, X^{(l)}_{m}]= -\frac{\hbar}{i}[\partial_{k}{\cal A}^{(l)}_{m}(P) - \partial_{m}{\cal A}^{(l)}_{k}(P)],
\end{eqnarray}
specific to the $l$-th level of the fast system, in contrast to  the starting canonical commutation relations \eqref{BO-standard commutator-2}. 
But this is {\em not} the deformation of the principle of quantum mechanics, since the Hamiltonian \eqref{BO effective Hamiltonian} consists of the covariant derivatives $X^{(l)}_{m}=X_{m}+{\cal A}^{(l)}_{m}(P)$ and thus manifestly gauge invariant with respect to the gauge symmetry of Berry's connection in the canonical Hamiltonian formalism, as was already mentioned in \eqref{gauge covariant Hamiltonian}. This is also confirmed by the fact that the commutation relations of $X^{(l)}_{m}=X_{m}+{\cal A}^{(l)}_{m}(P)$
 are dictated by the canonical commutation relations \eqref{BO-standard commutator-2} of the variables $X_{m}$ and $P_{k}$; every quantity expressed in terms of $X_{m}$ and $P_{k}$ is described by the conventional canonical commutation relations. 
 
We next comment on the electromagnetic gauge coupling.
 If the slower particle carries a charge $q=-e$, the electric current is given after the minimal replacement $P_{k}\rightarrow P_{k}+eA_{k}$ with an infinitesimal classical electromagnetic field $eA_{k}(X)$ in the total Hamiltonian \eqref{BO-1}
\begin{eqnarray}\label{electric gauge invariance}
 J_{k} (X,P)&=&-\frac{\delta}{\delta A_{k}(X)}[H_{0}(X, P_{k}+eA_{k}(X)) + H_{1}(x, p; P_{k}+eA_{k}(X))]|_
{A_{k}=0}\nonumber\\
&=&-e\frac{\delta}{\delta P_{k}}[H_{0}(X, P_{k}) + H_{1}(x, p; P_{k})]|_
{A_{k}=0},
 \end{eqnarray} 
where the last operation is identical to the time derivative of the slower coordinates in \eqref{BO-1}.
 We thus conclude 
 \begin{eqnarray}\label{gauge invariant current}
 J_{m}=-e \dot{X}_{m},
 \end{eqnarray} 
namely, the electromagnetic current is described by $-e\dot{X}^{k}$.
 
 On the other hand, if one would like to incorporate the transverse velocity in the quantum Hall effect in  \eqref{BO-6} dictated by the gauge invariance of the Berry's connection, one would choose 
 the electric current be  given by the current 
\begin{eqnarray}\label{covariant current}
 j_{m}\equiv -e\dot{X}^{(l)}_{m} = J_{m} - e\dot{{\cal A}}^{(l)}_{m} = J_{m} - e\left(\frac{\partial}{\partial P_{k}}{\cal A}^{(l)}_{m}\right)\dot{P}_{k}.
 \end{eqnarray}
although the electromagnetic gauge invariance chooses $ J_{m}$ \eqref{gauge invariant current} as the electromagnetic current \eqref{electric gauge invariance}
in the present model. 

The gauge invariance of Berry's phase and the gauge invariance of the electromagnetism are completely independent notions, and thus the compatibility of these
two gauge symmetries are not self-evident.
The term -$e\dot{{\cal A}}^{(l)}_{m} $ adds an extra  transverse velocity to obtain a desirable result of the anomalous Hall effect, but no strong argument for the term from the point of view of an electromagnetic gauge invariance.

\subsection{Algebraic incompatibility;  Berry's connection and electromagnetic vector potential}

The introduction of the electromagnetic vector potential $eA_{k}(X)$ into the adiabatic Hamiltonian with Berry's connection or the introduction of Berry's connection into the Hamiltonian containing the electromagnetic vector potential does not proceed in a simple manner in a Born-Oppenheimer approximation. We understand this complication as a cause of the appearance of the non-canonical system in \eqref{action-1}, since the Born-Oppenheimer approximation operates in a canonical Hamiltonian formalism and a natural formulation with both $eA_{k}(X)$ and ${\cal A}^{(l)}_{k}(P)$ would imply a canonical system and contradict the non-canonical system \eqref{action-1}. 

One may start with the adiabatic Hamiltonian  with a covariant derivative of Berry's connection, $ X_{k}+{\cal A}^{(l)}_{k}(P)$, as in \eqref{BO effective Hamiltonian}
\begin{eqnarray}\label{BO effective Hamiltonian-2}
H_{l}(P)=\frac{1}{2M}P_{k}^{2} +\frac{M\omega_{0}^{2}}{2}\left(X_{k}+{\cal A}^{(l)}_{k}(P) \right) \left(X_{k}+{\cal A}^{(l)}_{k}(P) \right) + E_{l}(P).
\end{eqnarray}
If one adds the covariant derivative of the electromagnetic vector potential 
\begin{eqnarray}
P_{k}\rightarrow P_{k}+ eA_{k}(X)
\end{eqnarray}
to this Hamiltonian, one obtains 
\begin{eqnarray}\label{BO-incompatibility}
H_{l}(P)&\rightarrow& \frac{1}{2M}(P_{k}+ eA_{k}(X))^{2} \nonumber\\
&+&\frac{M\omega_{0}^{2}}{2}\left(X_{k}+{\cal A}^{(l)}_{k}(P_{k}+ eA_{k}(X)) \right) \left(X_{k}+{\cal A}^{(l)}_{k}(P_{k}+ eA_{k}(X)) \right) \nonumber\\
&+& E_{l}(P_{k}+ eA_{k}(X)).
\end{eqnarray}
But this Hamiltonian is not written in the form of gauge invariance in terms of two covariant derivatives $X_{k}+{\cal A}^{(l)}_{k}(P)$ and $P_{k}+ eA_{k}(X)$, without any time derivative terms.

 In the Lagrangian formalism such as the path integral, one can re-write this system \eqref{BO effective Hamiltonian-2} to be invariant under both gauge symmetries, but now not canonical in a sense of a Hamiltonian formalism. One may start with the canonical system using \eqref{BO effective Hamiltonian-2}
 \begin{eqnarray}
&& \int {\cal D}P_{k}{\cal D}X_{k}\exp \{\frac{i}{\hbar}\int dt[P_{k}\dot{X}_{k} -H_{l}(P)]\}\\
&=&\int {\cal D}P_{k}{\cal D}X^{(l)}_{k}\exp \{\frac{i}{\hbar}\int dt[P_{k}\dot{X}^{(l)}_{k} + {\cal A}^{(l)}_{k}(P)\dot{P}_{k}-\frac{1}{2M}P_{k}^{2} -\frac{M\omega_{0}^{2}}{2}{X^{(l)}_{k}}^{2} - E_{l}(P)]\}\nonumber
 \end{eqnarray}
with $X^{(l)}_{k}=X_{k} + {\cal A}^{(l)}_{k}(P)$ and ${\cal D}P_{k}{\cal D}X_{k}={\cal D}P_{k}{\cal D}X^{(l)}_{k}$.  One may then add a formally electromagnetic gauge invariant term $\int dt [-eA_{k}(X^{(l)}_{k})\dot{X}^{(l)}_{k}]$ with a {\em time derivative} in the Lagrangian formalism,
 \begin{eqnarray}\label{BO path integral-1}
&&\int {\cal D}P_{k}{\cal D}X^{(l)}_{k}\exp \{\frac{i}{\hbar}\int dt[P_{k}\dot{X}^{(l)}_{k} + {\cal A}^{(l)}_{k}(P)\dot{P}_{k}-eA_{k}(X^{(l)}_{k})\dot{X}^{(l)}_{k}-\frac{1}{2M}P_{k}^{2} - E_{l}(P) \nonumber\\
&&\hspace{4.5cm} + e\phi(X^{(l)}_{k})]\}
 \end{eqnarray}
where we also replaced $\frac{M\omega_{0}^{2}}{2}\left(X^{(l)}_{k}\right)^{2}\rightarrow -e\phi(X^{(l)}_{k})$.
This last form of the action \eqref{BO path integral-1} agrees with \eqref{action-1} and thus one can satisfy the gauge invariances of both gauge symmetries simultaneously, but the action is no more quantized in a canonical manner in the Hamiltonian formalism \cite{Fujikawa-Umetsu-2022}.  The electromagnetic current as a source current of the field $A_{k}(X^{(l)}_{k})$ is given by \eqref{BO path integral-1} 
\begin{eqnarray}\label{covariant current-2}
 j_{m} = -e\dot{X}^{(l)}_{m}
 \end{eqnarray}
which is now consistent, unlike \eqref{covariant current}, but only in the framework of (non-canonical) classical theory.

The reason for the failure to satisfy the gauge invariances of Berry's phase and the electromagnetic vector gauge potential simultaneously in the canonical formalism, which are completely independent notions, is understood by looking at the total wave function of the Born-Oppenheimer approximation
\begin{eqnarray}
\Psi(x,P)=\sum_{n}\varphi_{n}(P)\phi_{n}(x,P).
\end{eqnarray} 
The gauge invariance of Berry's connection in the Born-Oppenheimer approximation, which is canonical Hamiltonian formalism, implies the gauge invariance  in each sector \eqref{state specific gauge symmetry}
\begin{eqnarray}
\Psi_{n}(x,P) = \varphi_{n}(P)\phi_{n}(x,P)
\end{eqnarray}
separately, while the electromagnetic gauge invariance acts universally on all the states in $\Psi(x,P)$ simultaneously. These two gauge symmetries are not compatible in the canonical Hamiltonian formalism \cite{Fujikawa-Umetsu-2022}.  (By putting the issue of $e\phi(\overline{X}_{k})$ aside.) 

The settings of the present problem and the ones 
in \eqref{action-1} are close to each other  and both are based on the approximately locally-defined (point-like)  Berry's connection (or phase). We thus conclude that the basic reason why we have no canonical formulation in the action \eqref{action-1}, which satisfies both gauge symmetries using extra time-derivatives, is that the two independent gauge symmetries are {\em not} compatible with each other in the canonical formulation where the time-derivative is not allowed to be used. In other words, if one attempts to quantize the action \eqref{action-1}, one would inevitably encounter the non-canonical system \cite{Fujikawa-Umetsu-2022}.

\subsection{Summary of the analyses of section 3}

The main purpose of this section is to understand the mechanism of the appearance of the anomalous classical Poisson brackets  in the adiabatic approximation starting with the system defined by canonical commutation relations. We assumed the fixed form of (a point-like) Berry's phase, even when we generally change the associated slower variables  in the analysis of commutation relations. 

In the Born-Oppenheimer approximation, the covariant derivative of Berry's connection 
$X^{(n)}_{k}=X_{k} +{\cal A}^{(n)}_{k}(P)$
automatically appears in a canonical formalism.  
 The use of the covariant derivative $X^{(n)}_{k}$ 
 constrains the slower variables to a specific state $\varphi_{n}(P)\phi_{n}(x,P)$.  The constraint of the slower canonical variables $X_{k}$ and $P_{l}$ to this specific state  $\varphi_{n}(P)\phi_{n}(x,P)$, which is a  constrained dynamics, causes a stricture on the electromagnetic gauge symmetry which operates {\em universally} on all the states in $\Psi(x,P)=\sum_{n} \varphi_{n}(P)\phi_{n}(x,P)$.
This stricture is recognized as an algebraic incompatibility of the electromagnetic covariant derivative $P_{k} +eA_{k}(X)$ and the covariant derivative $X^{(n)}_{k}$ in the canonical Hamiltonian formalism. We fail to define the Born-Oppenheimer approximation with the non-vanishing vector potential $eA_{k}(X)$ in a canonically invariant  manner. 

In the Lagrangian formalism, which allows the appearance of time-derivative terms in the action, we have more freedom and shown that we can satisfy both of gauge symmetries of ${\cal A}^{(n)}_{k}(P)$ and $eA_{k}(X)$ starting with the Lagrangian defined in terms of the Hamiltonian with ${\cal A}^{(n)}_{k}(P)$. But the resulting Lagrangian is no more quantized in a canonical manner \eqref{BO path integral-1}. This is perfectly consistent with the fact that one  cannot treat the Lagrangian \eqref{action-1}, which is gauge invariant under both of ${\cal A}^{(n)}_{k}(P)$ and $eA_{k}(X)$,  in a canonically invariant manner.

In the Born-Oppenheimer approximation, we can thus write the quantum mechanical equations \cite{Fujikawa-Umetsu-2022}
\begin{eqnarray}\label{BO-6-2}
\langle \dot{P}_{m}&=&e\frac{\partial}{\partial X^{(l)}_{m}}\phi(X^{(l)})\rangle, \nonumber\\
\langle \dot{X}^{(l)}_{m}&=& - \Omega^{(l)}_{mk}(P)\dot{P}_{k}
  +\partial_{m}\left[ \frac{P^{2}_{k}}{2M}+ E_{l}(P) \right]\rangle
\end{eqnarray}
with the canonical commutation relations
\begin{eqnarray}\label{canonical commutator in l-th level-3}
\langle[P_{k}, P_{l}]=0\rangle, \ \ \langle [P_{k}, X_{m}]=\frac{\hbar}{i}\delta_{kl}\rangle, \ \ 
\langle [X_{k}, X_{m}]= 0\rangle,
\end{eqnarray}
or  with $X_{m} = X^{(l)}_{m} - {\cal A}^{(l)}_{m}(P)$,
\begin{eqnarray}\label{canonical commutator in l-th level-2}
\langle[P_{k}, P_{l}]=0\rangle, \ \ \langle [P_{k}, X^{(l)}_{m}]=\frac{\hbar}{i}\delta_{kl}\rangle, \ \ 
\langle [X^{(l)}_{k}, X^{(l)}_{m}]= -\frac{\hbar}{i}[\partial_{k}{\cal A}^{(l)}_{m}(P) - \partial_{m}{\cal A}^{(l)}_{k}(P)]\rangle.
\nonumber\\
\end{eqnarray}
We have no anomalous Poisson bracket and no quantum mechanical Nernst effect.

If one should add a non-vanishing electromagnetic vector potential $eA_{k}(X)$, which is incompatible with Berry's connection in a Hamiltonian formalism, the system \eqref{BO-6-2} inevitably becomes non-canonical and thus one has the ``classical equations of motion''
\begin{eqnarray}\label{BO-6-3}
\dot{P}_{m}&=&-eF_{kl}(X^{(l)})\dot{X}^{(l)}_{m} +e\frac{\partial}{\partial X^{(l)}_{m}}\phi(X^{(l)}), \nonumber\\
\dot{X}^{(l)}_{m}&=& - \Omega^{(l)}_{mk}(P)\dot{P}_{k}
  +\partial_{m}\left[ \frac{P^{2}_{k}}{2M}+ E_{l}(P) \right].
\end{eqnarray}   
 If one should apply an extended canonical formalism to \eqref{BO-6-3}, one would then obtain the classical Poisson brackets \eqref{Poisson bracket}
and the classical Nernst effect. The (approximate) commutation relations by the background field method would then be given by \eqref{commutator-1}.

The deformation of Berry's phase itself by the variations of (contained) slower variables and their effects in defining commutation relations, which was discussed briefly in subsection 3.3 and discussed in detail in Appendix B, was not analyzed in the main part of this review. This is mainly because  this aspect, although essential, is not commonly discussed in the applications of Berry's phase, which is assumed to be a point-like object, to the anomalous Hall effect in the literature.

\newpage
\section{Chiral anomalies and fermions on the lattice}

Chiral anomalies, which were established in particle theory~\cite{Bertlmann, anomaly1,anomaly2,anomaly3}, have been used to elucidate the properties of ``Weyl fermions'' and other related phenomena in condensed matter and nuclear physics.  See, for example, \cite{nielsen2, fukushima, zyuzin, hosur, Wang, Venugopalan, Hongo-Hidaka}.  The present review article was motivated by these recent developments. We discuss the basic aspects of chiral anomalies in the conventional definition of particle physics, which may help understand these developments and clarify some controversial issues involved. 

Chiral anomalies are believed to be short distance effects~\cite{Wilson} and in fact  only the high frequency components of  fermion variables are essential in the (Euclidean) evaluation of chiral anomalies~\cite{fujikawa-suzuki}; thus the well-known relation for QED~\cite{anomaly1,anomaly2}
\begin{eqnarray}\label{identity0}
&&\partial_{\mu}\left(\bar{\psi}\gamma^{\mu}\gamma_{5}\psi\right)=
2im\bar{\psi}\gamma_{5}\psi+\frac{e^{2}}{2\pi^{2}}\vec{E}\cdot\vec{B}
\end{eqnarray}
holds for the fundamental electron in the condensed matter in an arbitrary small domain of space-time  independently of frequencies carried by the gauge field  $A_{\mu}$, which may include the Coulomb potential provided by surrounding charged particles in addition to the externally applied electromagnetic field. We often include the charge $e$ into the gauge field and suppress the explicit $e^{2}$. The basic relation \eqref{identity0}, which was established by a careful analysis of Feynman diagrams \cite{anomaly1, anomaly2}, is also known to be derived in the framework of the path integral as an extra Jacobian factor with the help of the Atiyah-Singer index theorem \cite{anomaly3}.

When one emphasizes the fundamental aspects, the chiral identity \eqref{identity0} is considered to hold irrespective of the effective description of the electron, since the chiral anomaly measures the electron in the fundamental level appearing in the Hamiltonian of the multi-electron Schr\"{o}dinger equation.  If one analyzes the chiral anomalies of quarks in the quark-gluon plasma, for example, the anomalies are given by the fundamental formula analogous to \eqref{identity0}. It is well-known that the Standard Model is consistent only when the (gauge) anomaly cancellation among quarks and leptons takes place. If the anomalies of quarks should be modified in the quark-gluon plasma while those of leptons are unchanged, for example, such a cancellation would be jeopardized.

 In condensed matter physics, on the other hand, one is usually interested in effective models defined in the Brillouin zone which is based on the underlying lattice structure but often without detailed specifications of gauge field couplings. One then analyzes the chiral properties of the electron using an effective theory defined on a lattice \footnote{One thus needs some expertize in condensed matter physics for the deep physical analyses.}. Thus the general properties of chiral anomalies in lattice gauge theory become relevant to understand the possible  anomalies or related phenomena in the effective theory in condensed matter physics. In the following, we are first going to explain the basic properties of chiral anomalies in  continuum theory and then in lattice gauge theory. Based on these preparations, we shall discuss some salient features of chiral anomalies which may be relevant to condensed matter physics and nuclear physics.

The chiral anomaly is also characterized by the anomalous commutation relations of composite operators. We  recall the anomalous commutation relations in the case of the continuum chiral anomaly. As the simplest example,  we mention the anomalous
commutation relation of the Gauss-law operator
\begin{eqnarray}
G=\overline{\psi}_{L}\gamma^{0}\psi_{L}(x)-\partial_{k}\dot{A}^{k}(x)
\end{eqnarray}
of  chiral Abelian gauge theory
\begin{eqnarray}\label{chiral Abelian gauge theory}
S=\int d^{4}x \{\overline{\psi}_{L}(x)i\gamma^{\mu}[\partial_{\mu}- iA_{\mu}(x)]\psi_{L}(x) - \frac{1}{4}F_{\mu\nu}F^{\mu\nu}\}
\end{eqnarray}
defined in the gauge $A_{0}=0$, namely,
\begin{eqnarray}\label{gauss}
\partial_{t}G(t,\vec{x})=\frac{i}{\hbar}[H, G(t,\vec{x})]=-\frac{1}{3}\frac{1}{4\pi^{2}}\vec{E}\cdot \vec{B}.
\end{eqnarray}     
This specific form of  anomalous commutation relation has been discussed in~\cite{ fujikawa-suzuki, fujikawa2}; one may note that $\partial_{t}G(t,\vec{x})=\partial_{\mu}(\overline{\psi}_{L}\gamma^{\mu}\psi_{L})(x)$ if one uses the equation of motion for $A^{k}(x)$ in the gauge $A_{0}=0$, and the anomaly relation \eqref{gauss} written in terms of the source current of the electromagnetic field
\begin{eqnarray}\label{anomaly in chiral Abelian theory}
\partial_{\mu}(\overline{\psi}_{L}\gamma^{\mu}\psi_{L})(x)=-\frac{1}{3}\frac{1}{4\pi^{2}}\vec{E}\cdot \vec{B}
\end{eqnarray}
 is known more commonly \cite{ faddeev, jo}. The relation  \eqref{gauss} shows that the Gauss operator, which is the generator of time independent gauge transformation (i.e., the residual gauge symmetry of the gauge condition $A_{0}(x)=0$) and thus responsible for the elimination of the longitudinal component of the massless gauge field,   is time dependent for chiral Abelian gauge theory; as an alternative interpretation of \eqref{gauss}, the  Hamiltonian $H$ is gauge non-invariant, and thus such a theory is {\em inconsistent} as a gauge theory. To have a consistent theory in continuum,  one  needs to increase the number of fermion flavors with appropriate charge assignments to ensure the (gauge) anomaly cancellation as in the Standard Model. We also note that the equal-time commutation relation of the Gauss operator itself for Abelian chiral gauge theory is known to be normal in continuum theory~\cite{fujikawa2, faddeev, jo, jackiw, adler}, namely, 
\begin{eqnarray}\label{commutator of Gauss operator}
[G(t,\vec{x}), G(t,\vec{y})]=0.
\end{eqnarray}

The factor $1/3$ in \eqref{gauss} is a characteristic of {\em consistent anomaly}. In comparison, a  combination of the axial current \eqref{identity0} with $m=0$ and a vector fermion-number current gives  the {\em covariant form} of anomaly (by including the charge into the gauge field),
\begin{eqnarray}\label{covariant  anomaly}
\partial_{\mu}j_{L}^{\mu}(x)=\partial_{\mu} \left[\overline{\psi}(x)\gamma^{\mu}(\frac{1-\gamma_{5}}{2})\psi(x) \right] = -
\frac{1}{4\pi^{2}}\vec{E}\cdot \vec{B},
\end{eqnarray}
 since the vector fermion-number current is conserved in (massless) QED.
This difference in the form of anomaly arises from the fact that $\overline{\psi}_{L}\gamma^{0}\psi_{L}(x)$ in \eqref{gauss} (and $\overline{\psi}_{L}\gamma^{\mu}\psi_{L}(x)$ in \eqref{anomaly in chiral Abelian theory}) couple to $A^{\mu}$, while $j_{5}^{\mu}=\bar{\psi}\gamma^{\mu}\gamma_{5}\psi$ in \eqref{identity0} and thus $j_{L}^{\mu}$ in \eqref{covariant anomaly} do not couple to $A^{\mu}$; the current $j_{L}^{\mu}$ in \eqref{covariant anomaly} is physically the lepton number current of the chiral gauge theory \eqref{chiral Abelian gauge theory} which is not conserved, namely, the lepton number coupled to chiral gauge fields is not conserved in the Standard Model~\cite{`t Hooft}. This lepton number non-conservation is related to the baryon number generation of the universe. The factor $1/3$ in \eqref{gauss} in the present Abelian gauge theory is understood as a Bose symmetrization factor of a triangle Feynman diagram.

It is often stated in the literature  that the chiral anomaly is characterized by two distinct aspects; the short-distance (high frequency) effects as described above and also the low-energy effects such as the picture of the spectral flow. It is, however, shown later that the ultra-violet behavior at large energies, i.e., the infinite depth of the Dirac sea is essential to understand the chiral anomaly in the picture of the spectral flow \cite{Ambjorn}; this fact implies that the chiral anomaly in the picture of the spectral flow is not well-defined on the lattice in which large frequencies are cut-off at  $\sim 1/a$. We are going to explain that the spectral flow in the Dirac sea with a finite depth corresponds to a pair production in a general sense rather than the chiral anomaly which generally implies the  (potential) non-conservation of the fermion number.  
Related to this observation, we are going to explain in the following that the chiral anomaly described by $\gamma_{5}$ for each species doubler  separately , which appears in a conventional lattice gauge theory and is not a local field by itself, is ill-defined for a finite lattice spacing $a\neq 0$. The notion of species doublers shall be explained in detail later.

In contrast, the Ginsparg-Wilson fermion \cite{Ginsparg-Wilson}, which is a new fermion theory defined by the block transformation in lattice theory,  describes a Dirac fermion without species doublers for a vector-like theory such as QED and QCD on the lattice using a new chiral operator $\Gamma_{5}$.  As is explained later, this new chiral operator $\Gamma_{5}$ gives rise to an  Atiyah-Singer-type index related to the chiral anomaly  even for a finite lattice spacing $a\neq0$.  In other words, the exact chiral symmetry of the lattice theory defined by $\Gamma_{5}$ is broken quantum mechanically by a Jacobian, which is given by the Atiyah-Singer-type index, in the path integral formulation. 

\subsection{N\"{o}ther's theorem and chiral symmetry}
We start with a review of the N\"{o}ther's theorem as a manifestation of symmetry principle, which is conveniently formulated in the path integral formalism, together  with an emphasis on the basic requirements on the path integral. 
N\"{o}ther's theorem is the basis of the relation between the symmetry and the conservation law in physics which  is represented by the Ward-Takahashi identities in quantum field theory.
We illustrate the N\"{o}ther's theorem and Ward-Takahashi identities using the Lagrangian of QED (Quantum Electrodynamics) but without writing the Maxwell Lagrangian explicitly
\begin{eqnarray}
{\cal L}&=&\overline{\psi}(x)i\gamma^{\mu}(\partial_{\mu} -ieA_{\mu}(x))\psi(x) -m\overline{\psi}(x)\psi(x)\nonumber\\
&=&\overline{\psi}(x)i\Dslash\psi(x) -m\overline{\psi}(x)\psi(x)
\end{eqnarray}
and the path integral is defined by 
\begin{eqnarray}
\int {\cal D}\overline{\psi}{\cal D}\psi \exp\left\{\frac{i}{\hbar}\int d^{4}x \left[\overline{\psi}(x)i\Dslash\psi(x) -m\overline{\psi}(x)\psi(x) \right] \right\}.
\end{eqnarray}
One may start with the basic identity, for example,
\begin{eqnarray}\label{identity1}
&&\int {\cal D}\overline{\psi}^{\prime}{\cal D}\psi \exp\left\{ \frac{i}{\hbar}\int d^{4}x \left[\overline{\psi}^{\prime}(x)i\Dslash\psi(x) -m\overline{\psi}^{\prime}(x)\psi(x)\right] \right\}\nonumber\\
&=&\int {\cal D}(\overline{\psi}+\overline{\epsilon}){\cal D}\psi \exp \left\{\frac{i}{\hbar}\int d^{4}x \left[ (\overline{\psi}+\overline{\epsilon})(x)i\Dslash\psi(x) -m(\overline{\psi}+\overline{\epsilon})(x)\psi(x) \right] \right\}\nonumber\\
&=&\int {\cal D}\overline{\psi}{\cal D}\psi \exp \left\{\frac{i}{\hbar}\int d^{4}x \left[\overline{\psi}(x)i\Dslash\psi(x) -m\overline{\psi}(x)\psi(x) \right] \right\}
\end{eqnarray}
with $\overline{\psi}^{\prime}(x)=\overline{\psi}(x)+\overline{\epsilon}(x)$. This identity means that the naming of the integration variables do not change the integral itself. We next make the basic assumption 
\begin{eqnarray}\label{fundamental requirement}
 {\cal D}(\overline{\psi}+\overline{\epsilon}){\cal D}(\psi +\epsilon)= {\cal D}\overline{\psi}{\cal D}\psi 
\end{eqnarray}
namely, the path integral measure is invariant under the translation in the functional space with $\psi(x)\rightarrow \psi(x)+ \epsilon(x)$ and $\overline{\psi}(x)\rightarrow \overline{\psi}(x)+\overline{\epsilon}(x)$ for $\epsilon(x)$ and $\overline{\epsilon}(x)$, which are {\em independent of each other and also independent of} $\psi(x)$ and $\overline{\psi}(x)$.

The last equality in \eqref{identity1} then implies 
\begin{eqnarray}\label{identity2}
&&\int {\cal D}\overline{\psi}{\cal D}\psi \exp\left\{\frac{i}{\hbar}\int d^{4}x \left[(\overline{\psi}+\overline{\epsilon})(x)i\Dslash\psi(x) -m(\overline{\psi}+\overline{\epsilon})(x)\psi(x) \right] \right\}\nonumber\\
&=&\int {\cal D}\overline{\psi}{\cal D}\psi \exp \left\{ \frac{i}{\hbar}\int d^{4}x \left[\overline{\psi}(x)i\Dslash\psi(x) -m\overline{\psi}(x)\psi(x) \right] \right\}
\end{eqnarray}
By expanding this relation in terms of $\overline{\epsilon}(x)$, one obtains 
\begin{eqnarray}\label{identity3}
&&\langle i\Dslash\psi(x) -m\psi(x)\rangle\nonumber\\
&&=\int {\cal D}\overline{\psi}{\cal D}\psi [i\Dslash\psi(x) -m\psi(x)]
\exp\left\{ \frac{i}{\hbar}\int d^{4}x \left[\overline{\psi}(x)i\Dslash\psi(x) -m\overline{\psi}(x)\psi(x) \right] \right\}\nonumber\\
&&=0.
\end{eqnarray}
The basic requirement \eqref{fundamental requirement} is thus equivalent to the quantum equations of motion.

We next examine the symmetry of the action defined in terms of the Lagrangian,
\begin{align}
S&=\int d^{4}x {\cal L}\nonumber\\
&=\int d^{4}x \left[\overline{\psi}(x)i\Dslash\psi(x) -m\overline{\psi}(x)\psi(x) \right].
\end{align}
 Besides the Lorentz invariance, it is invariant under the fermion number transformation 
\begin{eqnarray}
\psi(x)\rightarrow \psi^{\prime}(x) = e^{i\alpha}\psi(x), \ \ \ 
\overline{\psi}(x)\rightarrow \overline{\psi}^{\prime}(x) = \overline{\psi}(x)e^{-i\alpha}
\end{eqnarray}
with an arbitrary real number $\alpha$. We next consider an infinitesimal localized (space-time dependent) real parameter $\alpha(x)$
\begin{eqnarray}\label{vector transformation}
\psi(x)\rightarrow \psi^{\prime}(x) = e^{i\alpha(x)}\psi(x), \ \ \ 
\overline{\psi}(x)\rightarrow \overline{\psi}^{\prime}(x) = \overline{\psi}(x)e^{-i\alpha(x)}
\end{eqnarray}
for which we have 
\begin{align}
S^{\prime}&=\int d^{4}x \left[\overline{\psi}^{\prime}(x)i\Dslash\psi^{\prime}(x) -m\overline{\psi}^{\prime}(x)\psi^{\prime}(x) \right],\nonumber\\
&= S +\int d^{4}x \alpha(x)\partial_{\mu} \left[\overline{\psi}(x)\gamma^{\mu}\psi(x) \right].
\end{align}
 The identity associated with this symmetry is given by  
\begin{eqnarray}\label{identity4}
&&\int {\cal D}\overline{\psi}^{\prime}{\cal D}\psi^{\prime} \exp\left\{\frac{i}{\hbar}\int d^{4}x \left[\overline{\psi}^{\prime}(x)i\Dslash\psi^{\prime}(x) -m\overline{\psi}^{\prime}(x)\psi^{\prime}(x) \right] \right\}
\nonumber\\
&=&\int {\cal D}\overline{\psi}{\cal D}\psi \exp\left\{ \frac{i}{\hbar}\int d^{4}x \left[ \overline{\psi}(x)i\Dslash\psi(x) -m\overline{\psi}(x)\psi(x) \right] \right\}.
\end{eqnarray}
If one assumes that the path integral measure is invariant under this symmetry transformation
\begin{eqnarray}\label{Jacobian1}
{\cal D}\overline{\psi}^{\prime}{\cal D}\psi^{\prime} =
{\cal D}\overline{\psi}{\cal D}\psi
\end{eqnarray}
one has the identity (fermion-number conservation, or equivalently, the charge conservation in the present case ) by expanding the path integral in the order linear in $\alpha(x)$
\begin{eqnarray}\label{identity5}
&&\partial_{\mu}\langle \left[ \overline{\psi}(x)\gamma^{\mu}\psi(x) \right]\rangle\nonumber\\
&&=\int {\cal D}\overline{\psi}{\cal D}\psi \partial_{\mu} \left[ \overline{\psi}(x)\gamma^{\mu}\psi(x) \right]
\exp \left\{ \frac{i}{\hbar}\int d^{4}x \left[ \overline{\psi}(x)i\Dslash\psi(x) -m\overline{\psi}(x)\psi(x) \right] \right\}\nonumber\\
&&=0.
\end{eqnarray}

We next examine the chiral symmetry defined by 
\begin{eqnarray}
\psi(x)\rightarrow \psi^{\prime}(x) = e^{i\alpha\gamma_{5}}\psi(x), \ \ \ 
\overline{\psi}(x)\rightarrow \overline{\psi}^{\prime}(x) = \overline{\psi}(x)e^{i\alpha\gamma_{5}}
\end{eqnarray}
with an arbitrary (infinitesimal) real number $\alpha$, for which we have 
\begin{align}
S^{\prime}&=\int d^{4}x \left[ \overline{\psi}^{\prime}(x)i\Dslash\psi^{\prime}(x) -m\overline{\psi}^{\prime}(x)\psi^{\prime}(x) \right],\nonumber\\
&= S - \int d^{4}x \left[2i\alpha m\overline{\psi}(x)\gamma_{5}\psi(x) \right]
\end{align}
which shows that the chiral symmetry is (softly) broken by the mass term.
We next consider an infinitesimal localized (space-time dependent) real parameter $\alpha(x)$
\begin{eqnarray}\label{local chiral symmetry}
\psi(x)\rightarrow \psi^{\prime}(x) = e^{i\alpha(x)\gamma_{5}}\psi(x), \ \ \ 
\overline{\psi}(x)\rightarrow \overline{\psi}^{\prime}(x) = \overline{\psi}(x)e^{i\alpha(x)\gamma_{5}}
\end{eqnarray}
for which we have 
\begin{align}
S^{\prime}&=\int d^{4}x \left[ \overline{\psi}^{\prime}(x)i\Dslash\psi^{\prime}(x) -m\overline{\psi}^{\prime}(x)\psi^{\prime}(x) \right],\nonumber\\
&= S +\int d^{4}x \left\{ \alpha(x)\partial_{\mu} \left[\overline{\psi}(x)\gamma^{\mu}\gamma_{5}\psi(x) \right] - 2i\alpha(x)m\overline{\psi}(x)\gamma_{5}\psi(x) \right\}.
\end{align}
 The identity associated with this symmetry is given by  
\begin{eqnarray}\label{identity6}
&&\int {\cal D}\overline{\psi}^{\prime}{\cal D}\psi^{\prime} \exp \left\{\frac{i}{\hbar}\int d^{4}x \left[ \overline{\psi}^{\prime}(x)i\Dslash\psi^{\prime}(x) -m\overline{\psi}^{\prime}(x)\psi^{\prime}(x) \right] \right\}
\nonumber\\
&=&\int {\cal D}\overline{\psi}{\cal D}\psi \exp \left\{\frac{i}{\hbar}\int d^{4}x \left[\overline{\psi}(x)i\Dslash\psi(x) -m\overline{\psi}(x)\psi(x) \right] \right\}.
\end{eqnarray}
If one assumes that the path integral measure is invariant under this symmetry transformation
\begin{eqnarray}\label{Jacobian2}
{\cal D}\overline{\psi}^{\prime}{\cal D}\psi^{\prime} =
{\cal D}\overline{\psi}{\cal D}\psi
\end{eqnarray}
one has the identity (the partial conservation of the chiral current) 
\begin{align}\label{identity7}
&\partial_{\mu}\langle [\overline{\psi}(x)\gamma^{\mu}\gamma_{5}\psi(x)]\rangle - 2im\langle \overline{\psi}(x)\gamma_{5}\psi(x)\rangle \nonumber\\
=&\int {\cal D}\overline{\psi}{\cal D}\psi \left\{ \partial_{\mu} \left[\overline{\psi}(x)\gamma^{\mu}\gamma_{5}\psi(x) \right]- 2im \overline{\psi}(x)\gamma_{5}\psi(x) \right\}
\exp \{ \frac{i}{\hbar}S \}\nonumber\\
=&0.
\end{align}

\subsection{Evaluation of Jacobians (anomalies)}
We now examine if the relations \eqref{Jacobian1} and \eqref{Jacobian2} are justified. To analyze this issue in a reliable way, we define the Euclidean path integral \cite{fujikawa-suzuki, Fujikawa-1980}
\begin{eqnarray}
\int {\cal D}\overline{\psi}{\cal D}\psi \exp\left\{ \frac{1}{\hbar}\int d^{4}x \left[ \overline{\psi}(x)i\Dslash\psi(x) -m\overline{\psi}(x)\psi(x) \right] \right\}
\end{eqnarray}
with
\begin{eqnarray}
\Dslash=\gamma^{\mu}(\partial_{\mu}-ieA_{\mu}(x))=\sum_{\mu=1}^{4}\gamma^{\mu}(\partial_{\mu}-ieA_{\mu}(x))
\end{eqnarray}
where
\begin{eqnarray}
\gamma^{0}\rightarrow -i\gamma^{4}, \ \ \partial_{0}\rightarrow i\partial_{4}, \ \ 
A_{0}(x)\rightarrow iA_{4}(x)
\end{eqnarray}
and 
\begin{eqnarray}
\{\gamma^{\mu},\gamma^{\nu}\}=2\eta^{\mu\nu}
\end{eqnarray}
with $\eta^{\mu\nu}=(-1,-1,-1,-1)$. The operator $\Dslash$ is hermitian in the sense
\begin{eqnarray}
(\Psi, \Dslash\Psi)=\int d^{4}x \Psi(x)^{\dagger}\Dslash\Psi(x)=(\Dslash\Psi, \Psi).
\end{eqnarray}

We next expand the field variables as
\begin{eqnarray}
\psi(x)=\sum_{n}a_{n}\xi_{n}(x), \ \ \overline{\psi}(x)=\sum_{n}\bar{a}_{n}\xi^{\dagger}_{n}(x),
\end{eqnarray}
with
\begin{eqnarray}\label{eigenvalue equation} 
\Dslash\xi_{n}(x)=\lambda_{n}\xi_{n}(x), \ \ \int d^{4}x \xi^{\dagger}_{n}(x)\xi_{l}(x) =\delta_{nl}
\end{eqnarray}
and the path integral is written as 
\begin{eqnarray}
\int \prod_{n}d\bar{a}_{n}da_{n} \exp\left\{ \frac{1}{\hbar}\sum_{n}(i\lambda_{n}-m)\bar{a}_{n}a_{n} \right\}
\end{eqnarray}
which allows a reliable estimate of Jacobian factors; this specification implies a gauge invariant mode cut-off regularization of the Jacobian factor \cite{Bonora}.

We start with  \eqref{vector transformation} which is equivalent to
\begin{eqnarray}
&&a_{n}\rightarrow a^{\prime}_{n}=\sum_{l}\int d^{4}x \xi^{\dagger}_{n}(x)(1+i\alpha(x))\xi_{l}(x)a_{l}, \nonumber\\
&& \bar{a}_{n}\rightarrow \bar{a}^{\prime}_{n}=\sum_{l}\bar{a}_{l}\int d^{4}x \xi^{\dagger}_{l}(x)(1-i\alpha(x))\xi_{n}(x)
\end{eqnarray}
and remembering that $a_{n}$ and $\bar{a}_{n}$ are the Grassmann numbers and thus the integration is equivalent to the differentiation, we have
\begin{align}
\prod_{n}d\bar{a}^{\prime}_{n}da^{\prime}_{n}=&\det\left|\int d^{4}x \xi^{\dagger}_{n}(x)(1+i\alpha(x))\xi_{l}(x)\right|^{-1} \det\left| \int d^{4}x \xi^{\dagger}_{l}(x)(1-i\alpha(x))\xi_{n}(x) \right|^{-1}
\nonumber\\
&\times \prod_{n}d\bar{a}_{n}da_{n}
\end{align}
or in a regularized form, the Jacobian becomes (using $\det|M|=\exp[{\rm Tr}\ln M]$)
\begin{align}
J_{1}=\exp \left\{- \lim_{N\rightarrow \infty}\sum_{n=1}^{n=N} \left[ \int d^{4}x \xi^{\dagger}_{n}(x)i\alpha(x)\xi_{n}(x)-\int d^{4}x \xi^{\dagger}_{n}(x)i\alpha(x)\xi_{n}(x) \right] \right\} =1.
\end{align}
Thus the relation \eqref{Jacobian1} and consequently the identity \eqref{identity5} is justified.

We next examine the chiral symmetry \eqref{local chiral symmetry} which is equivalent to
\begin{eqnarray}
&&a_{n}\rightarrow a^{\prime}_{n}=\sum_{l}\int d^{4}x \xi^{\dagger}_{n}(x)(1+i\alpha(x)\gamma_{5})\xi_{l}(x)a_{l}, \nonumber\\
&& \bar{a}_{n}\rightarrow \bar{a}^{\prime}_{n}=\sum_{l}\bar{a}_{l}\int d^{4}x \xi^{\dagger}_{l}(x)(1+i\alpha(x)\gamma_{5})\xi_{n}(x)
\end{eqnarray}
and remembering that $a_{n}$ and $\bar{a}_{n}$ are the Grassmann numbers and thus the integration is equivalent to the differentiation, we have
\begin{align}
\prod_{n}d\bar{a}^{\prime}_{n}da^{\prime}_{n}=&\det\left| \int d^{4}x \xi^{\dagger}_{n}(x)(1+i\alpha(x)\gamma_{5})\xi_{l}(x)\right|^{-1}\det \left| \int d^{4}x \xi^{\dagger}_{l}(x)(1+i\alpha(x)\gamma_{5})\xi_{n}(x)\right|^{-1}
\nonumber\\
&\times \prod_{n}d\bar{a}_{n}da_{n}
\end{align}
or in a regularized form, the Jacobian becomes for an infinitesimal $\alpha(x)$
\begin{align}
J_{5}=&\exp \left\{ - \lim_{N\rightarrow \infty}\sum_{n=1}^{N} \left[ \int d^{4}x \xi^{\dagger}_{n}(x)2i\alpha(x)\gamma_{5}\xi_{n}(x) \right] \right\} \nonumber\\
=&\exp\left\{ - \lim_{M\rightarrow \infty}\sum_{n} \left[\int d^{4}x \xi^{\dagger}_{n}(x)2i\alpha(x)\gamma_{5}e^{-\lambda_{n}^{2}/M^{2}}\xi_{n}(x) \right]\right\} 
\end{align}
where we replaced the mode cut-off by  the eigenvalue cut-off. Since the eigenvalue equations \eqref{eigenvalue equation} are gauge covariant,  both of the mode cut-off and eigenvalue cut-off preserve the gauge invariance \cite{Bonora}; this procedure allows an explicit evaluation of the Jacobian by preserving the manifest gauge invariance \cite{Witten}.
The quantity appearing in the exponential of the Jacobian $J_{5}$ for $\alpha(x)={\rm constant}$ corresponds to the Atiyah-Singer index which is related to the Chern-Pontryagin character for a non-Abelian gauge theory defined  in a compact space-time such as $S^{4}$. This Atiyah-Singer index theorem has been checked for a simple case in the Euclidean space-time~\cite{Jackiw-Rebbi}. A recent review of the Atiyah-Singer index theorem is found in \cite{Fukaya}. The eigenvalue cut-off (as well as the mode cut-off)  preserves this index. Intuitively,
\begin{eqnarray}\label{Atiyah-Singer index}
\sum_{n}\int d^{4}x \xi^{\dagger}_{n}(x)\gamma_{5}e^{-\lambda_{n}^{2}/M^{2}}\xi_{n}(x)= (n_{+}-n_{-})
\end{eqnarray}
where the indices $n_{\pm}$ correspond to the number of eigenstates with $\lambda_{n}=0$ and $\gamma_{5}\xi_{n}(x)=\pm\xi_{n}(x)$. We here used the fact that $\int d^{4}x \xi^{\dagger}_{n}(x)\gamma_{5}e^{-\lambda_{n}^{2}/M^{2}}\xi_{n}(x)=0$ for $\Dslash\xi_{n}=\lambda_{n}\xi_{n}(x)$ with $\lambda_{n}\neq 0$ since $\Dslash\gamma_{5}\xi_{n}=-\lambda_{n}\gamma_{5}\xi_{n}(x)$, namely, $\gamma_{5}\xi_{n}(x)$ is an eigenstate of a different eigenvalue from $\xi_{n}(x)$.  It is known that any smooth function $f(\lambda_{n}^{2}/M^{2})$, which satisfies $f(0)=1$ and thus preserves the index relation and goes to zero rapidly $f(\infty)=0$, gives rise to the  correct chiral anomaly. Thus our choice of the gauge invariant  regularization of the Jacobian is general.

We thus evaluate the quantity which may be called a ``local index'' explicitly
\begin{eqnarray}\label{anomaly1}
&&\lim_{M\rightarrow \infty}{\rm Tr}\{ 2i\alpha(x)\gamma_{5}\exp[-\Dslash^{2}/M^{2}]\}\nonumber\\
&\equiv&\lim_{M\rightarrow \infty}\sum_{n}\int d^{4}x \xi^{\dagger}_{n}(x)2i\alpha(x)\gamma_{5}\exp[-\Dslash^{2}/M^{2}]\xi_{n}(x)\nonumber\\
&=&\lim_{M\rightarrow \infty}\int d^{4}x 2i\alpha(x)tr \int\frac{d^{4}k}{(2\pi)^{4}}e^{-ikx}\gamma_{5}\exp[-\Dslash^{2}/M^{2}]e^{ikx}\nonumber\\
&=&\lim_{M\rightarrow \infty}\int d^{4}x 2i\alpha(x)tr\gamma_{5} \int\frac{d^{4}k}{(2\pi)^{4}}\exp[-(i\kslash +\Dslash)^{2}/M^{2}]\nonumber\\
&=&\lim_{M\rightarrow \infty}\int d^{4}x 2i\alpha(x)tr\gamma_{5}M^{4} \int\frac{d^{4}k}{(2\pi)^{4}}\exp[-(i\kslash +\Dslash/M)^{2}]
\end{eqnarray}
where the trace of the well-regularized operator is converted to the trace using the plane waves in the third line, $\sum_{n}\rightarrow tr \int\frac{d^{4}k}{(2\pi)^{4}}$ with the $tr$ standing for the trace over Dirac indices,  and then let the plane wave $e^{ikx}$ move through the differential operator.  In the final step, we re-scaled the momentum variable $\kslash\rightarrow M\kslash$.
We next note
\begin{eqnarray}
&&\exp[-(i\kslash +\Dslash/M)^{2}]\nonumber\\
&=&\exp[\kslash^{2}-i(\kslash\Dslash +\Dslash\kslash)/M + \Dslash^{2}/M^{2}]\nonumber\\
&=&\exp[-|k|^{2}-ik^{\mu}D_{\mu}/M+ D^{\mu}D_{\mu}/M^{2} +\frac{1}{4}[\gamma^{\mu},\gamma^{\nu}][D_{\mu},D_{\nu}]/M^{2}]\nonumber\\
&=&\exp[-|k|^{2}-ik^{\mu}D_{\mu}/M+ D^{\mu}D_{\mu}/M^{2} +\frac{-ie}{4}[\gamma^{\mu},\gamma^{\nu}]F_{\mu\nu}/M^{2}]\nonumber\\
&\Rightarrow&\exp[-|k|^{2}+\frac{-ie}{4}[\gamma^{\mu},\gamma^{\nu}]F_{\mu\nu}/M^{2}]
\end{eqnarray}
where the last step is based on the examination of  \eqref{anomaly1}, namely, one needs at least 4 $\gamma$-matrices to survive the trace $tr\gamma_{5}$ with $\gamma_{5}$ and only the terms larger than or equal to $1/M^{4}$, when one expands the exponential factor in powers of $1/M$, survive in the limit $M\rightarrow\infty$. 

We thus have
\begin{eqnarray}\label{anomaly2}
&&\lim_{M\rightarrow \infty}{\rm Tr} \left\{ 2i\alpha(x)\gamma_{5}\exp\left[ -\Dslash^{2}/M^{2} \right] \right\}\nonumber\\
&=&\lim_{M\rightarrow \infty}\int d^{4}x~2i\alpha(x)tr\gamma_{5}M^{4} \int\frac{d^{4}k}{(2\pi)^{4}}\exp \left[-|k|^{2}+\frac{-ie}{4}[\gamma^{\mu},\gamma^{\nu}]F_{\mu\nu}/M^{2}\right]\nonumber\\
&=&\int d^{4}x ~2i\alpha(x)\frac{-e^{2}}{32}tr\gamma_{5}([\gamma^{\mu},\gamma^{\nu}]F_{\mu\nu})^{2}\int\frac{d^{4}k}{(2\pi)^{4}}\exp[-|k|^{2}]\nonumber\\
&=&\int d^{4}x ~2i\alpha(x)\frac{e^{2}}{32} \left( \frac{1}{16\pi^{2}} \right) 16i\epsilon_{\mu\nu\alpha\beta}F^{\mu\nu}F^{\alpha\beta}\nonumber\\
&=&\int d^{4}x ~\alpha(x)\frac{-e^{2}}{16\pi^{2}}\epsilon_{\mu\nu\alpha\beta}F^{\mu\nu}F^{\alpha\beta}
\end{eqnarray}
with $\epsilon_{1230}=1$. Thus the chiral Jacobian is given by \footnote{The relations \eqref{Atiyah-Singer index} and \eqref{Chiral Jacobian} imply the formal relation $n_{+} - n_{-}=
\int d^{4}x\frac{e^{2}}{32\pi^{2}}\epsilon_{\mu\nu\alpha\beta}F^{\mu\nu}F^{\alpha\beta}$ with $\epsilon_{1234}=1$ which, when extended to the Yang-Mills theory,
corresponds to the Atiyah-Singer index theorem.}
\begin{eqnarray}\label{Chiral Jacobian}
J_{5}=\exp \left[  \int d^{4}x ~\alpha(x)\frac{e^{2}}{16\pi^{2}}\epsilon_{\mu\nu\alpha\beta}F^{\mu\nu}F^{\alpha\beta} \right]
\end{eqnarray}
namely, the chiral identity \eqref{identity6}, which is naively \eqref{identity7} without a nontrivial Jacobian,  is now  replaced by the one with the correct anomaly
\begin{eqnarray}\label{identity8}
\partial_{\mu}\langle [\overline{\psi}(x)\gamma^{\mu}\gamma_{5}\psi(x)]\rangle - 2im\langle \overline{\psi}(x)\gamma_{5}\psi(x)\rangle = -\frac{e^{2}}{16\pi^{2}}\epsilon_{\mu\nu\alpha\beta}F^{\mu\nu}F^{\alpha\beta} 
\end{eqnarray}
with $\epsilon_{1230}=1$ (if one chooses  $\epsilon_{0123}=1$,  the signature of the anomaly term is changed.).

The crucial property in the above evaluation is that  finite frequencies smaller than any finite $\Lambda^{2}$ in the fourth line of \eqref{anomaly1} gives a vanishing contribution to the final result,
\begin{align}
&\lim_{M\rightarrow \infty}\int d^{4}x ~2i\alpha(x)tr\gamma_{5} \int_{|k^{2}|<\Lambda^{2}}\frac{d^{4}k}{(2\pi)^{4}}\exp\left[ -(i\kslash +\Dslash)^{2}/M^{2} \right] \nonumber\\
=&\lim_{M\rightarrow \infty}\int d^{4}x ~2i\alpha(x)tr\gamma_{5}\nonumber\\
&\times M^{4}\int_{|k^{2}|<\Lambda^{2}/M^{2}}\frac{d^{4}k}{(2\pi)^{4}}\exp\left[-|k^{2}| -(i\kslash \Dslash/M +\Dslash i\kslash/M+\Dslash^{2}/M^{2}) \right]
=0,
\end{align}
since
\begin{eqnarray}
\lim_{M\rightarrow \infty}M^{4}\int_{|k^{2}|<\Lambda^{2}/M^{2}}\frac{d^{4}k}{(2\pi)^{4}}\exp[-|k^{2}|]
= O(M^{0})
\end{eqnarray}
and thus any term which may survive the trace operation with $\gamma_{5}$, tr $\gamma_{5}$, needs to contain the inverse powers of $M$, and thus vanishes.  In this sense, the local chiral anomaly comes from the {\em short distances} (or large frequencies)  in the (Euclidean) space-time \cite{Wilson}, while the integrated Atiyah-Singer index (for non-Abelian theory) arises from the sector of vanishing eigenvalues. This correlation of the short-distance and long-distance properties is an essential aspect of the topological chiral anomaly.

The above prescription of the gauge invariant mode cut-off of the Jacobian factor  becomes subtle for the evaluation of anomalies associated with gauge symmetry. In chiral gauge theory, such as 
\begin{eqnarray}
{\cal L}=\overline{\psi}(x)i\Dslash \left(\frac{1-\gamma_{5}}{2} \right)\psi(x)
\end{eqnarray}
one generally encounters the quantum breaking of gauge invariance, namely, gauge anomalies \cite{Bardeen}. The evaluation of the Jacobian based on the above gauge invariant mode cut-off using $\{\Dslash (\frac{1-\gamma_{5}}{2})\}^{\dagger}\Dslash (\frac{1-\gamma_{5}}{2})$ gives rise to the so-called covariant form of anomalies. The covariant gauge anomalies are sufficient to analyze the anomaly cancellation in chiral gauge theory such as in the Standard Model \cite{Gross-Jackiw} and in string theory \cite{Witten}, since the gauge anomaly cancellation implies that one can impose the gauge invariance on all the vertices in the context of Feynman diagrams. The covariant anomaly is based on the evaluation of anomalies by imposing the gauge invariance on all the vertices except for the specific vertex specified by the current appearing in the Ward-Takahashi identities. If one can cancel the covariant anomalies among different flavors of fermions, it means that one can impose the gauge invariance on all the vertices coupled to the gauge field in the sense of Feynman diagrams; this is the statement of the cancellation of gauge anomalies. 

A way to evaluate directly the so-called consistent form of anomalies in the path integral, which has a form of gauge non-invariance in general, is to use a Pauli-Villars regularization with a bosonic fermion by defining a fermion operator of the form $D =\Dslash (\frac{1-\gamma_{5}}{2}) + \delslash (\frac{1+\gamma_{5}}{2})$ \cite{Bardeen}. The Jacobians then completely cancel between the physical fermion and the bosonic Pauli-Villars regulator fields; the Ward-Takahashi identity then contains both the physical fermion and the Pauli-Villars regulator but without anomalies. One reproduces the so-called consistent form of anomalies from the Pauli-Villars mass term in the limit of $m_{PV}\rightarrow \infty$, in which the Pauli-Villars regulator field (except for the regulator mass term) decouples from the physical Hilbert space and only the physical field survives the limit together with anomalies. We illustrate a mechanism similar to the Pauli-Villars regularization in the anomaly evaluation in lattice gauge theory later.    

To summarize the analysis in this subsection, we can treat all the known chiral anomalies in the path integral formalism.

\subsection{Fermions in lattice gauge theory}
Fermions defined on the lattice exhibit some novel properties related to chiral symmetry as was noticed by K. Wilson in his original formulation of the lattice gauge theory~\cite{Wilson-1974}. We discuss the novel properties from the point of view of chiral anomalies.
\subsubsection{Chiral symmetry and species doubling}
We explain the basic aspects of fermions defined on the lattice.
We use the simplest Lagrangian of QED defined on the 4-dimensional hypercubic lattice. We choose the hypercubic lattice since the simplest (Euclidean) hypercubic lattice with the lattice spacing $a$ is considered to be required to restore the Lorentz or $O(4)$ symmetry in the continuum limit.
The action is defined by 
\begin{align}\label{lattice QED}
\frac{S_{F}}{a^{4}} =&\sum_{x}\Bigg\{\sum_{\mu}\left(\frac{i}{2a}\right)[\overline{\psi}(x)\gamma^{\mu}U_{\mu}(x)\psi(x+a^{\mu})-\overline{\psi}( x+a^{\mu})\gamma^{\mu}U_{\mu}(x)^{\dagger}\psi(x)] \nonumber\\
&-m_{0}\overline{\psi}(x)\psi(x)
+ \sum_{\mu}\left(\frac{r}{2a}\right)[\overline{\psi}(x)U_{\mu}(x)\psi(x+a^{\mu})+\overline{\psi}(x+a^{\mu})U_{\mu}(x)^{\dagger}\psi(x) \nonumber\\
&-2\overline{\psi}(x)\psi(x)] \Bigg\}
\end{align}
with
\begin{eqnarray}
U_{\mu}(x)=e^{-iaeA_{\mu}(x)}.
\end{eqnarray}
This basic structure is valid for a non-Abelian gauge theory also.
The fermions are placed on the lattice point and the gauge field is placed on the link connecting the lattice points.
We sum $x$ over all the lattice points and sum over $\mu=1\sim 4$ with $a^{\mu}=a\hat{\mu}$, where $\hat{\mu}$ stands for the unit vector in the direction of $\mu$.  The action is invariant under the gauge transformations
\begin{eqnarray}
&&U_{\mu}(x)\rightarrow U(\omega(x))U_{\mu}(x)U(\omega(x+a^{\mu}))^{\dagger},\nonumber\\
&&\psi(x)\rightarrow U(\omega(x))\psi(x), \ \ \overline{\psi}(x)\rightarrow \overline{\psi}(x)U(\omega(x))^{\dagger}.
\end{eqnarray}
The first kinetic term in \eqref{lattice QED} is invariant under the (global) chiral symmetry
\begin{eqnarray}\label{lattice chiral symmetry}
\psi(x)\rightarrow e^{i\gamma_{5}\alpha}\psi(x), \ \ \  \overline{\psi}(x)\rightarrow\overline{\psi}(x) e^{i\gamma_{5}\alpha}
\end{eqnarray}
with a real constant parameter $\alpha$, while the mass term $m_{0}$ and the term with $r/a$ break this chiral symmetry.
In the naive continuum limit $a\rightarrow 0$, 
we have 
\begin{eqnarray}\label{Wilson fermion near continuum}
S_{F}\simeq \int d^{4}x \left\{ \frac{i}{2}\left[ \overline{\psi}(x)\gamma^{\mu}D_{\mu}\psi(x)-D_{\mu}\overline{\psi}(x)\gamma^{\mu}\psi(x) \right] -m_{0}\overline{\psi}(x)\psi(x) \right\} 
\end{eqnarray}
and thus the term with $r/a$, which is called the Wilson term, appears to vanish in the naive continuum limit. 

To understand the physical meaning of the Wilson term, we examine the free propagator given by the lattice Lagrangian \eqref{lattice QED} with $r=0$
\begin{eqnarray}\label{naive lattice propagator}
\Big[ \sum_{\mu}\frac{1}{a}\gamma^{\mu}\sin ak_{\mu} - m_{0} \Big]^{-1}.
\end{eqnarray}
The momenta on the lattice theory are chosen in the fundamental Brillouin zone
\begin{eqnarray}\label{fundamental Brillouin zone}
-\frac{\pi}{2a}\leq k_{\mu}<\frac{3\pi}{2a}.
\end{eqnarray}
If one chooses the momenta near $k_{\mu}\simeq 0$ for each direction in \eqref{naive lattice propagator}, one recovers the continuum propagator
\begin{eqnarray}
[\gamma^{\mu}k_{\mu} - m_{0}]^{-1}
\end{eqnarray} 
in the limit $a\rightarrow 0$.
  On the other hand, if one chooses 
  $k_{1}=\pi/a +k_{1}^{\prime}$ with small $k_{1}^{\prime}$, for example, one has in the limit $a\rightarrow 0$
 \begin{eqnarray}
 \Big[\sum_{\mu=2}^{4}\gamma^{\mu}k_{\mu}
 -\gamma^{1}k_{1}^{\prime}-m_{0}\Big]^{-1}
\end{eqnarray}  
which after the change of the signature of 
$k_{1}^{\prime}$  gives another pole with the mass $m_{0}$. We thus have two particle poles in each momentum direction and in total $2^{4}=16$ fermions. This phenomenon is called the {\em species doubling}.

If one chooses $r\neq 0$, which breaks chiral symmetry (strongly),
\begin{eqnarray}
\Big[\sum_{\mu}\frac{1}{a}\gamma^{\mu}\sin ak_{\mu} - m_{0} -\sum_{\mu}\frac{r}{a}(1-\cos ak_{\mu})\Big]^{-1}
\end{eqnarray}
then all the extra fermion poles except for the one at $k_{\mu}\simeq 0$ are eliminated from the physical spectrum in the limit $a\rightarrow0$, since the extra poles have the masses $\sim m_{0} + r/a$.
  
 To understand the species doubling from a symmetry point of view, one may define following Karsten and Smit~\cite{Karsten-Smit}
  \begin{eqnarray}\label{shift operator}
  T_{\mu}(x) =\gamma^{\mu}\gamma^{5}\exp[i\pi (x^{\mu}/a)]
  \end{eqnarray}
which satisfies the Clifford algebra $T_{\mu}T_{\nu}+T_{\nu}T_{\mu}=2\delta_{\mu\nu}$. One then defines 16 operators
  \begin{eqnarray}\label{even operator}
  1, T_{1}T_{2}, T_{1}T_{3}, T_{1}T_{4}, T_{2}T_{3}, T_{2}T_{4}, T_{3}T_{4}, T_{1}T_{2}T_{3}T_{4},
  \end{eqnarray}
  and 
  \begin{eqnarray}\label{odd operator}
T_{1}, T_{2}, T_{3}, T_{4}, T_{1}T_{2}T_{3}, T_{2}T_{3}T_{4}, T_{3}T_{4}T_{1}, T_{4}T_{1}T_{2}.
\end{eqnarray}  
If one denotes any one of these 16 operators by $T$, one can confirm that the action \eqref{lattice QED} with $r=0$ (but $m_{0}\neq 0$) is invariant under 
\begin{eqnarray}
\psi(x)\rightarrow T(x)\psi(x), \ \ \overline{\psi}(x)\rightarrow \overline{\psi}(x)T^{-1}(x).
\end{eqnarray}
 The operator \eqref{shift operator} adds a momentum $\pi/a$ in the direction $\mu$, and thus one recognizes that these operators except for $T=1$ generates 15 extra poles starting with the pole at $k_{\mu}=0$. A crucial property  is that all $T$ in \eqref{even operator} commute with $\gamma_{5}$ and all $T$ in \eqref{odd operator} anti-commute with $\gamma_{5}$. 
 
 If one attempts to define a left-handed massless fermion $[(1-\gamma_{5})/2]\psi(x)$ and $\overline{\psi}(x)[(1+\gamma_{5})/2]$, which means one sets $m_{0}=r=0$ in \eqref{lattice QED} to ensure the exact chiral symmetry,
 \begin{eqnarray}\label{lattice chiral QED}
\frac{S_{0}}{a^{4}} &=&\sum_{x}\sum_{\mu}\left(\frac{i}{2a}\right)[\overline{\psi}(x)\gamma^{\mu}U_{\mu}(x)[(1-\gamma_{5})/2]\psi(x+a^{\mu})\nonumber\\
&&-\overline{\psi}( x+a^{\mu})\gamma^{\mu}U_{\mu}(x)^{\dagger}[(1-\gamma_{5})/2]\psi(x)],
\end{eqnarray} 
the species doubling and the symmetry operators in \eqref{even operator} and \eqref{odd operator} imply that one inevitably has 8 left-handed fermions and 8 right-handed fermions. (Starting with the right-handed fermion $[(1+\gamma_{5})/2]\psi(x)$ and $\overline{\psi}(x)[(1-\gamma_{5})/2]$, one similarly finds 8 right-handed massless fermions and 8 left-handed massless fermions.)  This statement, which is based on the hypercubic lattice, is generic since the Lorentz invariance in the continuum limit is believed to require the hypercubic lattice. An equivalent statement has been given by Nielsen and Ninomiya which is based on a topological consideration~\cite{Nielsen-Ninomiya}. It is important that this analysis of species doubling itself is not directly based on the analysis of chiral anomalies.
 
We next analyze the above statement of species doubling from a point of view of  chiral anomalies in lattice gauge theory.  The chiral anomaly on the lattice in the limit $a\rightarrow 0$, which is used as a means to examine to what extent  the lattice theory realizes the continuum theory, has been discussed by many authors by evaluating Feynman diagrams in the past~\cite{Karsten-Smit}. We here instead discuss the same issue by an analysis of the chiral Jacobian in the path integral.

From the Lagrangian \eqref{lattice QED}, we identity the Euclidean latticized (hermitian) Dirac operator for the simple case with $m_{0}=r=0$ and by setting $\hat{p}_{\mu}=-i\partial_{\mu}$
\begin{eqnarray}
\Dslash \equiv \frac{i}{2a}[\gamma^{\mu}U_{\mu}(x)e^{(-ia\hat{p}_{\mu})}  -e^{(-ia\overleftarrow{\hat{p}}_{\mu})}\gamma^{\mu}U^{\dagger}_{\mu}(x)] 
\end{eqnarray}  
where $e^{(-ia\hat{p}_{\mu})}$ stands for the difference operator. 
The lattice version of the regularized Jacobian \eqref{anomaly1} for the localized chiral transformation $\psi(x)\rightarrow e^{i\gamma_{5}\alpha(x)}\psi(x)$ and $\overline{\psi}(x)\rightarrow\overline{\psi}(x) e^{i\gamma_{5}\alpha(x)}$, under the global version of which (i.e., global chiral symmetry) the action \eqref{lattice chiral QED} is invariant,  is given for an infinitesimal $\alpha(x)$
\begin{eqnarray}\label{lattice anomaly1}
\lim_{M\rightarrow \infty}a^{4}\sum_{x} 2i\alpha(x)\sum_{n=1}^{16}tr T^{-1}_{n}\gamma_{5}T_{n} \int_{-\pi/2a\leq k^{\mu}<\pi/2a}\frac{d^{4}k}{(2\pi)^{4}}e^{-ikx}\exp[-(\Dslash)^{2}/M^{2}]e^{ikx}\nonumber\\
\end{eqnarray}
where we used the 16 symmetry operators in \eqref{even operator} and \eqref{odd operator} and restricted the momentum range to the physical domain
\begin{eqnarray}
-\pi/2a\leq k^{\mu}<\pi/2a
\end{eqnarray}
starting with the fundamental Brillouin zone \eqref{fundamental Brillouin zone}. Note that $\Dslash^{\dagger}\Dslash=\Dslash^{2}$. If one takes the sum over the 16 operators first, then the Jacobian factor \eqref{lattice anomaly1} vanishes identically
since $\sum_{n=1}^{16} T^{-1}_{n}\gamma_{5}T_{n}=0$, namely, the lattice massless or massive  theory with 
$r=0$ has no chiral anomalies due to 16 species doublers altogether   
; anomalies vanish for any $a\neq 0$ and thus smoothly vanish in the limit $a\rightarrow 0$ also. We comment on this fact from a different point of view later.

Coming back to $S_{F}$ in \eqref{lattice QED}, it is important to examine  the contributions to chiral anomaly from the 15 species doublers, which acquire the effective masses of the order $m_{0} + r/a$. 
In the small $a$ limit, $a\rightarrow 0$ but $a\neq 0$, the 16 species with the effective masses in each subdomain of the fundamental Brillouin zone 
\begin{eqnarray}\label{effective masses}
&&m_{0}\ \ (1,1), \ \ m_{0}+2r/a\ \ (4, -1), \ \ m_{0}+4r/a\ \ (6, 1), \ \ m_{0}+6r/a\ \ (4, -1), \nonumber\\ &&m_{0}+8r/a\ \ (1, 1) 
\end{eqnarray}
 appear. Here we indicate inside the bracket, (multiplicity, chiral charge); for example, fermions with the effective mass $m_{0}+2r/a$  appear 4 times with the chiral charge $-1$.  
Mathematically, it is better defined to consider the $a\rightarrow0$ limit first {\em with $r/a$ kept fixed} and later consider the limit $r/a \rightarrow \infty$, for which one can identify the fermion species with the kinetic energy term $\Dslash$ of a continuum fermion with masses of the order $m_{0}+r/a$ in \eqref{effective masses}. In this limit one obtains the Jacobian (including the effective mass terms for the sake of book-keeping)
\begin{align}\label{lattice anomaly4}
&\lim_{M\rightarrow \infty}\int d^{4}x ~2i\alpha(x)\sum_{n=1}^{16}tr T^{-1}_{n}\gamma_{5}T_{n}\nonumber\\
&\times \lim_{a\rightarrow 0}\int_{-\pi/2a\leq k^{\mu}<\pi/2a}\frac{d^{4}k}{(2\pi)^{4}}\exp[-((i\kslash+\Dslash)^{2} +M^{2}_{n})/M^{2}]
\end{align}
with $M_{n}$ standing for the effective mass of the $n$-th species in \eqref{effective masses}. This anomaly factor vanishes for $M\rightarrow\infty$ with $M_{n}$ kept fixed; each $M_{n}$ produces the continuum anomaly up to a sign factor of chirality, but the sum $\sum_{n=1}^{16} T^{-1}_{n}\gamma_{5}T_{n}$ gives a vanishing result.

To write the Ward-Takahashi identity for the localized chiral transformation $\psi(x)\rightarrow e^{i\gamma_{5}\alpha(x)}\psi(x)$ and $\overline{\psi}(x)\rightarrow\overline{\psi}(x) e^{i\gamma_{5}\alpha(x)}$, one first evaluates the Jacobian factor, which is given by 
\eqref{lattice anomaly4}, that vanishes as we have explained above. Namely, the chiral identity in lattice gauge theory in the present setting is {\em anomaly-free}. 
To write the Ward-Takahashi identity, we first write the action in the form in the limit $a\rightarrow 0$ but with $r/a$ kept finite (in a symbolic notation),
\begin{eqnarray}
{\cal L}&\simeq& \sum_{n}\int_{V_{n}}\frac{d^{4}k}{(2\pi)^{4}} \frac{d^{4}q}{(2\pi)^{4}}\overline{\psi}(k+q)\Aslash(q)\psi(k) + \sum_{n}\int_{V_{n}}\frac{d^{4}k}{(2\pi)^{4}}\overline{\psi}(k)(-\kslash+M_{n})\psi(k)\nonumber\\
&\simeq&\sum_{n}\int d^{4}x ~\overline{\psi}_{n}(x)[i\Dslash + M_{n}]\psi_{n}(x)
\end{eqnarray}
where we first consider the Fourier transform of the field and then rewrite it by the effective fields representing the species doublers. The symbol $V_{n}$ indicates the $n$-th subdomain in the Brillouin zone.
 The Ward-Takahashi identity is then written in the form which is anomaly-free
\begin{eqnarray}\label{Ward-Takahashi for Wilson}
\sum_{n}\partial_{\mu}\langle J_{5n}^{\mu}(x)\rangle
=\sum_{n}2iM_{n}\langle \overline{\psi}_{n}(x)T^{-1}_{n}\gamma_{5}T_{n}\psi_{n}(x)\rangle 
\end{eqnarray}
where the right-hand side is the contribution of all the species doublers with the effective masses $M_{n}$ in \eqref{effective masses} which break chiral symmetry. In the limit $r/a \rightarrow \infty$, the 15 species doublers with large effective masses decouple from the physical space;  at the same time,
the effective large mass terms of 15 species doublers give rise to the chiral anomalies
\begin{eqnarray}
-[4\times (-1) + 6\times (1) +4\times (-1) +1\times (1)]
\frac{1}{2\pi^{2}}\vec{E}\cdot\vec{B} =\frac{1}{2\pi^{2}}\vec{E}\cdot\vec{B}
\end{eqnarray}
and thus
\begin{eqnarray}
\partial_{\mu}\langle J_{51}^{\mu}(x)\rangle
=2im_{0}\langle \overline{\psi}_{1}(x)\gamma_{5}\psi_{1}(x)\rangle + \frac{1}{2\pi^{2}}\vec{E}\cdot\vec{B}
\end{eqnarray}
which agrees with the perturbative analysis. Readers are referred to the very detailed diagrammatic analysis in \cite{Karsten-Smit}.
This mechanism is the same as the Pauli-Villars regularization in continuum theory, in which the physical fermion and the (bosonic) Pauli-Villars regulator together give the vanishing Jacobian but the large Pauli-Villars mass term, which breaks chiral symmetry strongly, gives the ordinary chiral anomaly in the large mass limit $m_{PV}\rightarrow\infty$ of the regulator field.

 As for the fermion number transformation $\psi(x)\rightarrow e^{i\alpha}\psi(x)$ and $\overline{\psi}(x)\rightarrow\overline{\psi}(x) e^{-i\alpha}$, which is an exact symmetry of \eqref{lattice QED}, it is confirmed that one obtains the (trivially) vanishing Jacobian and thus the fermion number is conserved in the vector-like QED.  
 
 The absence of the chiral anomaly in the limit $a\rightarrow 0$ but $a\neq 0$ is indicated by the explicit form of the above Jacobian factor \eqref{lattice anomaly1} and supported by the short-distance idea of anomaly by K. Wilson, namely, the anomaly does not appear when one cuts-off all the short distances.  On the one hand, the inevitable appearance of the chiral anomaly for a single fermion defined in the continuum is known, as in the conventional analysis. These two facts combined imply that any chiral symmetric theory once defined on the lattice automatically re-arranges itself to have the spectrum with species doublers, which is overall anomaly-free to ensure the  smooth limit $a\rightarrow 0$. This view of the inevitable appearance of 16 species doublers (in the $O(4)$ symmetric Euclidean theory) is physically attractive.
We emphasize that the phenomenon of  species doublers arises in the present scheme from the  {\em chiral symmetry} generated by $\gamma_{5}$ in the fundamental Lagrangian defined on the lattice.

\subsubsection{Spectral flow in a Dirac sea with a finite depth}

We would like to comment further on the issue of the  species doublers and spectral flow.  A precise Lagrangian formulation of species doubling, which appears inevitably when one imposes the $\gamma_{5}$ symmetry,  has been discussed in $d=4$ space-time already. This is symbolically stated as  the absence of a (well-defined left-handed) massless ``neutrino'' on the lattice~\cite{Nielsen-Ninomiya}, and the related {\em inevitable} appearance of species doublers seems to be often assumed in condensed matter physics also~\cite{hosur} \footnote{ The mechanism of the appearance of species doublers in condensed matter physics does not have a solid physical basis.}.

The notion of species doublers means that a single ``local'' fermion and thus a single ``local'' current defined on the lattice actually describes the multiple species of fermions, i.e., doublers  in momentum space in the limit $a\rightarrow 0$, if the chiral symmetry $\gamma_{5}$ is imposed on the underlying lattice theory.  (One may recall that the Wilson fermion \eqref{lattice QED} without chiral symmetry does not generate species doublers.)
For example, if one attempts to define a left-handed massless fermion
\begin{eqnarray}
\psi_{L}(x)=[(1-\gamma_{5})/2]\psi(x)
\end{eqnarray}
on the lattice, the species doubling implies that one inevitably obtains (at least) both a left-handed fermion and a right-handed fermion in momentum space in the limit $a\rightarrow 0$.  Similarly, $\psi_{R}(x)=[(1+\gamma_{5})/2]\psi(x)$ implemented on the lattice inevitably induces (at least) both  a right-handed fermion and a left-handed fermion in momentum space in the limit $a\rightarrow 0$.  As another aspect of the species doubling,
the chiral fermion such as $\psi_{L}(x)$, which would contain chiral gauge anomaly  \eqref{gauss} when coupled to a gauge field  in continuum theory and thus inconsistent by itself, contains no chiral gauge anomaly when placed on the lattice which cuts-off the short distances and thereby avoiding the inconsistency induced by the anomaly. Instead, we are forced to have species doublers for $a\neq0$ (and as a result, you may say that it becomes anomaly-free).  A massless Dirac fermion in continuum consists of  $\psi_{L}$ and $\psi_{R}$ and thus anomaly-free in (vector-like) gauge symmetry, but placed on the lattice both $\psi_{L}$ and $\psi_{R}$ would require their species doublers; for $a\rightarrow 0$, one would then have multiple species of vector-like Dirac  fermions (and thus, for example, the massless ``electron'' and  ``muon'' appear, although one originally intended to define only the ``electron''), and in this sense the construction is inconsistent.  In any case, the notion of species doubling on the lattice implies that one would obtain at least twice as many fermions in momentum space  in the limit $a\rightarrow 0$ than originally intended, if the chiral symmetry defined by $\gamma_{5}$ is imposed on the lattice Lagrangian. 

 On the other hand, a  recent progress of the Ginsparg-Wilson fermion, which arises from the idea of block transformation on the lattice, allows a definition of  a single Dirac fermion without doublers on the lattice~\cite{Neuberger1,Neuberger2}. See \cite{fujikawa-suzuki} for further references. The fermion field in such a theory is {\em exponentially local} ~\cite{Neuberger3, Hernandez} (stated intuitively, the construction does not appear to be manifestly local but actually becomes local in the limit $a\rightarrow 0$)  without any species doublers. In contrast, the original construction of the lattice fermion by K. Wilson is called {\em ultra-local} for which species doublers appear when $\gamma_{5}$ symmetry is imposed.  In the formulation of Ginsparg and Wilson, the exact chiral symmetry is realized by a new effective chiral operator $\Gamma_{5}$ that is regarded as a deformation of $\gamma_{5}$ for finite $a$. One then obtains a non-trivial index and a Jacobian as a symmetry breaking factor under chiral transformation using $\Gamma_{5}$ in the path integral formulation of chiral identity~\cite{Hasenfratz, luescher}. One thus obtains a correct anomaly in a natural manner in the continuum limit~\cite{Fujikawa-NP1999} and the path integral formulation of chiral anomalies become logically more consistent than in the case of the continuum theory (in the sense that the Jacobian is well-defined for a finite theory). A more detailed account of the Ginsparg-Wilson fermion shall be given later.
 
\begin{figure}[H]
\centering
\includegraphics[width=8cm]{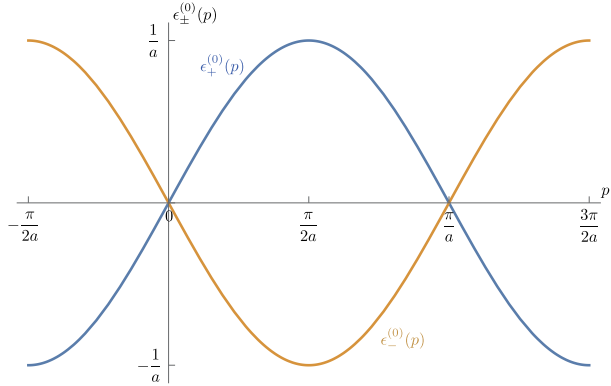}
\caption{\small  The energy spectrum of a $d=2$ massless Dirac fermion. $\epsilon^{(0)}_{\pm}(p)$ corresponds to chirality $\gamma_{5}=\pm 1$, respectively.}\label{Fig41}
\end{figure}

To understand what is going on in the species doubling intuitively, it is instructive to consider a $d=1+1$ dimensional lattice fermion.  We thus consider a Hamiltonian (i.e., continuous time) formulation of a free Dirac fermion based on the simplest discretization without the electromagnetic field for the moment
\begin{eqnarray}\label{two-dimension}
H=\sigma_{3}\frac{\sin ap}{a}+\sigma_{1}m_{0}+\sigma_{1}\frac{r}{a}(1-\cos ap)
\end{eqnarray}
with a constant $r$ which is called the Wilson parameter. See also Ambjorn et al. \cite{Ambjorn} and Nielsen and Ninomiya \cite{nielsen2} for a related analysis; we here comment briefly on this problem from our point of view.  Note that $H(p)$ in \eqref{two-dimension} has a period $2\pi/a$ in $p$. If one recalls that chiral $\gamma_{5}$ is given by $\sigma_{3}$ in this notation,
the first term is chiral invariant but the second and third terms do not commute with $\gamma_{5}$ and thus break chiral symmetry. The chiral symmetric Hamiltonian is thus given by setting $m_{0}=r=0$,
\begin{eqnarray}
H_{0}=\sigma_{3}\frac{\sin ap}{a}.
\end{eqnarray} 
We have the energy spectrum of $H_{0}(p)$ as
\begin{eqnarray}
\epsilon^{(0)}_{\pm}(p)=\pm \frac{\sin ap}{a}
\end{eqnarray} 
where $\epsilon^{(0)}_{\pm}(p)$ correspond to chirality $\gamma_{5}=\pm 1$, respectively.  Note that $\epsilon^{(0)}_{\pm}(p+2\pi/a)=\epsilon^{(0)}_{\pm}(p)$. See Fig.\ref{Fig41}. In terms of Weyl fermions, this exhibits the spectrum of  Weyl fermions (in $d=1+1$) for small $|\epsilon^{(0)}_{\pm}(p)|$, and $\epsilon^{(0)}_{\pm}(p)=0$ at $p=0$ and $p=\pi/a$ in the Brillouin zone. Moreover, $\epsilon^{(0)}_{+}(p)$ with $\gamma_{5}=1$ near $p=\pi/a$ has the same  structure as $\epsilon^{(0)}_{-}(p)$ with $\gamma_{5}=-1$ near $p=0$, and similarly, $\epsilon^{(0)}_{-}(p)$ near $p=\pi/a$ has the same  structure as $\epsilon^{(0)}_{+}(p)$ near $p=0$. 

\vspace{1mm}

 Namely, we have {\em species doubling} of the Weyl fermion for chiral invariant theory $H_{0 L}=-\frac{\sin ap}{a}$, for example, in the limit $a\rightarrow 0$. (In the conventional continuum theory, a single massless Dirac fermion in $d=2$ will contain one right-handed Weyl fermion and one left-handed Weyl fermion both at $p=0$.) 
 
 We may write the solution corresponding to $\epsilon^{(0)}_{-}(p)$ in the form
\begin{eqnarray}\label{doublers}
\psi_{L}(x)&=&\int_{-\pi/2a}^{\pi/2a}\frac{dp}{(2\pi)}e^{-ipx}\psi_{L}(p)+\int_{\pi/2a}^{3\pi/2a}\frac{dp}{(2\pi)}e^{-ipx}\psi_{L}(p)\nonumber\\
&=&\int_{-\pi/2a}^{\pi/2a}\frac{dp}{(2\pi)}e^{-i\epsilon^{(0)}_{-}(p)t+ipx^{1}}\psi_{L}(p)\nonumber\\
&+&e^{i\pi x^{1}/a}\int_{-\pi/2a}^{\pi/2a}\frac{dp}{(2\pi)}e^{-i\epsilon^{(0)}_{+}(p)t+ipx^{1}}\psi_{L}(p+\pi/a)\nonumber\\
&\equiv&e_{L}(x)+e^{i\pi x^{1}/a}\sigma_{1}\sigma_{3}\mu_{R}(x)
\end{eqnarray}
by choosing the Brillouin zone $-\pi/2a\leq p<3\pi/a$. We defined formally two fields $e_{L}(x)$ and $\mu_{R}(x)$ although they are actually part of a single field $\psi_{L}(x)$; when one discusses short distance properties  such as the chiral anomaly of $\psi_{L}(x)$, which imply the maximum extension in momentum space by uncertainty principle,  one cannot separate $e_{L}(x)$ and $\mu_{R}(x)$ for finite $a\neq 0$. We emphasize that $e_{L}(x)$ and $\mu_{R}(x)$ separately cannot define  well-defined local fields for $a\neq 0$ since half of the Brillouin zone is missing in them. Nevertheless, this notation is useful to understand the following discussions. 

We note that the chiral current and the fermion number current are anomaly-free
in the lattice regularized
\begin{eqnarray}
{\cal L}=\overline{\psi}_{L}(x)\frac{i}{2a}[\gamma^{\mu}U_{\mu}(x)e^{(-ia\hat{p}_{\mu})}  -e^{(-ia\overleftarrow{\hat{p}}_{\mu})}\gamma^{\mu}U^{\dagger}_{\mu}(x)] \psi_{L}(x)
\end{eqnarray}
in $d=2$ Euclidean space-time for $a\neq 0$. (In fact, one can show that each species doubler, which contains two fermion poles, is anomaly free for $a\neq 0$ by evaluating the Jacobian factor for each species doubler separately in momentum space; this is analyzed in the next subsection.) This is also consistent with the idea of chiral anomaly as a short-distance effect by K. Wilson \cite{Wilson}; the lattice theory which cuts-off the short distance effect does not contain the anomaly. The present lattice model  is thus completely anomaly-free, and all the symmetries specified by $\gamma_{5}$ of the Lagrangian or Hamiltonian hold in the naive form.

\begin{figure}[H]
\centering
\includegraphics[width=8cm]{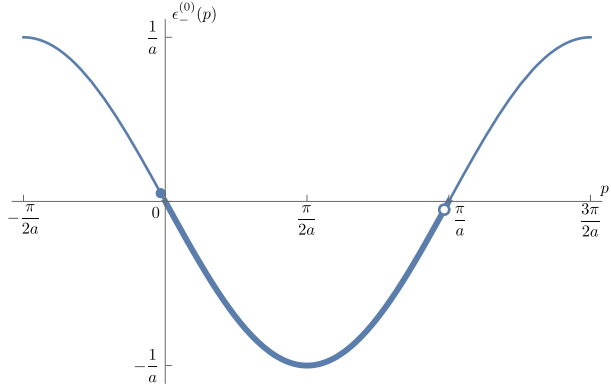}
\caption{\small  A schematic picture of the spectral flow in $d=1+1$ dimensions for the state with the energy spectrum $\epsilon^{(0)}_{-}(p)$. A particle is produced near $p=0$ and a hole is produced near $p=\pi/a$. }\label{Fig42}
\end{figure}

One may introduce an infinitesimal external gauge potential $A_{1}(t)$ which is spatially constant in the gauge $A_{0}=0$. One may then examine the semi-classical movement of the electron under the external uniform electric field $eE=e\partial_{t}A_{1}$ starting with the configuration $\epsilon^{(0)}_{-}(p)$ with all the negative energy states $0\leq p \leq \pi/a$ being filled initially as in Fig.\ref{Fig42}, in analogy with the motion of the electron in condensed matter physics.
\vspace{1mm}
\noindent A particle creation at the momentum close to $p=0$, driven by the Lorentz force $eE$ of the external weak field, implies a hole creation close to $p=\pi/a$ as in Fig.\ref{Fig42}, namely, we have a pair production
\begin{eqnarray}\label{d=2-pairproduction}
\psi_{L}+ \bar{\psi}_{L}\ \ \ {\rm or}\ \ \ e_{L}+ \bar{\mu}_{R}.
\end{eqnarray}
with the fermion number balance
\begin{eqnarray}\label{fermion-number balance}
\Delta N&=&eE\Delta t/(2\pi/L) - eE\Delta t/(2\pi/L)\nonumber\\
&=&\frac{e}{2\pi}\int_{0}^{L} dx\int_{0}^{\Delta t}dt \partial_{t}A_{1} -
\frac{e}{2\pi}\int_{0}^{L} dx\int_{0}^{\Delta t}dt \partial_{t}A_{1}=0
\end{eqnarray}
where the momentum increase $eE\Delta t$  by the exerted Lorentz force was divided by the level spacing $2\pi/L$ in momentum space assuming the size of the box to be $L$ to count the fermion number. We have no net fermion number production. 

\begin{figure}[H]
\centering
\includegraphics[width=8cm]{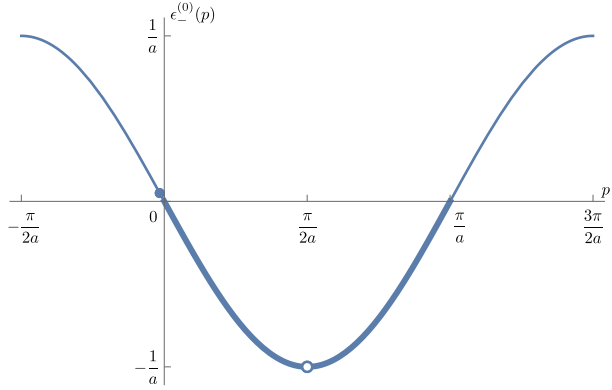}
\caption{\small  A schematic picture of a spectral flow in $d=1+1$ dimensions with a particle production near $p=0$ and a hole production near $p=\pi/2a$. }\label{Fig43}
\end{figure}

Since  
\begin{eqnarray}
\frac{e}{2\pi}\partial_{t}A_{1} =\frac{e}{4\pi}\epsilon_{\mu\nu}F^{\mu\nu}
\end{eqnarray}
in the $A_{0}=0$ gauge agrees with the chiral anomaly in the present $d=1+1$ case, one might be tempted to interpret the above formula \eqref{fermion-number balance} by saying that each species doubler produces the standard chiral anomaly but these anomalies are canceled among the species doublers, although the present lattice model is anomaly-free and thus all the considerations above are normal and naive ones; besides, no local field on the lattice with $a\neq 0$ is defined for each species doubler separately. It is shown that the above naive  expectation is {\em not} realized; if one wants to understand the anomaly of the species doubler in $-\pi/2a <p\leq \pi/2a $ in the present manner, one needs to consider the configuration in Fig.\ref{Fig43}.
  If one considers a particle production near the Fermi level $\epsilon_{-}=0$, one may consider this time that a hole is generated in the deep inside the Dirac sea at $p=\pi/2a$ by the spectral flow. \\

\vspace{1mm}

 \noindent Thus a particle-hole pair creation $e_{L} + \bar{e}_{L}$ with
 \begin{eqnarray}
\Delta N
&=&\frac{e}{2\pi}\int_{0}^{L} dx\int_{0}^{\Delta t}dt \partial_{t}A_{1} -
\frac{e}{2\pi}\int_{0}^{L} dx\int_{0}^{\Delta t}dt \partial_{t}A_{1}=0
\end{eqnarray} 
takes place driven by the Lorentz force of the external electric field $eE$ in $d=2$ rather than a net particle production which would be induced by the anomaly.  If one considers the other species doubler defined in $\pi/2a<p\leq 3\pi/2a$ together, the spectral flow fills the hole at $p=\pi/2a$ in Fig.\ref{Fig43} and produces a hole near $p = 3\pi/2a$, and  obtains $\Delta N=0$ again. Combined together one recovers the result in Fig.\ref{Fig42}, which is a well-defined and naive relation. In essence, one has basically a {\em single} fermion from a point of view of anomalies in the presence of species doubling for $a\neq 0$, although formally one has two fermions.

   To have a possible net particle production associated with the chiral anomaly of a Weyl fermion defined in $-\pi/2a<p\leq \pi/2a$, one has to generate the infinitely deep Dirac sea by setting $a\rightarrow 0$ first in Fig.\ref{Fig43} to have an infinite number of degrees of freedom and then a potential net particle production $\Delta N\neq 0$ would be allowed in the picture of spectral flow. Our view is that the spectral flow in a finite lattice theory cannot realize the chiral anomaly. Besides, to analyze the actual particle production, one would need to ensure the gauge anomaly cancellation by postulating some extra flavors.

Incidentally, the customary assumption of the inevitable  appearance of a pair of "Weyl fermions" (species doublers) in the band diagram in condensed matter physics, which are based on the pseudo-chiral symmetry (implied by the two neighboring bands), is not a prediction of the chiral symmetry of any basic theory such as a fundamental  Hamiltonian (i.e., a multi-electron Schr\"{o}dinger Hamiltonian)~\cite{nielsen2}. It is a consequence of {\em simulating} the spectrum in condensed matter physics by a chiral gauge theory on the lattice. 

\subsubsection{No well-defined anomaly for each species doubler for $a\neq 0$} 

In view of the anomaly calculation as a Jacobian and the analysis of simple examples of the spectral flow for $a\neq 0$ on the lattice in the preceding subsection  combined with the observation of K. Wilson as to anomalies as the  short-distance effects, we argue that the anomaly for each species doubler separately is not well-defined (or absent) for $a\neq 0$.  
We  re-examine the Jacobian calculation in \eqref{lattice anomaly1}.
The lattice version of the regularized Jacobian \eqref{anomaly1} for the localized chiral transformation $\psi(x)\rightarrow e^{i\gamma_{5}\alpha(x)}\psi(x)$ and $\overline{\psi}(x)\rightarrow\overline{\psi}(x) e^{i\gamma_{5}\alpha(x)}$ is given by 
\begin{eqnarray}\label{lattice anomaly-x}
\lim_{M\rightarrow \infty}a^{4}\sum_{x} 2i\alpha(x)\sum_{n=1}^{16}tr T^{-1}_{n}\gamma_{5}T_{n} \int_{-\pi/2a\leq k^{\mu}<\pi/2a}\frac{d^{4}k}{(2\pi)^{4}}e^{-ikx}\exp[-(\Dslash)^{2}/M^{2}]e^{ikx}\nonumber\\
\end{eqnarray}
where $
\Dslash = \frac{i}{2a}[\gamma^{\mu}U_{\mu}(x)e^{(-ia\hat{p}_{\mu})}  -e^{(-ia\overleftarrow{\hat{p}}_{\mu})}\gamma^{\mu}U^{\dagger}_{\mu}(x)] $. In this expression,
we used the 16 symmetry operators in \eqref{even operator} and \eqref{odd operator} and restricted the momentum range to the physical domain
$-\pi/2a\leq k^{\mu}<\pi/2a$
starting with the fundamental Brillouin zone \eqref{fundamental Brillouin zone}. 
If one adopts a view that the continuum limit defines a  physical theory, one would say that each species doubler gives the conventional chiral anomaly in the limit $a\rightarrow0$ (and $M\rightarrow \infty$ later), but because of the species doubling these anomalies cancel completely. Thus the chiral symmetric theory is anomaly free because of the species doubling on the lattice (actually, the species-wise cancellation of anomalies takes place even for $a\neq0$ in the present case due to the symmetry of the hypercubic lattice). This is the common interpretation of the absence of chiral anomaly on the lattice.

On the other hand, from a view point of adopting the theory with finite $a\neq 0$ to be physical, one may first examine the contribution of the physical domain specified by $T_{1}$  for fixed $a\neq 0$ in the above anomaly formula. For fixed $a$, the operator $\Dslash$ is a bounded operator, 
$||\Dslash|| = O(1/a)$. Thus the exponential factor goes away in the limit $M\rightarrow\infty$. In this limit, the volume of the momentum space is finite $(\pi/a)^{4}/(2\pi)^{4}$ for $a\neq0$, but the trace operation with $\gamma_{5}$
vanishes ${\rm tr}\gamma_{5}=0$. Namely, the contribution of each species doubler separately vanishes.  This is consistent with the picture of K. Wilson \cite{Wilson}, since the finite $a$ cuts off the short distance contributions  completely and thus no anomaly appears. 

As long as the notion of species doublers plays a central role, it implies that the spectrum is continuously connected beyond the boundaries of species doublers and the height of the boundaries of species doublers is finite. Thus the notions of species doublers, which imply $a\neq0$, and the well-defined chiral anomaly for each species doubler separately are not compatible. (To our knowledge, no explicit evaluation of the chiral anomaly for each species doubler separately has been performed in lattice gauge theory with fixed $a$ at $a\neq 0$.)  It is a safe statement that the chiral anomaly is not well-defined for each species doubler separately, which is defined in a part of the Brillouin zone and thus not a local field in space-time for $a\neq 0$.

\subsection{Modified chiral symmetry on the lattice: Ginsparg-Wilson fermion}

A definition of chiral symmetric fermion on the lattice using the idea of a block transformation has been proposed by Ginsparg and Wilson~\cite{Ginsparg-Wilson, Neuberger1, Neuberger2, Neuberger3, Hernandez}. This scheme utilizes the chiral symmetry generated by a new $\Gamma_{5}$ which is different from the conventional $\gamma_{5}$. This construction may  
be useful, not only conceptually but also practically, in considering an effective fermion such as a massless Dirac fermion (and possibly a Weyl fermion) without species doublers in condensed matter and related fields.
Although the construction of chiral fermions in this scheme is still not completed, some aspects of this construction are interesting. We would like to explain the basic aspects of this construction in this subsection. See also D. Kaplan \cite{Kaplan} for  more general schemes of chiral fermions and anomalies on the lattice.

\subsubsection{Dirac fermion theory without species doublers 
}

 The  Ginsparg-Wilson fermion is defined by 
\begin{eqnarray}\label{GWfermion}
S=\sum_{x,y}\bar{\psi}(x)D(x,y)\psi(y)
\end{eqnarray}   
on the Euclidean $d=4$ hypercubic lattice, for example; $D$ is a lattice generalization of the covariant Dirac derivative 
${\not \!\! D}$.
This fermion operator $D$ satisfies the Ginsparg-Wilson relation which is suggested by the block transformation~\cite{Ginsparg-Wilson}
\begin{eqnarray}\label{GWrelation}
\gamma_{5}D + D\gamma_{5} = aD\gamma_{5}D
\end{eqnarray} 
where $a$ stands for the lattice spacing. 
We also impose the hermiticity condition common in Euclidean theory
\begin{eqnarray}
(\gamma_{5}D)^{\dagger}=(\gamma_{5}D).
\end{eqnarray}
It is known that one can construct the operator $D$ that satisfies the Ginsparg-Wilson relation with the hermiticity condition  and  contains a single species in the fundamental Brillouin zone, which is mainly concentrated  in the physical  sub-domain
 \begin{eqnarray}
-\pi/2a\leq p_{\mu}<\pi/2a
\end{eqnarray}
without any species doublers. See Neuberger in \cite{Neuberger1,Neuberger2} for an explicit construction of the operator $D$ and the locality analysis \cite{Neuberger3, Hernandez}; see also ~\cite{Hasenfratz, luescher} and \cite{Fujikawa-NP1999} for the detailed analysis of the construction in connection with the anomaly evaluation. The following discussions are, however, understood without having the explicit form of $D$, as the analysis is mostly based on the formal relation \eqref{GWrelation} and the fact that the species doubling is absent. Intuitively, $D$ contains the Dirac operator appearing in the Wilson action \eqref{lattice QED} inside it (and thus no doubling with $2\pi/a$ periodic in momentum space) and one uses a modified chiral symmetry operator to define the exact chiral symmetry for $D$, since the chiral symmetry generated by $\gamma_{5}$ is broken in \eqref{lattice QED}.

One can then define the massless Dirac fermion for the vector-like gauge theory such as in QED and QCD (quantum chromodynamics) by a symbolic notation on the lattice
\begin{eqnarray}\label{vector-like Dirac fermion}
 \int {\cal D}\overline{\psi}{\cal D}\psi \exp\left[ \sum_{x,y}\bar{\psi}(x)D(x,y)\psi(y) \right]=
 \int {\cal D}\overline{\psi}{\cal D}\psi \exp\left[ \int \overline{\psi}D\psi \right].
 \end{eqnarray}
 One may define the operator 
\begin{eqnarray}
\hat{\gamma}_{5}=\gamma_{5}(1-aD), \ \ \ \hat{\gamma}_{5}^{2}=1
\end{eqnarray}
where we used the Ginsparg-Wilson relation \eqref{GWrelation}.
One may then define a (modified) chiral transformation
\begin{eqnarray}\label{chiral-transformation}
\psi\rightarrow e^{i\alpha \hat{\gamma}_{5}}\psi, \ \ \bar{\psi}\rightarrow \bar{\psi}e^{i\alpha \gamma_{5}}
\end{eqnarray}  
under which the action is invariant since
\begin{eqnarray}
\gamma_{5}D+D\hat{\gamma}_{5}=0
\end{eqnarray}
due to the Ginsparg-Wilson relation.
Under the transformation \eqref{chiral-transformation}, we have a Jacobian of the form (this natural appearance of the Jacobian is an advantage of the Ginsparg-Wilson construction, compared to the continuum case discussed in the proceeding subsection)
\begin{eqnarray}\label{lattice-jacobian}
\exp \left[ -2i{\rm Tr}\alpha\frac{1}{2}(\gamma_{5}+\hat{\gamma}_{5}) \right]=\exp\left[-2i{\rm Tr}\alpha\Gamma_{5} \right],
\end{eqnarray} 
where 
\begin{eqnarray}\label{new gamma5}
\Gamma_{5}\equiv \gamma_{5}\left( 1-\frac{1}{2}aD \right).
\end{eqnarray}
The above chiral symmetry \eqref{chiral-transformation} may formally be replaced by \cite{luescher} 
\begin{eqnarray}\label{chiral-transformation2}
\psi\rightarrow e^{i\alpha \Gamma_{5}}\psi, \ \ \bar{\psi}\rightarrow \bar{\psi}e^{i\alpha 
(1-\frac{1}{2}aD)\gamma_{5} }
\end{eqnarray}  
using the relation
\begin{eqnarray}
(1-\frac{1}{2}aD)\gamma_{5}D+D\Gamma_{5}=0
\end{eqnarray}
arising from the Ginsparg-Wilson relation and resulting in the same Jacobian factor \eqref{lattice-jacobian}.

To evaluate the Jacobian \eqref{lattice-jacobian}, one  may classify all the normalizable eigenstates 
of the hermitian operator on a finite lattice~\cite{Chiu} 
\begin{eqnarray}
H\equiv \gamma_{5}D,
\end{eqnarray}
namely,
\begin{eqnarray}
H\phi_{n}=\lambda_{n}\phi_{n}
\end{eqnarray}
with
\begin{eqnarray}
(\phi_{n},\phi_{m})=\phi^{\dagger}_{n}\phi_{m}\equiv \sum_{x}\phi^{\dagger}_{n}(x)\phi_{m}(x)=\delta_{nm}.
\end{eqnarray}
All the states on the lattice with a finite spacing $a$ are categorized into the following 3 classes
using the basic relation derived from \eqref{GWrelation}, which is an analogue of $\gamma_{5}\Dslash+\Dslash\gamma_{5}=0$ in continuum theory,
\begin{eqnarray}\label{chiral-algebra}
\Gamma_{5}H + H\Gamma_{5}=0
\end{eqnarray}
with $\Gamma_{5}=\gamma_{5}-\frac{1}{2}aH$ in \eqref{new gamma5}:\\
(i) Zero modes ($n_{\pm}$ states),
\begin{eqnarray}
H\phi_{n} = 0,\  \gamma_{5}\phi_{n} = \pm \phi_{n}, 
\end{eqnarray}
since such $\phi_{n} $ can be chosen as the eigenstate of $\Gamma_{5}$, $\Gamma_{5}\phi_{n} = \pm \phi_{n}$.\\
(ii) Highest states ($N_{\pm}$ states) with $\Gamma_{5}\phi_{n}=0$,
\begin{eqnarray}\label{higheststates}
H\phi_{n} = \pm\frac{2}{a}\phi_{n},\ \gamma_{5}\phi_{n} = \pm \phi_{n},
\end{eqnarray}
respectively. \\
(iii) Remaining paired states with $0 < |\lambda_{n}|< 2/a$,
\begin{eqnarray}\label{paired states}
H\phi_{n}=\lambda_{n}\phi_{n},\   H(\Gamma_{5}\phi_{n})=-\lambda_{n}(\Gamma_{5}\phi_{n}),
\end{eqnarray}
with $\Gamma_{5}\phi_{n}\neq 0$
and  $(\phi_{n}, \Gamma_{5}\phi_{n})=0$. Note that $\lambda_{n}$ and $-\lambda_{n}$ appear in pairs. Sandwiching the Ginsparg-Wilson relation \eqref{GWrelation} by $\phi^{\dagger}_{n}\gamma_{5}$ 
and $\phi_{n}$, one obtains $(\phi_{n}\gamma_{5},\phi_{n})=\frac{a}{2}\lambda_{n}$ for $\lambda_{n}\neq0$, and thus $2/a \geq |\lambda_{n}|$, namely, the cases analyzed above cover all the possible eigenvalues of $H$.   

 The sum rule 
\begin{eqnarray}
n_{+}+N_{+}=n_{-}+N_{-}
\end{eqnarray}
holds, which  is a result of
\begin{eqnarray}\label{sumrule}
{\rm Tr}\gamma_{5}=\sum_{n} (\phi_{n}, \gamma_{5}\phi_{n})=n_{+}+N_{+}-(n_{-}+N_{-})=0
\end{eqnarray}
that holds even for non-Abelian Yang-Mills fields in the finite dimensional functional space; this relation shows that no chiral anomaly for the symmetry generated by the fundamental $\gamma_{5}$ in the form of the Jacobian. Note that we used
\begin{eqnarray}
\sum_{n} (\phi_{n}, \gamma_{5}\phi_{n})=\sum_{n} (\phi_{n}, \Gamma_{5}\phi_{n}) + \sum_{n} (\phi_{n}, \frac{1}{2}H\phi_{n})=0
\end{eqnarray}
 for the states in \eqref{paired states}.
The quantum breaking of effective chiral symmetry (non-trivial Jacobian) in \eqref{lattice-jacobian} is described by the Atiyah-Singer-type index~\cite{Chiu}
\begin{eqnarray}\label{index}
{\rm Tr}\Gamma_{5}=\sum_{n} (\phi_{n}, \Gamma_{5}\phi_{n})=n_{+}-n_{-},
\end{eqnarray}
which implies the chiral asymmetry for $n_{+}-n_{-}\neq 0$.
The difference of $\Gamma_{5}$ and $\gamma_{5}$ is very important; the effective chiral symmetry $\Gamma_{5}$ projects out the highest states $N_{\pm}$ in \eqref{higheststates} on a finite lattice and realizes the chiral asymmetry \eqref{index}.  It may be natural to define the Hilbert space for $a\neq 0$ in the present scheme using $\Gamma_{5}$, which is a good symmetry of the action, then those  $N_{\pm}$ states are projected out from the evaluation of the index as in \eqref{higheststates} (and also in the smooth limit $a\rightarrow 0$). 

The index is also written with a regulator as
\begin{eqnarray}\label{index2}
\sum_{n} (\phi_{n}, \Gamma_{5}e^{-H^{2}/M^{2}}\phi_{n})=n_{+}-n_{-},
\end{eqnarray}
which is independent of the values of $M$. If one first considers $a\rightarrow 0$ and then $M\rightarrow \infty$, 
the index $n_{+}-n_{-}$ is given by the Chern-Pontryagin number, $\sim\int d^{4}x F\tilde{F}$ in the continuum limit for non-Abelian gauge theory, which is an analogue of the integral of $\frac{1}{4\pi^{2}}\vec{E}\cdot\vec{B}$ in Abelian theory. 
The index vanishes for Abelian theory, but we assume the formal index relation since a local version of ${\rm Tr}\Gamma_{5}$ gives the correct anomaly in the continuum limit; we define a local version of \eqref{index2} using the plane waves and the trace with respect to Dirac indices (and Yang-Mills  indices  in the case of non-Abelian theory)~\cite{Fujikawa-NP1999},
\begin{align}\label{local-index}
&\lim_{M\rightarrow \infty}\lim_{a\rightarrow 0}\int_{B} \frac{d^{4}p}{(2\pi)^{4}}{\rm tr} \left\{e^{-ipx}\Gamma_{5}\exp\left[ -\frac{(\gamma_{5}D)^{2}}{M^{2}} \right]e^{ipx} \right\}\nonumber\\
=&\frac{1}{4\pi^{2}}\vec{E}\cdot\vec{B}
\end{align} 
where the momentum integral is over the fundamental Brillouin zone; the eigenvalues of $H=\gamma_{5}D$ blow up outside the physical domain $-\pi/2a\leq p_{\mu}<\pi/2a$ in the limit $a\rightarrow 0$ first and this information relies on the specific construction of $D$ in addition to the Ginsparg-Wilson relation. In the physical domain, $\gamma_{5}D\rightarrow \gamma_{5}\Dslash$ in the limit $a\rightarrow 0$. Also, $\Gamma_{5}\sim \gamma_{5} -\frac{1}{2}aM \rightarrow \gamma_{5}$ inside the expression \eqref{local-index} in the limit $a\rightarrow 0$ with $M$ kept finite, and one thus recovers the continuum formula of the Jacobian evaluation for  $\gamma_{5}$,
\begin{eqnarray}\label{anomaly3}
\lim_{M\rightarrow \infty}{\rm Tr}\{ 2i\alpha(x)\gamma_{5}\exp[-\Dslash^{2}/M^{2}]\}
=\int d^{4}x~\alpha(x)\frac{1}{4\pi^{2}}\vec{E}\cdot\vec{B}.
\end{eqnarray}
The (formally integrated) relation, which is an analogue of the Atiyah-Singer index theorem, 
\begin{eqnarray}
n_{+}-n_{-}=\int d^{4}x~\frac{1}{4\pi^{2}}\vec{E}\cdot\vec{B}
\end{eqnarray}
 is concluded from \eqref{index2}.
 
  As for the relation 
$ {\rm Tr}\gamma_{5}=\sum_{n} (\phi_{n}, \gamma_{5}\phi_{n})=n_{+}+N_{+}-(n_{-}+N_{-})=0$, this is a specific relation for a lattice with $a\neq 0$. But it contains information on an anomaly relation as above in the limit $a\rightarrow 0$ with a suitable regularization, which replaces $N_{+}-N_{-}$ by the anomaly factor in the limit $a\rightarrow 0$. In fact, if one starts with ${\rm Tr}\Gamma_{5}$,
\begin{align}
{\rm Tr}\Gamma_{5}&={\rm Tr} \left(\gamma_{5}-\frac{a}{2}H \right)\nonumber\\
&={\rm Tr} \left( -\frac{a}{2}H \right)\nonumber\\
&=-(N_{+}-N_{-})\nonumber\\
&=\left( -\frac{1}{2} \right) \int d^{4}x ~tr \gamma_{5} \int_{B} \frac{d^{4}k}{(2\pi)^{4}}e^{-ikx}aDe^{ikx}
\end{align}
and using the explicit formula for the Ginsparg-Wilson operator $D$, one can show that 
\begin{eqnarray}
-(N_{+}-N_{-})=\lim_{a\rightarrow 0} \left( -\frac{1}{2} \right) tr \gamma_{5} \int_{B} \frac{d^{4}k}{(2\pi)^{4}}e^{-ikx}aDe^{ikx}= -\int d^{4}x ~\frac{1}{4\pi^{2}}\vec{E}\cdot\vec{B}.
\end{eqnarray}

We have thus demonstrated that the massless Dirac fermion for vector-like gauge theory such as QCD and QED can be formulated using the Ginsparg-Wilson fermion without any species doublers by maintaining the {\em exact chiral symmetry} defined by $\gamma_{5}$ and $\hat{\gamma}_{5}$ (or essentially $\Gamma_{5}$), and the chiral anomaly is evaluated as a Jacobian factor  as in continuum theory and in fact in a more logically consistent manner. 
This is a major achievement of the Ginsparg-Wilson fermion.

This final result may be compared with the formulation with the Wilson action using $\gamma_{5}$ in \eqref{lattice QED}; the same anomaly relation is obtained by the chiral symmetry defined by $\gamma_{5}$ in the continuum limit $a\rightarrow 0$, which is however strongly broken by the large effective masses  of the order $r/a$  of species doublers for any finite $a\neq 0$ in the Wilson action.  The major issue is the compatibility of the presence of exact chiral symmetry and the absence of species doublers.

\subsubsection{Lattice Chiral fermions without species doublers}
      
It is interesting to examine if one can incorporate the chiral fermions (Weyl fermions) in the scheme of Ginsparg and Wilson. 
 We define two projection operators,
 \begin{eqnarray}\label{projectors}
P_{\pm}=\frac{1}{2}(1\pm \gamma_{5}),\ \ \ \hat{P}_{\pm}=\frac{1}{2}(1\pm \hat{\gamma}_{5})
\end{eqnarray}  
and define the chiral components by 
\begin{eqnarray}\label{chiral-components}
\psi_{R,L}=\hat{P}_{\pm}\psi, \ \ \bar{\psi}_{L,R}=\bar{\psi}P_{\pm}
\end{eqnarray}
which satisfy (in a simplified notation)
\begin{eqnarray}\label{GWfermion2}
S=\int \bar{\psi}_{L}D\psi_{L}+\int \bar{\psi}_{R}D\psi_{R}
\end{eqnarray} 
using  the identity arising from the Ginsparg-Wilson relation
\begin{eqnarray}\label{decomposition of Ginsparg-Wilson operator}
D=P_{+}D\hat{P}_{-}+P_{-}D\hat{P}_{+}.
\end{eqnarray}
We thus define the left-handed massless fermion on the lattice by 
\begin{eqnarray}\label{chiral fermion}
\int {\cal D}\overline{\psi}_{L}{\cal D}\psi_{L} \exp \left[ \int \overline{\psi}_{L}D\psi_{L} \right]
\end{eqnarray}
which satisfies the exact chiral symmetry generated by $P_{+}$ and $\hat{P}_{-}$ and no species doublers appear.
The present definition of the chiral fermion slightly differs from the conventional definition in continuum theory, in that we do not define the generator of chiral gauge symmetry in the present construction. We instead project the vector-like theory by  gauge field dependent chiral projection operators to define a chiral fermion theory. 

In the interacting case with gauge fields included in $D(U)$, one may tentatively assume 
\begin{eqnarray}\label{chiral fermion2}
\int {\cal D}U{\cal D}\overline{\psi}_{L}{\cal D}\psi_{L} \exp\left[\int \overline{\psi}_{L}D\psi_{L}\right].
\end{eqnarray}
As for the fermion number anomaly associated with
\begin{eqnarray}
\psi_{L}\rightarrow e^{i\alpha(x)}\psi_{L}, \ \ \  \overline{\psi}_{L}\rightarrow \overline{\psi}_{L}e^{-i\alpha(x)}
\end{eqnarray}
one has the Jacobian
\begin{eqnarray}
J=\exp[-{\rm Tr}i\alpha(x)\hat{P}_{-} +{\rm Tr} i\alpha(x) P_{+}]=\exp[i {\rm Tr} \alpha(x) \Gamma_{5}].
\end{eqnarray}
This anomaly may be evaluated in the continuum limit just as \eqref{local-index}. In the continuum limit $a\rightarrow 0$ for a smooth gauge field configurations, it is confirmed that one recovers the continuum path integral formula with the correct fermion number anomaly without encountering the species doublers. In this sense, the present scheme has (partially) achieved an important task to define a lattice chiral fermion without species doublers with a desired properties of a chiral fermion in the continuum limit.

However, the truly satisfactory definition of the path integral for a chiral fermion is still missing. The remaining issue is to define the path integral measure to analyze the anomalies associated with gauge symmetry. Simply stated, it is not easy to generate the consistent form of gauge anomalies in the path integral defined on the lattice. Since the lattice theory is supposed to define everything finite and thus consistently, the final formulation needs to produce the consistent form of gauge anomaly by a Jacobian factor.
This task has been achieved for an Abelian gauge theory  ~\cite{luescher2, suzuki}, but the formulation for non-Abelian theory is still missing. In continuum theory, one usually defines the so-called covariant form of anomaly which is sufficient to analyze the anomaly cancellation in the Standard Model \cite{Gross-Jackiw, Witten}; in this case, one can treat the Jacobian for the gauge anomaly by imitating the analysis of Atiyah-Singer-type index, namely, as in the manner we have done so far in this article. There is a well-defined prescription to convert the covariant form of anomaly, which is easy to evaluate, to the consistent form of anomaly in continuum theory~\cite{fujikawa-suzuki}; this fact shows that the anomaly cancellation condition is the same for both forms of anomalies in continuum theory.

 Technically, the Ginsparg-Wilson operator is exponentially local \cite{Neuberger3, Hernandez} while the conventional lattice Lagrangian is ultra-local; this difference may be another reason why one evades the appearance of species doubling in the Ginsparg-Wilson scheme.

\subsubsection{Comment on Ginsparg-Wilson fermion and related issues}

It is interesting to examine if one can use the Ginsparg-Wilson fermion \cite{Ginsparg-Wilson, Neuberger1, Neuberger3, Hernandez}, which contains no species doubling, to {\em simulate} an effective fermion  in the context of practical applications \footnote{The idea of species doublers asserts that even a massless Dirac fermion (not a chiral fermion) such as in massless QED cannot avoid the appearance of species doublers in the conventional lattice formulation.}.  
This is compared with the original proposal of Wilson \eqref{lattice QED} which gives the correct chiral anomaly but the chiral symmetry defined in terms of $\gamma_{5}$ is broken  strongly in the Lagrangian.
The massless Dirac fermion, which is defined without technical complications, is given by \eqref{vector-like Dirac fermion}
\begin{eqnarray}\label{vector-like Dirac fermion-2}
 \int {\cal D}\overline{\psi}{\cal D}\psi \exp\left[ \int \overline{\psi}D\psi \right].
 \end{eqnarray}
This construction, when one should suitably rotate back to the Minkowski theory,  may be useful in the applications to the condensed matter physics, for example. The conventional chiral $U(1)$ anomaly and consistent chiral anomaly, if necessary, are evaluated.  
One may simulate a massless Dirac fermion (such as Dirac semi-metal) coupled to the electromagnetic field without worrying about the species doubling.

The Ginsparg-Wilson fermion gave an impressive progress in the lattice gauge theory, but not complete yet. In this respect, one may still examine other constructions \cite{Kaplan}, and also the staggered fermions \cite{Kogut-Susskind, Susskind} and its generalizations, for example. The situation in the staggered fermion is not rosy, as was forcefully criticized by M. Creutz \cite{Creutz} among others. But the issue of constructing satisfactory lattice fermions is still alive and remains interesting.

\section{Some examples in nuclear physics and related fields}
\subsection{Species doubling and chiral anomaly}
We first comment on  the general aspects of the proposal of Nielsen and Ninomiya \cite{nielsen2} on the possible implications of chiral anomalies in condensed matter physics. This paper contains many stimulating ideas and  it has been very influential in the recent developments, although this paper does not mention Berry's phase explicitly. The main theme of this paper is the use of a chiral fermion in lattice gauge theory with the notion of species doublers to {\em simulate} the Weyl fermion in condensed matter physics.

The authors also made important contributions to the subject of species doubling \cite{Nielsen-Ninomiya}. They argued that the species doubling is an inevitable consequence of the definition of fermions on a general class of lattice. They make use of 
the torus structure of the momentum space description of lattice fermions, for example, 
\begin{eqnarray}
-\frac{\pi}{2a}\leq p_{\mu}<\frac{3\pi}{2a}, 
\end{eqnarray} 
in the case of the hypercubic lattice (in their $1+3$ lattice setting \cite{Kogut-Susskind}, they use the 3-dimensional torus) and the important assumption of the existence of the well-defined chiral charge or charge density specified by $\gamma_{5}$. The existence of the well-defined chiral charge implies the absence of chiral anomaly, as is mentioned below. 
They also analyze an interesting construction of chiral fermions starting with the two-level crossing in  lattice theory, which is similar to our derivation of Berry's phase in the two-level crossing problem in Section 2.
Interested readers are referred to the original work \cite{Nielsen-Ninomiya}. 

Stated simply for an explicit example, the notion of species doublers means that the chiral fermion which is invariant under the chiral symmetry generated by $\gamma_{5}$
\begin{eqnarray}\label{lattice chiral fermionS5}
S=\int d^{4}x ~\overline{\psi}(x) \left[ \gamma^{\mu}i(\partial_{\mu} -ieA_{\mu}) \right]\left( \frac{1-\gamma_{5}}{2} \right) \psi(x),
\end{eqnarray}
inevitably contains multiple species of chiral fermions when transcribed to a theory defined on the lattice by preserving $\gamma_{5}$ symmetry, which is a generalization of the analysis of K. Wilson\cite{Wilson-1974}. In the present case defined on a 4-dimensional hypercubic lattice, which is Lorentz invariant in the continuum limit, one in fact obtains 8 left-handed fermions and 8 right-handed fermions  as has been confirmed by an elementary analysis in Section 4. The crucial assumption is the existence of chiral symmetry generated by $\gamma_{5}$. We have already seen in Section 4 that the original lattice model of K. Wilson \cite{Wilson-1974}, which does not have the chiral symmetry, does not contain species doublers  for $a\rightarrow0$. 

The action \eqref{lattice chiral fermionS5} transcribed to a theory on a lattice satisfies the naive chiral identity, namely, 
\begin{eqnarray}\label{No anomalyS5}
\partial_{\mu}J^{\mu}_{L}=0
\end{eqnarray}
for a chiral current $J^{\mu}_{L}$ defined on the lattice. The theory defined on the lattice is well-regularized, and if chiral symmetry is preserved, leads to an anomaly-free Ward-Takahashi identities when combined with ${\rm Tr}\gamma_{5}=0$, which ensures the vanishing Jacobian. The above conservation implies that both the charge conservation and the fermion number conservation are satisfied in a naive form.

Coming back to the analysis of possible implications of anomalies in condensed matter physics in \cite{nielsen2}, their basic assumption is the realization of chiral anomalies in terms of spectral flow. This is partly because it is not simple to evaluate chiral anomalies directly in their scheme of Weyl fermions on the lattice. As we emphasized in Section 4, a local field in space-time is not defined for each species doubler
separately, since the species doubler is defined only in a part of the Brillouin zone in momentum space; if one should be able to define a local field in space-time using only a part of the Brillouin zone, one would define such a field from the beginning instead of using the entire fundamental Brillouin zone.  This loss of locality of each species doubler separately suggests that the chiral anomaly, which is a short-distance effect \cite{Wilson}, is not well-defined for
each species doubler separately \footnote{Very intuitively, a left-handed chiral fermion on the lattice in the scheme of species doubling is left-handed in a half of the Brillouin zone and right-handed in the other half of the Brillouin zone.}. In fact, we illustrated this fact for a simple $1+1$ dimensional model in Fig.4.2 and Fig.4.3 in Section 4. A semi-classical analysis with the assumed classical equations of motion in momentum space works to explain some aspects of chiral anomaly such as the particle creation of a left-handed fermion in the case of continuum theory, for which one can confirm the existence of the chiral anomaly by the conventional method. But it does not work in lattice theory with $a\neq 0$ where the chiral charge is assumed to be well-defined with the appearance of species doublers; the well-defined chiral charge implies the absence of chiral anomaly.  The spectral flow instead implies the pair production in a general sense in the case of a  smooth spectrum covering the neighboring species doublers in lattice theory; in a theory with well-defined chiral charges and thus with no chiral anomalies \eqref{No anomalyS5}, it is natural that the spectral flow does not generate chiral anomalies. A statement equivalent to this fact is also found in \cite{nielsen2}.

We thus understand that the analysis in \cite{nielsen2} suggests  that the analogue of chiral anomaly in condensed matter physics is the pair production in a general sense in lattice gauge theory. In fact, they state in Abstract of their paper \cite{nielsen2}, `` 
For such materials, in the presence of parallel electric and strong magnetic
fields, there exists an effect similar to the ABJ  (Adler-Bell-Jackiw) anomaly that is the movement of the electrons in the energy-momentum
space from the neighborhood of one degeneracy point to another one'', which is a very accurate and interesting observation, to be consistent with \eqref{No anomalyS5}. This effect may be termed properly as {\em phenomenon similar} to the chiral anomaly, although  in their lattice model one has no chiral anomaly for a local current due to the appearance of species doublers.  

In the case of the Ginsparg-Wilson fermion \cite{Ginsparg-Wilson, Neuberger1, Neuberger3, Hernandez} discussed in Section  4, we have shown that one can construct a suitable latticized vector-like QCD (or QED) Lagrangian with chiral symmetry but  without species doublers, which evades the no-go argument of Nielsen and Ninomiya, but with modified chiral operator $\Gamma_{5}$ and the action is defined with a more general notion of locality. As for the chiral fermion, one can construct a single Weyl fermion without species doublers for the Abelian gauge theory using the Ginsparg-Wilson fermion. But one then encounters the chiral anomaly even for the lattice fermion arising from ${\rm Tr}\Gamma_{5}\neq 0$ and has to add other ``flavors'' to cancel the gauge anomaly, just as in the Standard Model in continuum  where the anomaly cancellation among leptons and quarks is essential. Thus the doubling of Weyl fermions in a  different context is required if one asks the absence of chiral gauge anomalies in the framework of  the Ginsparg-Wilson fermion in lattice theory. The difference between the Ginsparg-Wilson fermion and the conventional lattice fermion with species doublers is that we have a well-defined anomaly for the chiral fermion in the Ginsparg-Wilson fermion and thus it should be canceled by other flavors, while in the construction with species doublers with $\gamma_{5}$,  the absence of anomaly is ensured by the appearance of species doublers which are not local fermion fields separately for $a\neq 0$.

\subsection{Chiral magnetic effect}

We comment on  the ``chiral magnetic effect'' first suggested by Nielsen and Ninomiya in condensed matter physics \cite{nielsen2}, although the term ``chiral magnetic effect'' was not used. The early use of the term ``chiral magnetic effect'' appears in  Kharzeev, McLerran and Warringa \cite{Kharzeev} with an emphasis on the effect of the sphaleron. The notion of the sphaleron is crucial and central to the chiral magnetic effect in nuclear physics, while no notion of the sphaleron appears in the condensed matter contents. In this sense, these two phenomena should better be distinguished, although we use the term ``chiral magnetic effect'' for both phenomena for simplicity.

 We  first mention the chiral magnetic effect in condensed matter physics, as one of the concrete physical suggestions \cite{nielsen2}. This effect is generically written in the form
\begin{eqnarray}
J_{k}= \frac{e^{2}}{4\pi^{2}}(\mu_{R}-\mu_{L}) B_{k}
\end{eqnarray}
where $J_{k}$ stands for the electromagnetic current induced by a strong external magnetic field $B_{k}$ and  the difference $(\mu_{1}-\mu_{2})$ of the chemical potentials of two states with different chirality in the thermal equilibrium \cite{nielsen2}.  The crucial aspect of this relation is that the induced current is proportional to the direction of the external magnetic field. The current thus satisfies $\vec{\nabla}\cdot \vec{J}=0$ naively.
The  chiral magnetic effect in the quark-gluon plasma is written in a similar form under a suitable assumption and can be tested by
experiments~\cite{fukushima}.
Our interest in the context of the present review is whether this relation is regarded as a consequence of  quantum anomalies and possibly Berry's phase.

\subsubsection{Chiral magnetic effect in condensed matter physics}
We examine the analysis of Nielsen and Ninomiya \cite{nielsen2}. As we have already emphasized, their setting of a lattice model does not contain any chiral anomalies in the conventional sense. They assume that the idea of chiral invariant species doublers in lattice gauge theory is applicable to simulate the condensed matter physics defined in the Brillouin zone, and thus the species doublers ensure the absence of chiral anomalies. Intuitively speaking, their analysis corresponds to the chiral Lagrangian
\begin{eqnarray}\label{chiral fermion on the latticeS5}
S=\int dt d^{3}x ~\overline{\psi}(x) \left[\gamma^{0}i(\partial_{t} -ieA_{0}) + \gamma^{k}i(\partial_{k} -ieA_{k}) \right] \left(\frac{1-\gamma_{5}}{2} \right)\psi(x)
\end{eqnarray}
{\em placed on a lattice}. This Lagrangian is invariant under the chiral transformation defined by $\gamma_{5}$ and that all the naive operations are well-defined in a theory defined on the discrete lattice, and thus the chiral identities hold without any anomalies if one uses ${\rm Tr}\gamma_{5}=0$. 
Moreover, the local chiral fermion number current and the local chiral electromagnetic current are proportional to each other. One thus obtains the single relation on the lattice
\begin{eqnarray}\label{chiral current conservationS5}
\partial_{\mu}J^{\mu}_{L}(x)=0
\end{eqnarray}
as already mentioned in \eqref{No anomalyS5}. One naively encounters the  left- and right-handed fermions in momentum space  because of the assumed species doubling on the lattice, but these species doublers are not local fields in the real space-time for $a\neq0$. One thus has  no anomaly relations, and only the conserved local current as in \eqref{chiral current conservationS5}. 
This is different from the massless Dirac fermion {\em in continuum} where one encounters two Weyl fermions
\begin{align}\label{continuum massless DiracS5}
S=&\int d^{4}x \overline{\psi}(x) \left[\gamma^{0}i(\partial_{t} -ieA_{0}) + \gamma^{k}i(\partial_{k} -ieA_{k}) \right] \left(\frac{1-\gamma_{5}}{2} \right)\psi(x) \nonumber\\
&+\int d^{4}x ~\overline{\psi}(x) \left[\gamma^{0}i(\partial_{t} -ieA_{0}) + \gamma^{k}i(\partial_{k} -ieA_{k}) \right] \left(\frac{1+\gamma_{5}}{2}\right)\psi(x)
\end{align}
for which one can define two local currents with an anomaly \cite{anomaly1, anomaly2}
\begin{eqnarray}
\partial_{\mu}J^{\mu}(x)=0, \ \ \ \partial_{\mu}J^{\mu}_{5}(x)= \frac{e^{2}}{2\pi^{2}}\vec{E}\cdot\vec{B}.
\end{eqnarray}
From the point of view of chiral anomalies, these two theories are very different. Besides, one  encounters more than two Weyl-type fermions in the momentum space in \eqref{chiral fermion on the latticeS5} and two Weyl fermions in \eqref{continuum massless DiracS5}.

They then introduce the idea of the spectral flow in the setting of \eqref{chiral fermion on the latticeS5} \cite{nielsen2}. The spectral flow can describe the essence of the chiral anomaly appearing in a theory which gives rise to chiral anomalies in the conventional evaluation, by analyzing the local properties in momentum space, namely, on the basis of the behavior near the origin of the momentum space.  But the spectral flow does not generate the chiral anomaly for a theory which has no anomalies in the conventional sense. The spectral flow is based on the behavior near the origin, while the chiral anomaly is sensitive to the global behavior of the spectrum, in particular, the ultraviolet behavior of the spectrum. 

In their analysis, the spectral flow thus induces the pair production in a general sense, as we already emphasized (and also mentioned in their paper \cite{nielsen2}); the species doubler separately, which is not a local field for $a\neq 0$, does not give well-defined anomalies in the conventional sense.  The analysis in \cite{nielsen2} is thus regarded as a semi-classical analysis in a theory which is intrinsically anomaly-free.
Thus technical details of their analysis aside, their results do not contain the chiral anomaly as we understand in field theory (except for the limit $a\rightarrow0$ for which species doublers become dis-connected chiral fermions, but then their interpretation in the context of condensed matter physics is not obvious.). But the pair production in a general sense gives rise to interesting effects such as the chiral magnetic effect.

One thus starts with the conservation condition of the chiral current in \eqref{chiral current conservationS5}.
This conservation means two facts. Firstly, the possible gauge anomaly does not appear, and one can thus ensure the charge conservation condition. Secondly, the  chiral fermion number current has no anomaly, namely, no net particle production. As we have emphasized, each species doubler is not a local field in space-time and thus the notion of anomaly for each species doubler separately is ill-defined for $a\neq 0$.  Nevertheless they apply a classical spectral flow to each species doubler separately and infer the particle creation rate which is the same as the chiral anomaly naively assumed for each species doubler separately, namely, the particle creation rate under the parallel $E$ and $B$
\begin{eqnarray}
\pm \frac{e^{2}}{4\pi^{2}}EB
\end{eqnarray}
for the right-handed and left-handed species doubler, respectively; this value is sensible in the limit $a\rightarrow 0$, and may be that it is a reasonable guess without any other ways to estimate it. Assuming that the external electric force $E$ pushes up the Fermi surface of the right-handed doubler  $\mu_{R}$ and pushes down the left handed doubler $\mu_{L}$, they obtain the energy balance condition for the generated electromagnetic current $J_{A}$
\begin{eqnarray}
EJ_{A}=\frac{e^{2}}{4\pi^{2}} EB (\mu_{R} - \mu_{L}),
\end{eqnarray}
namely,
\begin{eqnarray}\label{chiral magnetic effect1}
J_{A}=\frac{e^{2}}{4\pi^{2}} B (\mu_{R} - \mu_{L})
\end{eqnarray}
which is a generic form of the proposed chiral magnetic effect in a theory of two species doublers of Weyl-type fermions.
They also define the relaxation time $\tau_{1}$ for the transfer from the right-handed Weyl cone into the left-handed  Weyl cone, which are described by species doublers,  by the external electric field $E$
\begin{eqnarray}\label{chemical potentialS5}
(\mu_{R} - \mu_{L})=evE\tau_{1}
\end{eqnarray}
with the effective velocity $v$ of the Weyl fermion appearing at the level crossing point 
\begin{eqnarray}
\epsilon(p)=\pm v\vec{p}\cdot\vec{\sigma}.
\end{eqnarray}
Readers are referred to the original paper for more details \cite{nielsen2}.

If one combines \eqref{chemical potentialS5} with \eqref{chiral magnetic effect1}, the magnitude of anomalous vector current $J_{A}$ is formally proportional to $EB$ and in this sense  analogous to the anomalous particle creation itself (or more precisely, the pair production rate in their model) but now dealing with an electromagnetic  current.

 The derivation of \eqref{chiral magnetic effect1} is based on a semi-classical analysis in an anomaly-free (lattice) theory by assuming a specific form of the particle creation rate for each species doubler separately implied by the idea of spectral flow. The creation rate is reasonable for a semi-classical flow of the particle, i.e., a pair production rate in a general sense induced by  the applied $EB$. The notion of quantum anomaly we know in field theory does not appear,  but nevertheless an analogue of the chiral anomaly appears and leads to a  suggestion of the stimulating phenomenon \eqref{chiral magnetic effect1}.

\subsubsection{Chiral magnetic effect in quark-gluon plasma}

We next discuss the derivation of the chiral magnetic effect in the quark-gluon plasma. This effect has been studied in connection with the effect of the sphaleron such as in~\cite{Kharzeev}. More recent analyses of the effect in the context of nuclear physics are found in 
\cite{Moore, Mace, Mueller}, for example.
We comment on the derivation of the chiral magnetic effect in the quark-gluon plasma following Fukushima, Kharzeev and Warringa (FKW) \cite{fukushima}.  They assume that the chiral chemical potential $\mu_{5}$, which gives rise  to the left-right asymmetry, is generated by some mechanism (such as the spharelon) in the basic QCD; their scheme is thus logically more transparent and differs in an important way from the chiral magnetic effect we discussed in the context of condensed matter physics. They then examine the mechanism of the chiral magnetic effect induced by the external strong magnetic field $B$ in great detail; they present several different ways to derive the formula of the proposed chiral magnetic effect and show that all of them lead essentially to the same conclusion. 

They apply first the idea of the (justified) spectral flow to the Dirac fermion in \eqref{continuum massless DiracS5}. They then infer the right-hand and left-handed particle creation  at 
\begin{eqnarray}\label{spectral flowS5}
\pm \frac{p_{F}}{2\pi}\cdot \frac{eB}{2\pi} =\pm \frac{e^{2}}{4\pi^{2}}\vec{E}\cdot\vec{B}t
\end{eqnarray}
using the transverse Landau level density $eB/2\pi$ and the longitudinal momentum change near the Fermi surface by the applied electric field $p_{F}=eEt$.  The rate of the change of the chiral particle number $N_{5}=N_{R}-N_{L}$ is then given by 
\begin{eqnarray}\label{spectral flow2S5}
\frac{d^{4}N_{5}}{dtd^{3}x}=\frac{e^{2}}{2\pi^{2}}\vec{E}\cdot\vec{B}
\end{eqnarray}
which agrees with the rate given by the chiral anomaly for the Dirac fermion \eqref{continuum massless DiracS5} in continuum.  When one writes the chiral chemical potential $2\mu_{5}=\mu_{R}-\mu_{L}$, the energy balance of the particle creation is given by 
\begin{eqnarray}\label{energy balanceS5}
\int d^{3}x ~\vec{E}\cdot\vec{j}=\mu_{5}\frac{dN_{5}}{dt}=\frac{e^{2}\mu_{5}}{2\pi^{2}}\int d^{3}x ~\vec{E}\cdot\vec{B}
\end{eqnarray}
 where $\vec{j}$ is a current which is supposed to be generated by $\mu_{5}$ and the chiral anomaly. By choosing $\vec{E}$ constant and parallel to $\vec{B}$, one obtains the formula of the chiral magnetic effect
 \begin{eqnarray}\label{integrated chiral magnetic effect}
 \vec{J}=\int d^{3}x ~\vec{j}= \frac{e^{2}\mu_{5}}{2\pi^{2}}\int d^{3}x ~\vec{B}. 
\end{eqnarray}
They then set $\vec{E}=0$ in the final formula. 

This derivation is close to that of Nielsen and Ninomiya \cite{nielsen2}, but there are important differences; namely, the chiral anomaly for two Weyl fermions separately exists in the present case, which one may evaluate in the conventional manner also, and thus one may confirm that the estimate by the spectral flow agrees with the chiral anomaly. Secondly, they set $\vec{E}=0$ in the end since $\mu_{5}$ is assumed to be generated by QCD; this makes the chiral magnetic effect  induced purely by the magnetic field. This derivation may show that the chiral magnetic effect is the consequence of the chiral anomaly for the Dirac fermion, provided that $\mu_{5}$ is generated by QCD separately. 

In this derivation, two technical questions come to our mind: Firstly, if the effect arises from the genuine chiral anomaly, which is valid for any space-time dependence of $A_{\mu}(t,\vec{x})$ allowed in the model,  the general (very rapid) time dependence of $A_{\mu}(t,\vec{x})$ might influence the final result. Secondly, the presence of $\mu_{5}$ in the action may influence the chiral anomaly evaluation itself. As for the first question, one may choose $B$ as $F_{12}(t,\vec{x})$ and $E$ as $F_{03}(t,\vec{x})$, and then $\vec{E}$ and $\vec{B}$ are chosen almost independently. One may thus conclude the local relation
\begin{eqnarray}\label{local version of chiral magnetic effect}
 \vec{j}(t,\vec{x})= \frac{e^{2}\mu_{5}}{2\pi^{2}} \vec{B}(t,\vec{x}) 
\end{eqnarray}
from \eqref{energy balanceS5} instead of the integrated version \eqref{integrated chiral magnetic effect}, and one may conclude 
\begin{eqnarray}
 \vec{\nabla}\cdot\vec{j}(t,\vec{x})= 0
 \end{eqnarray}
 which is a strong constraint. This may imply that the relation \eqref{energy balanceS5}, which is not manifestly Lorentz invariant, is valid for the static cases; this is natural at (near) thermal equilibrium.

As for the second question of the possible effect of the chemical potential on the anomaly evaluation, FKW analyze the effective action 
\begin{eqnarray}\label{effective action with chemical potential}
S=\int d^{4}x ~\overline{\psi}(x)[i\Dslash -m]\psi(x)
\end{eqnarray}
with
\begin{eqnarray}
\Dslash =\gamma^{\mu}(\partial_{\mu} -ieA_{\mu} -ie A^{5}_{\mu}\gamma_{5}).
\end{eqnarray}
FKW then adopt the anomaly relation for the vector current $j^{\mu}(x)$,
\begin{eqnarray}\label{V-A anomaly}
\partial_{\mu}j^{\mu}(x)= -\frac{e^{3}}{4\pi^{2}}F^{A}_{\mu\nu} 
\tilde{F}^{V, \mu\nu}
\end{eqnarray}
with $F^{A}_{\mu\nu} =\partial_{\mu}A^{5}_{\nu}-\partial_{\nu}A^{5}_{\mu}$ and $F^{V}_{\mu\nu} =\partial_{\mu}A_{\nu}-\partial_{\nu}A_{\mu}$. The evaluation of this relation proceeds as follows; one may  start with the evaluation of the divergence of the gauge invariant vector current, namely, the gauge invariant fermion number current $J^{\mu}(x)=\overline{\psi(x)}\gamma^{\mu}\psi(x)$.
This is evaluated in perturbation theory in the path integral, for example, by 
\begin{align}
\frac{e^{2}}{2!}\left(\frac{i}{\hbar} \right)^{2}\int d^{4}yd^{4}z\left[\langle T^{\star}J^{\mu}(x)\hat{j}^{\alpha}(y)j_{5}^{\beta}(z)\rangle
+\langle T^{\star}J^{\mu}(x)j_{5}^{\beta}(z)\hat{j}^{\alpha}(y)\rangle\right] A_{\alpha}(y)A^{5}_{\beta}(z) 
\end{align}
where $\hat{j}^{\alpha}(x)$ has formally the same form as $J^{\mu}(x)$ and $j_{5}^{\beta}(x)$ is the corresponding axial current. By applying the divergence $\partial^{x}_{\mu}$, one obtains
\begin{eqnarray}\label{anomaly1-S5}
\partial_{\mu}J^{\mu}(x)= -\frac{e^{2}}{4\pi^{2}}F^{A}_{\mu\nu} 
\tilde{F}^{V, \mu\nu}
\end{eqnarray}
which is obtained from (the twice of ) a triangle diagram\footnote{Note that the standard chiral anomaly \eqref{identity0} is written in the form $-\frac{e^{2}}{8\pi^{2}}F^{V}_{\mu\nu} 
\tilde{F}^{V, \mu\nu}$.} 
\begin{eqnarray}
\langle T^{\star}J^{\mu}(x)\hat{j}^{\nu}(y)j_{5}^{\sigma}(z)\rangle
\end{eqnarray}
with the conservation conditions $\partial_{\nu}\hat{j}^{\nu}=0$ and $\partial_{\sigma}j_{5}^{\sigma}=0$. The anomaly \eqref{anomaly1-S5} agrees with the anomaly \eqref{V-A anomaly} used by FKW, and this shows that the vector current used in their analysis is the gauge invariant (specific) fermion number current times the charge $e$ (instead of the source current of the electromagnetic field). We thus set
\begin{eqnarray}
j^{\mu}(x)= eJ^{\mu}(x).
\end{eqnarray}
From the relation \eqref{V-A anomaly}, FKW  infer using an argument of the derivative expansion
\begin{eqnarray}\label{induced currentS5}
j^{\mu}(x)= -\frac{e^{3}}{4\pi^{2}}\epsilon^{\mu\nu\sigma\rho}A^{5}_{\nu} 
F^{V}_{\sigma\rho}  
\end{eqnarray}
 which is the leading term for the very slowly varying $A^{5}_{\mu}$. If one sets $eA^{5}_{0}=\mu_{5}$, which is assumed to be constant, and $eA^{5}_{k}=0$, one finally obtains
\begin{eqnarray}\label{derivative expansion}
\vec{j} =\frac{e^{2}\mu_{5}}{2\pi^{2}}\vec{B}
\end{eqnarray}
which agrees with \eqref{integrated chiral magnetic effect} in the static limit with constant $\vec{B}$ in the spirit of the derivative expansion.

This anomaly interpretation \eqref{derivative expansion} of the chiral magnetic effect is mathematically transparent, and it is important that they use the gauge invariant (specific) fermion number current instead of the source current of the electromagnetic current, which differs from the fermion number current by a Bose symmetrization factor of $1/2$ in the present model. The anomaly in the electromagnetic source current spoils the gauge invariance and thus needs to be canceled by some means \footnote{One may recall that gauge anomalies are canceled among leptons and quarks in the Standard Model, while the gauge invariant lepton and quark number currents have physical meanings satisfying $B-L=0$.}.  The anomaly in the gauge invariant fermion number current is consistent with the spectral flow argument used in \eqref{spectral flow2S5}. How to measure the fermion number current directly is however an interesting issue. See \cite{Hongo-Hidaka} for an analysis of  the related  problem together with an extensive list of references.

FKW also gives other derivations of their suggested relation, which are indirectly related to the anomaly consideration. As a result. they present a general formula of the chiral magnetic effect, which is applicable to QCD plasma,
\begin{eqnarray}
J=N_{c}\sum_{f} q_{f}\left[\frac{q_{f}\Phi}{2\pi}\right]_{F}\frac{L_{z}\mu_{5}}{\pi}
\end{eqnarray}
where  $N_{c}=3$ is the color factor, $q_{f}= \frac{2e}{3}, \frac{-e}{3},\ {\rm and} \ \frac{-e}{3}$ are the charges of quarks. The flux $\Phi$ stands for the magnetic flux and $L_{z}$ is the size of the system in the direction of the magnetic field. $\left[x\right]_{F}$ is the floor function which
is the largest integer smaller than x.

This last formula, which contains the (indirect) effect  of the chiral anomaly in the presence of $\mu_{5}$,  provides a definite prediction and can be tested readily by experiments.  The presence of $\mu_{5}$ provided by QCD, which makes the derivation of the chiral magnetic effect transparent, is different from the other scheme of the chiral magnetic effect already mentioned \cite{nielsen2}; the  chiral magnetic effect thus becomes completely independent of $\vec{E}$. 

As for the derivation of $\mu_{5}$ from QCD, which is discussed in detail in \cite{fukushima, Kharzeev,Moore, Mace, Mueller}, it is beyond the scope of the present review. See also \cite{Nair}.


\subsection{Chiral anomaly from Berry's phase}

A similarity of Berry's phase in some physical situations with chiral anomalies has been recognized by various people in the past \cite{Stone, Jackiw-IJMP, Sonoda, Niemi}. As a well-known example among early suggestions, Stone mentioned the similarity of the monopole-type term induced by Berry's phase with the Wess-Zumino term in quantum anomalies \cite{Stone}. This problem was analyzed in detail in \cite{Fujikawa-2006}; the essence 
 is summarized as follows.  The chiral anomaly appears in the shoulder of the functional determinant of the fermion operator 
\begin{eqnarray}\label{functional determinant}
{\rm det}\Dslash \rightarrow e^{i\alpha(A)}{\rm det}\Dslash 
\end{eqnarray}
with the chiral anomaly factor $\alpha(A)$ such as the Wess-Zumino term. On the other hand, Berry's phase is defined level-wise, namely, a different form of Berry's phase for each energy level of the Dirac operator $\Dslash$ separately. Thus two notions are very distinct in this sense.

Recently, a direct derivation of chiral gauge anomaly 
 from Berry's phase via the anomalous Hall effect has been suggested in \cite{Son-Yamamoto, Stephanov}.
Their analysis is partly regarded as a variant of Stone's analysis, and it is associated with each energy level of the system instead of the total determinant in the context mentioned above \eqref{functional determinant}, as is shown later. Their derivations themselves are interesting as a derivation of an object similar to chiral anomaly (but the final result does not quite agree). In the following we briefly sketch the essence of the derivation in the paper by Son and Yamamoto \cite{Son-Yamamoto} as a representative one among related attempts to derive the chiral  anomaly from the anomalous Hall effect, and we add critical comments from our point of view. 

 Their derivation \cite{Son-Yamamoto} contains many logical steps which are finally combined in the derivation of $U(1)$ gauge anomaly for a theory defined in the Brillouin zone, although the final result does not agree with what one expects. The crucial ingredient in their derivation is the classical anomalous Poisson brackets induced by the anomalous Hall effect and the simulation of band crossing in condensed matter physics by a chiral lattice gauge theory with species doublers, which are used to ensure the anomaly cancellation at the end.  In the analysis of the anomaly associated with species doublers in lattice theory, we already emphasized that the chiral gauge anomaly of each species doubler by itself is ill-defined for a finite lattice spacing $a\neq0$ in Section 4, since each species doubler is defined in a part of the fundamental Brillouin zone and thus gives rise to a {\em non-local field} in space-time.

\subsubsection{Various forms of chiral $U(1)$ anomaly}
In the analysis of $U(1)$-type gauge anomalies, the distinction of the covariant form of gauge anomaly and the consistent form of gauge anomaly is important, both in the technical sense and in physical implications. 
We thus explain the difference between the gauge invariant fermion number current and the gauge non-invariant electromagnetic source current in chiral Abelian gauge theory defined in  continuum. It becomes  transparent if one uses the perturbation approach.
One may start with the chiral fermion in the presence of the background gauge field
\begin{eqnarray}\label{chiral Abelian gauge theoryS5}
S=\int d^{4}x \left\{ \overline{\psi}_{L}(x)i\gamma^{\mu}[\partial_{\mu}- iA_{\mu}(x)]\psi_{L}(x) - \frac{1}{4}F_{\mu\nu}F^{\mu\nu} \right\}.
\end{eqnarray}
One may define the fermion number current by the variational derivative with $\psi(x)\rightarrow e^{i\alpha(x)}\psi(x)$ and $\overline{\psi_{L}}(x)\rightarrow \overline{\psi_{L}}(x)e^{-i\alpha(x)}$, namely,
\begin{eqnarray}\label{chiral fermion-number current}
J_{L}^{\mu}(x)=\overline{\psi}_{L}(x)\gamma^{\mu}\psi_{L}(x) 
\end{eqnarray}
which contains an anomaly
\begin{eqnarray}\label{chiral fermion-number anomaly}
\partial_{\mu}J_{L}^{\mu}(x)=-\frac{1}{4\pi^{2}}\vec{E}\cdot \vec{B}.
\end{eqnarray}
Thus the gauge invariant fermion number current in a chiral Abelian gauge theory  is not conserved. 
This is evaluated in perturbation theory in the path integral, for example, by 
\begin{eqnarray}
\frac{1}{2!}(\frac{i}{\hbar})^{2}\int d^{4}yd^{4}z\langle T^{\star}J_{L}^{\mu}(x) j_{L}^{\alpha}(y) j_{L}^{\beta}(z)\rangle A_{\alpha}(y)A_{\beta}(z)
\end{eqnarray}
and applying the divergence $\partial^{x}_{\mu}$. To define $J_{L}^{\mu}(x)$ to be gauge invariant is to define the chiral currents $j_{L}^{\alpha}(y)$  and $j_{L}^{\beta}(z)$ coupled to gauge fields, which formally have the same form as \eqref{chiral fermion-number current},  to be invariant under the gauge transformations of $A_{\alpha}(y)$ and $A_{\beta}(z)$, respectively. One thus evaluates the triangle diagram 
\begin{eqnarray}\label{triangle diagram}
 \langle T^{\star}J_{L}^{\mu}(x) j_{L}^{\alpha}(y) j_{L}^{\beta}(z)\rangle
 \end{eqnarray}
by imposing the conservation (gauge invariance) conditions on $j_{L}^{\alpha}(y)$  and $j_{L}^{\beta}(z)$,
\begin{eqnarray}\label{current conservation condition}
\partial_{\alpha}j_{L}^{\alpha}(y)=\partial_{\beta}j_{L}^{\beta}(z)=0.
\end{eqnarray} 
One then finds the anomalous relation for $J_{L}^{\mu}(x)$ in \eqref{chiral fermion-number anomaly} \cite{anomaly1, anomaly2}.

To define the gauge anomaly, one examines a vertex of three gauge fields and thus  needs to treat all the currents in \eqref{triangle diagram} on an equal-footing (to ensure the Bose symmetry), namely,
\begin{eqnarray}\label{consistent anomalyS5}
\langle T^{\star}\hat{j}_{L}^{\mu}(x)\hat{j}_{L}^{\alpha}(y) \hat{j}_{L}^{\beta}(z)\rangle
&=& \frac{1}{3}[\langle T^{\star}J_{L}^{\mu}(x) j_{L}^{\alpha}(y) j_{L}^{\beta}(z)\rangle\nonumber\\
&+&\langle T^{\star} j_{L}^{\mu}(x) j_{L}^{\alpha}(y)J_{L}^{\beta}(z)\rangle+\langle T^{\star} j_{L}^{\mu}(x)J_{L}^{\alpha}(y) j_{L}^{\beta}(z)\rangle]
\end{eqnarray}
where $\hat{j}_{L}^{\mu}(x)$ is the source current of the electromagnetic field appearing at a vertex of three gauge fields.
Note that the three currents $\hat{j}_{L}^{\mu}(x)$, $J_{L}^{\mu}(x)$ and $ j_{L}^{\mu}(x)$ are identical if there should be no anomaly, but actually very different due to the anomaly. The differences of these three currents are specified  {\em operationally} when one evaluates the triangle diagram; if one imposes the Bose symmetry on three currents, one obtains $\hat{j}_{L}^{\mu}(x)$ on the left-hand side of \eqref{consistent anomalyS5}. If one treats two of the three currents in a Bose symmetric manner together with the gauge invariance condition as in \eqref{current conservation condition}, the currents are specified to be  $ j_{L}^{\mu}(x)$. As for $J_{L}^{\mu}(x)$, no specification other than the Lorentz and chiral properties in the present context.

One then finds from \eqref{consistent anomalyS5} by noting \eqref{current conservation condition}
\begin{eqnarray}
\partial^{x}_{\mu}\langle T^{\star}\hat{j}_{L}^{\mu}(x)\hat{j}_{L}^{\alpha}(y) \hat{j}_{L}^{\beta}(z)\rangle =
\frac{1}{3}\partial^{x}_{\mu}[\langle T^{\star}J_{L}^{\mu}(x) j_{L}^{\alpha}(y) j_{L}^{\beta}(z)\rangle]
\end{eqnarray}
which means in terms of the operator notation using \eqref{chiral fermion-number anomaly}
\begin{eqnarray}\label{chiral gauge anomaly}
\partial_{\mu}\hat{j}_{L}^{\mu}(x) =-\frac{1}{3}\frac{1}{4\pi^{2}}\vec{E}\cdot \vec{B}.
\end{eqnarray}
This is the chiral gauge anomaly \eqref{gauss} we discussed in Section 4
\begin{eqnarray}\label{gauss2}
\partial_{t}G(t,\vec{x})=\frac{i}{\hbar}[H, G(t,\vec{x})]=-\frac{1}{3}\frac{1}{4\pi^{2}}\vec{E}\cdot \vec{B}.
\end{eqnarray}
Note that the anomaly discussed  by Nielsen and Ninomiya \cite{nielsen2} using the argument of fermion creation corresponds to the chiral fermion number anomaly \eqref{chiral fermion-number anomaly} in the context of continuum theory \footnote{In the lattice theory with species doublers, the gauge anomaly does not appear for $a\neq 0$. Thus the electric charge conservation is ensured, and the particle production takes place under this constraint. Thus only the pair production in a general context is allowed for $a\neq 0$.}, while the method to derive the anomaly discussed by Son and Yamamoto corresponds to the calculational scheme of the (consistent) gauge anomaly \eqref{chiral gauge anomaly} (but the result gives the covariant  anomaly), as we explain below. Thus two anomalies are quite different even in the context of continuum theory; the gauge invariant anomaly \eqref{chiral fermion-number anomaly} has a physical meaning, while the gauge non-invariant anomaly \eqref{chiral gauge anomaly} which needs to be canceled  has no physical meaning by itself. 

\subsubsection{Derivation of chiral anomaly from anomalous Hall effect} 
Coming back to  the possible derivation of chiral gauge anomaly in \cite{Son-Yamamoto} which contains many logical steps,  we mention the aspects which are related to Berry's phase and the characteristic features of chiral anomalies in the context of the present review. The effective equations of the anomalous Hall effect, which 
incorporate Berry's phase  near the level crossing point \cite{Niu,Nagaosa}, play a central role in their analysis. The equations of motion \eqref{semi-classical equation} are commonly summarized by
 the effective action \eqref{action-1}~\cite{Duval} 
\begin{eqnarray}\label{action-1S5}
S=\int dt[p_{k}\dot{x}_{k} - eA_{k}(\vec{x},t)\dot{x}_{k} + {\cal A}_{k}(\vec{p})\dot{p}_{k}-\epsilon_{n}(\vec{p})+e\phi(\vec{x},t)],
\end{eqnarray}  
and the equations of motion are  derived from the (classical) action for which one  assumes a variational principle.  Without the electromagnetic vector potential $eA_{k}=0$, the action is reduced to the canonical action \eqref{gauge covariant Hamiltonian} and thus no anomalous {\em quantum} commutation relations and no Nernst effect are induced.  Without $eA_{k}$, one still has the anomalous Hall effect as in the Born-Oppenheimer approximation \eqref{BO-6}, but no quantum anomaly is  induced. Namely, within the framework of quantum mechanics, the anomalous Hall effect does not induce the quantum anomaly. 

If one adds a non-vanishing electromagnetic vector potential $eA_{k}\neq 0$, the system \eqref{action-1S5} becomes non-canonical and thus classical due to the incompatibility of Berry's phase and the electromagnetic vector potential  $eA_{k}$, as was emphasized in Section 3.  When one reduces the quantum system to a classical system by adding  $eA_{k}$, one induces the Poisson brackets  if  one assumes  
an extended phase space formalism\cite{Faddeev-Jackiw},
\begin{eqnarray}\label{Poisson bracketS5} 
&&\{x_{k},x_{l}\}=\frac{\epsilon^{klm}\Omega_{m}}{1+e\vec{B}\cdot\vec{\Omega}} , \ \ \ \{p_{k},x_{l}\}=-\frac{\delta_{kl}+e\Omega_{k}B_{l}}{1+e\vec{B}\cdot\vec{\Omega}},\nonumber\\
&&\{p_{k},p_{l}\}=- \frac{\epsilon^{klm}eB_{m}}{1+e\vec{B}\cdot\vec{\Omega}},
\end{eqnarray}
for an assumed genuine Dirac monopole $\Omega_{kl}=\epsilon^{klm}\Omega_{m}$.

Son and Yamamoto \cite{Son-Yamamoto} operate precisely in this classical scheme with Berry's phase and the electromagnetic vector potential. In the second paper of theirs \cite{Son-Yamamoto-2} a more abstract argument is given, but we follow the detailed presentation in their original paper \cite{Son-Yamamoto}.  To be specific, they start with the action 
\begin{eqnarray}\label{action-2S5}
S=\int dt[p_{k}\dot{x}_{k} - A_{k}(\vec{x},t)\dot{x}_{k} + {\cal A}_{k}(\vec{p})\dot{p}_{k}-H(\vec{x},\vec{p})],
\end{eqnarray}
with $H(\vec{x},\vec{p})$ unspecified and assuming that ${\cal A}_{k}(\vec{p})\dot{p}_{k}$ is given by the genuine Dirac monopole and that the same Poisson brackets in \eqref{Poisson bracketS5} hold (the electric charge $e$ is suppressed).
They then evaluate the classical commutation relations of the density operator $n(\vec{x})$ of the (chiral) fermion number  
\begin{eqnarray}\label{anomalous commutatorS5}
[n(\vec{x}), n(\vec{y})] = -i\left( \vec{\nabla}\times \vec{\sigma} +\frac{k}{4\pi^{2}}\vec{B}\right)\cdot \vec{\nabla}\delta(\vec{x}-\vec{y})
\end{eqnarray}
using the above anomalous Poisson bracket with 
\begin{eqnarray}
\sigma_{i}(\vec{x})= -\int \frac{d^{3}p}{(2\pi)^{3}}p_{i}\Omega_{k}\frac{\partial n_{\vec{p}}(\vec{x})}{\partial p_{k}}
\end{eqnarray}
where $n_{\vec{p}}(\vec{x})$ is the fermion number density in the Fermi liquid description, and 
\begin{eqnarray}\label{flux of Berry's phase}
k=\frac{1}{2\pi}\int d\vec{S}\cdot \vec{\Omega}.
\end{eqnarray}
Eq.\eqref{anomalous commutatorS5} may be compared with \eqref{commutator of Gauss operator} in the related continuum theory.

They then consider a conserved energy operator $H$ with the static  electromagnetic potential $A_{k}$ (which does not generate the electric field)  but without the scalar potential, and evaluate
\begin{eqnarray}
\dot{n}=i[H, n] =-\vec{\nabla}\cdot \vec{j}
\end{eqnarray}
with
\begin{eqnarray}\label{Berry phase induced current}
j_{k}=\int \frac{d^{3}p}{(2\pi)^{3}}\left[ -\epsilon_{\vec{p}}\frac{\partial n_{\vec{p}}}{\partial p_{k}} -(\Omega_{i}\frac{\partial n_{\vec{p}}}{\partial p_{i}})\epsilon_{\vec{p}}B_{k} -\epsilon_{\vec{p}}\epsilon^{klm}\Omega_{l} \frac{\partial n_{\vec{p}}}{\partial p_{m}} \right].
\end{eqnarray}
They next turn on the electric field by adding a term to the Hamiltonian
\begin{eqnarray}\label{total HamitonianS5}
H^{\prime} = H + \int d^{3}x \phi(\vec{x})n(\vec{x})
\end{eqnarray}
 and 
\begin{eqnarray}\label{derivation of anomaly relationS5}
\dot{n}&=&i[H^{\prime}, n] \nonumber\\
&=&-\vec{\nabla}\cdot \vec{j} - \left(\vec{\nabla}\times\vec{\sigma} +\frac{k}{4\pi^{2}}\vec{B}\right)\cdot \vec{\nabla}\phi(\vec{x})
\end{eqnarray} 
which is rewritten as an anomalous conservation law
\begin{eqnarray}\label{anomalyS5}
\partial_{t}n + \vec{\nabla}\cdot \vec{j}^{\prime} = \frac{k}{4\pi^{2}}\vec{E}\cdot \vec{B},
\end{eqnarray}
with
\begin{eqnarray}
\vec{j}^{\prime}=\vec{j} +\vec{E}\times \vec{\sigma}.
\end{eqnarray}
This is a sketch of the derivation of the relation \eqref{anomalyS5} from the Poisson brackets associated with the anomalous Hall effect. The readers are referred to the original paper for further details \cite{Son-Yamamoto}.

This is an interesting derivation of the chiral relation from the anomalous Hall effect. We now want to add comments on their derivation from our point of view. To be precise, their derivation of \eqref{anomalyS5} is based on the Poisson brackets and thus classical. 
 Their derivation is based on the Hamiltonian \eqref{total HamitonianS5} and the Poisson brackets \eqref{derivation of anomaly relationS5}. The Hamiltonian \eqref{total HamitonianS5}
shows that the operator $n(\vec{x})$ is the source of the electromagnetic field $\phi(\vec{x})$, while the value of their anomalous term \eqref{anomalyS5} shows a specific form of the anomalous term for the gauge invariant fermion number density \eqref{chiral fermion-number anomaly}.  To be more explicit, the equal time Poisson brackets used in  \eqref{derivation of anomaly relationS5} 
\begin{eqnarray}\label{commuttorS5}
\dot{n}(\vec{x})=i[\int d^{3}y \phi(\vec{y})n(\vec{y}), n(\vec{x})] + ...
\end{eqnarray}
shows that it is natural to identify the operator $n(\vec{x})$ appearing on the left-hand side being equal to the operator $n(\vec{x})$ coupled to the gauge field  
$\phi(\vec{y})$ in the Hamiltonian. Namely, the evaluation method is the anomaly contained in the {\em source current} of the electromagnetic field, i.e., corresponds to \eqref{chiral gauge anomaly} in the continuum theory. Unlike the Feynman diagrams, there is no means to distinguish two $n(\vec{x})$ appearing in the commutation relations \eqref{commuttorS5}; the relation  \eqref{commutator of Gauss operator} in continuum theory, for example, is based on this consideration. Mathematically, this fact shows that the anomalous term \eqref{anomalyS5} should be the Bose symmetric gauge anomaly in \eqref{chiral gauge anomaly}, namely, the correct result implied by their derivation should be
\begin{eqnarray}\label{anomaly2S5}
\partial_{t}n + \vec{\nabla}\cdot \vec{j}^{\prime} = \frac{1}{3}\frac{k}{4\pi^{2}}\vec{E}\cdot \vec{B},
\end{eqnarray}
but, nevertheless, their anomaly \eqref{anomalyS5} corresponds to the covariant anomaly.  The relation \eqref{anomalyS5} looks like a chiral anomaly but the derivation method and the result do not correspond to {\em any known anomaly}.

Physically, this implies that  both quantities $n(\vec{x})$ and $\vec{j}^{\prime}$, which are the source currents of the electromagnetic field, (and $H^{\prime}$ also) are gauge non-invariant (i.e., not conserved) and thus unphysical quantities, just as $G$ and $H$ in \eqref{gauss2}. If the anomaly \eqref{anomaly2S5} is canceled by another species doubler, the Hamiltonian $H^{\prime}$ would become physical, as far as the  quantum anomaly is concerned.

In the derivation of possible chiral anomaly in \cite{Son-Yamamoto}, their basic assumption is that the classical anomalous Poisson brackets \eqref{Poisson bracketS5}  contain all the information of chiral anomalies and all the remaining operations can be done naively. But this assumption is not warranted to be valid when one has  anomalies. Note that all the currents in \eqref{consistent anomalyS5}, $\hat{j}_{L}^{\mu}(x)$, $J_{L}^{\mu}(x)$ and $ j_{L}^{\mu}(x)$, are identical without quantum anomalies (and also classically)  but different due to quantum anomalies in continuum; the identification and the distinction of current operators are subtle in the analysis of quantum anomalies.   The crucial factor $1/3$ is obtained most simply by a Bose symmetric Feynman diagram evaluation as in \eqref{consistent anomalyS5}.

The present  analysis implies that  the specification of the  anomalous Poisson brackets of the semi-classical density $n(\vec{x})$ \eqref{anomalous commutatorS5} alone is not sufficient to analyze the chiral anomaly. To refute the criticism with regard to \eqref{anomaly2S5}, it would be necessary to specify how to distinguish the covariant current from the consistent form of current based on \eqref{commuttorS5}.  So far we treated the anomaly evaluation as the one in continuum theory.
If one should assume the species doubling in the Brillouin zone to cancel the anomaly \eqref{anomalyS5} in \cite{Son-Yamamoto}, one would also need to analyze how to define anomaly for each non-local species doubler {\em separately} for $a\neq0$; we, however, emphasized that no anomaly appears  for each non-local species doubler separately.

\subsubsection{Remarks on the anomalous Hall effect and chiral anomalies}

The local gauge symmetry of Berry's connection enforces each $\varphi_{n}(P)\phi_{n}(x,P)$ in the total  wave function $\Psi(x,P)=\sum_{n}\varphi_{n}(P)\phi_{n}(x,P)$ to be independent in the Born-Oppenheimer approximation. In contrast, the electromagnetic gauge symmetry associated with the vector potential $eA_{k}$ acts universally on all the states in $\Psi(x,P)$. These two independent gauge symmetries cannot be compatible, and they render the action of  the anomalous Hall effect non-canonical and classical if one adds $eA_{k}$. The conventional applications of Berry's phase to the anomalous Hall effect with anomalous Poisson brackets are thus classical and not quantum \cite{Niu}. In contrast, the quantum mechanical anomalous Hall effect as described by the Born-Oppenheimer approximation, which cannot incorporate $eA_{k}$,  does not produce any quantum anomaly nor the Nernst effect \eqref{BO-6-2}.

The distinction of the covariant form of anomaly and the consistent form of anomaly is crucial in the analysis of anomalies. It would be nice if one would analyze this issue and specify what kind of anomaly one is obtaining by the classical Poisson bracket approach to possible anomalies. In their presentation \cite{Son-Yamamoto}, the mathematical formulation
 corresponds to the consistent form of anomaly but the actual value of the anomaly is the covariant anomaly; their formula  as it stands does not correspond to any known anomaly.

The use of the modified Poisson brackets associated with the anomalous Hall effect implies that one is analyzing physics in the specific $n$-th level, which may be natural in condensed matter physics, namely,  each possible (classical) anomalous Poisson bracket is associated with each level  $n$ of the fast system. The analysis in
\cite{Son-Yamamoto} is thus consistent with the past analysis of possible anomalies by Stone \cite{Stone}, but
it differs from the conventional
understanding of the chiral anomaly defined universally for the field, which includes all the levels such as ${\rm det}\Dslash$ in \eqref{functional determinant} \cite{Fujikawa-2006} \footnote{This aspect is supported by the detailed evaluations of various chiral anomalies \cite{Yee}.}. See also \cite{Mueller-2}.\ \ 
If one evaluates the fast system exactly without adiabatic approximations, no anomalous Poisson brackets appear as was emphasized in Section 3. The derivation of the anomaly from anomalous Poisson brackets \eqref{Poisson bracketS5} thus differs from the common belief of chiral anomalies as the exact property of  field theory. In other words, the anomalous Poisson brackets  (and the possible anomaly in \cite{Son-Yamamoto}) arise when one discards most terms linear in time-derivative in the exact expression such as in \eqref{adiabatic Stone phase2}.


In other attempts to derive chiral anomalies from Berry's phase such as in Stephanov and Yin \cite{Stephanov}, the notion of chiral kinetic theory is fundamental. This notion, which is beyond the scope of the present review, is widely used in nuclear physics. The fact that they attempt to derive the chiral anomaly from a specific band $n$ on the basis of classical Poisson brackets is still the  features common with other approaches to the derivation of chiral anomalies from Berry's phase.

Our conclusion of the subject of the possible derivation of chiral anomalies from Berry's phase is that it is interesting to recognize phenomena ``similar'' to the chiral anomaly on the basis of  Berry's phase, which are however not quite the same as the chiral anomaly.
  
\subsubsection{Generic Berry's phase and  commutation relations}

It was argued in \cite{Deguchi-Fujikawa-AP} that Berry's phase as an adiabatic quantum effect  does not introduce the anomalous commutation relations containing a monopole. The analysis is based on the adiabatic properties of Berry's phase, namely, the adiabatic Berry's phase is defined in the adiabatic approximation of level crossing without specifying any interaction terms and not a soliton-like localized object away from the precise adiabatic limit. In addition, the commutation relations are the short-time effect as is clearly shown by the BJL prescription which is explained in Appendix A; this is also the case in the canonical quantization which uses the symplectic  forms {\em retaining} all terms with time derivatives. The adiabatic  Berry's phase is thus ``resolved'' in the short-time limit. This fact is intuitively illustrated by \eqref{exact non-adiabatic2} and \eqref{adiabatic Stone phase2} in the Weyl model in subsection 3.3 ; the exact solution of Berry's phase discussed in Appendix B  covers  these two (snapshot) solutions. The exact solution thus interpolates from the adiabatic Berry's phase represented by a monopole \eqref{adiabatic Stone phase2} to the non-adiabatic trivial solution \eqref{exact non-adiabatic2} which is crucially important in the analysis of commutation relations. 

We thus argued in \cite{Deguchi-Fujikawa-AP}  that the quantum effect of adiabatic Berry's phase, which is defined in the framework of quantum mechanics, does not deform canonical commutation relations by the   monopole.   
But this analysis of the exact Berry's phase is not required by the main example chosen in the present review, namely, the common analysis of the anomalous Hall effect based on the classical anomalous Poisson bracket  with an assumed point-like monopole.
 As long as one analyzes the anomalous Hall effects approximately in terms of  the classical Poisson brackets with an assumed point-like monopole as in the literature, a different aspect of the problem  involving Berry's phase is emphasized; we followed this practice in the main part of this review. 
 
 The above fact is the main reason why we did not discuss much the contents related to Appendix B, namely, Berry's phase as a non-local object implied by the quantum mechanical adiabatic approximation, although we believe it to be fundamental and presented it in Appendix B in detail.  The vanishing of Berry's phase terms in all the Poisson brackets such as \eqref{Poisson bracket}, which is the view in \cite{Deguchi-Fujikawa-AP},  corresponds to the point of view (i) in subsection 3.2.3.

\newpage
\section{Discussions and conclusion}
We have discussed two main themes, Berry's phase and chiral anomalies, in this review. We emphasized the basic aspects of these notions rather than their practical applications.  We here add a comment on the topological aspects of Berry's phase and then summarize the essential aspects of these two notions. 
\subsection{Berry's phase and topology}

 Historically, the topological properties of Berry's phase were first recognized by the observation of the Longuet-Higgins' phase change rule \cite{Higgins}, namely, the phase change of a particle when it encircles the level crossing point. The topological aspects of Berry's phase themselves were not emphasized much in the present review. This phase change rule is important when one discusses the issue if the phase change rule implies the level crossing in general on the basis of topological considerations. The analysis based on an exactly solvable model in Appendix B clearly shows that the (potential) level crossing in the {\em adiabatic limit} generally implies the appearance of both topological phase and the consequential phase change rule.
But one cannot argue the inverse, namely, that the topology inevitably enforces the level crossing in a general situation; one cannot argue that the topology of Berry's phase  enforces the breakdown of the adiabatic theorem and thus leading to level crossing.
On the contrary, it is natural to consider that the topology arises from the suppressed level crossing combined with the precise adiabatic assumption.
It is an interesting issue to understand the basic mechanism why the topological phase appears in the level crossing phenomenon in the adiabatic limit. One might understand that it  is related to the adiabatic theorem which states that no level crossing takes place in the ideal adiabatic limit; this may imply an appearance of some kind of  obstruction to the level crossing, which may be identified with the monopole singularity in the precise adiabatic limit, that deflects the approaching ``particle'' from any direction away to avoid the collision at the crossing point. Further illustrations on the Longuet-Higgins phase change rule and topology change are found  in Appendix B.5, which are  based on an exactly solvable model.

\subsection{Berry's phase and its universal property}

Berry's phase is a product of the adiabatic approximation, which does not specify the interaction term, for example. It depends only on the notion of adiabaticity. For example, Berry's phase in a quantum mechanical  system and Berry's connection in the Born-Oppenheimer approximation could be very different in practice. But it is remarkable that they share an important  property in common; Berry's phase in an anomalous Hall effect cannot co-exist with the electromagnetic vector potential in a canonical Hamiltonian formalism. We clarified the qualitative reason why it happens to be so in the Born-Oppenheimer approximation, and we utilized  this specific universal property in the present analysis. 

Berry's phase itself assumes various forms when one changes the parameters which characterize Berry's phase even in the two-dimensional model. See Appendix B. The topology of a specific two-dimensional model is characterized by a Dirac monopole-like configuration in the adiabatic limit, but when one analyzes the commutation properties of Berry's phase, the completely different (non-adiabatic) parameter domains are important. In the present review, however,  we mainly utilized the above universal property in the anomalous Hall effect.

\subsection{Species doubling and absence of anomalies on the lattice}

As for the chiral gauge theory defined on the lattice, which is used to simulate the ``Weyl fermion'' in condensed matter physics, one uses a chiral $\gamma_{5}$ invariant  action (in a continuum notation) such as 
\begin{eqnarray}
S =\int dtd^{3}x \left\{ \overline{\psi}(t,\vec{x})\left[ i\gamma^{\mu}\partial_{\mu}-ieA_{\mu}(t,\vec{x}) \right] \left(\frac{1-\gamma_{5}}{2} \right)\psi(t,\vec{x}) \right\}
\end{eqnarray}
defined on a heypercubic lattice, for simplicity. One then encounters the so-called species doublers. The species doublers are similar to conventional chiral fermions but different in some important aspects.  
 On the lattice, the momentum is limited in the Brillouin zone which one may choose $-\frac{\pi}{2a} \leq p_{\mu}\leq \frac{3\pi}{2a}$ for finite $a\neq0$. One of the 16 species doublers in the present context  is defined in, for example, 
\begin{eqnarray}
\frac{\pi}{2a} \leq p_{\mu}\leq \frac{3\pi}{2a}
\end{eqnarray}
which clearly shows that each species doubler is defined in a part of the Brillouin zone and thus it is {\em not a local field} in space-time.
 According to the idea of K. Wilson, the chiral anomaly is a short distance phenomenon. Generally, the chiral anomaly for the chirally symmetric fermion defined by $\gamma_{5}$ on the lattice vanishes, since the short-distance of space-time is completely cut-off for $a\neq0$.
Thus  the chiral anomaly is not defined for the lattice fermion in general and for each non-local species doubler either. 
 In fact, to our knowledge, no explicit evaluation of chiral anomaly for each species doubler separately in a $\gamma_{5}$ invariant theory has been performed with fixed $a\neq0$. 
 One may use the idea of the spectral flow to estimate the chiral anomaly for each species doubler separately, but it is not well-defined as stated above.

To realize the chiral anomaly on the lattice, one needs to modify the definition of the chiral generator to $\Gamma_{5}$ (and the notion of locality) such as in the Ginsparg-Wilson fermion. In this modified scheme, one can show that a non-vanishing Jacobian is produced by the chiral transformation generated by $\Gamma_{5}$, which correctly gives rise to the conventional chiral anomaly ${\rm Tr}\Gamma_{5}\neq 0$ explicitly for 
the limit $a\rightarrow 0$.  We have sketched how to define a massless Dirac fermion {\em without species doublers} on the lattice in the framework of the Ginsparg-Wilson fermion.\\

In conclusion, it is hoped that the analyses of the basic aspects of Berry's phase and chiral anomalies  in this review will be useful in the future applications of these two important notions.

\section*{Acknowledgements}
We thank Kenji Fukushima for suggesting us to write the present review.  One of us (KF) thanks Naoto Nagaosa for helpful comments on the basic aspects of the anomalous Hall effect in condensed matter physics. KF also thanks Shinichi Deguchi for the collaborations on Berry's phase which provided a part of the basic materials of this review, and Yoshio Kikukawa and Hiroshi Suzuki for  helpful comments on chiral fermions on the lattice. We also thank Masaru Hongo for useful comments on the presentation of the manuscript.
The present work is supported in part by JSPS KAKENHI (Grant No.18K03633).

\newpage
\appendix 
\newpage
\section{Bjorken-Johnson-Low  prescription}
In this Appendix, we briefly summarize the basic rules of the Bjorken-Johnson-Low (BJL) prescription \cite{BJL} 
 which becomes useful in the present discussions. The BJL prescription was introduced to understand the equal-time commutation relations of two current operators (current algebras) from a point of view of the short-time limit of the product of two operators. It works for the analysis of 
current operators appearing in the anomalous Ward-Takahashi identities for which the conventional canonical formulation does not work in general. The early use of the BJL prescription was given by Jackiw and Johnson \cite{jackiw} and Adler \cite{adler}. The prescription presented here is arranged so that it applies to  systems where the canonical quantization may not be well-defined as well as to conventional systems.

The canonical quantization of the Hamiltonian
\begin{eqnarray}\label{Lagrangian-1}
H=\frac{p_{k}^{2}}{2m} +V(x_{k})
\end{eqnarray}
or an equivalent Lagrangian
\begin{eqnarray}\label{Lagrangian-2}
L=p_{k}\dot{x}_{k} - \left[\frac{p_{k}^{2}}{2m} +V(x_{k}) \right]
\end{eqnarray}
are defined by the equal-time commutation relations  \footnote{We use the same notation $x_{k}(t)$, for example, for both classical and quantum variables, but our notational convention will not cause any confusions in the present paper.} 
\begin{eqnarray}\label{canonical commutators}
[x^{k}(t), p^{l}(t)]= i\hbar \delta_{kl}, \ \ [x^{k}(t), x^{l}(t)]= 0, \ \ [p^{k}(t), p^{l}(t)]= 0
\end{eqnarray}
supplemented by the equations of motion
\begin{eqnarray}
\dot{x_{k}}=\frac{p_{k}}{m}, \ \ \dot{p}_{k} = -\frac{\delta}{\delta x_{k}} V(x_{l}).
\end{eqnarray}
These commutation relations 
are not modified by the size or shape of the potential $V(x_{k})$ and the mass parameter (including higher order corrections in $\hbar$), as long as they are not singular. The canonical commutation relations are very universal and only the equations of motion are case-dependent. At first sight, it might look physically  strange that even a very large potential does not modify the canonical commutation relations.  Our analysis of the no-deformation of the canonical commutation relations even by the appearance of adiabatic Berry's phase is related to this universality, although the low energy effective equations of motion may be modified.

Our understanding of the above aspect of the canonical commutation relations is based on the fact that they are defined by the motion with very large ``frequencies'' of phase space variables, of which meaning is explained below. Thus any large but finite non-singular potential or even the kinetic energy term does not modify the canonical commutation relations. 
A definition of canonical commutation relations which explicitly incorporates this fact is convenient when one analyzes
general commutation relations  in quantum mechanics and quantum field theory, in particular, in the  cases where the conventional canonical formulation is not applicable. The BJL prescription  is convenient in these respects. 

We first illustrate this prescription for the free Lagrangian without $V(x)$ and $\frac{p^{2}}{2m}$. We thus start with 
\begin{eqnarray}
L_{(0)}=p^{k}(t)\dot{x^{k}}(t)
\end{eqnarray}
and the phase space path integral of the form~\footnote{ The path integral is originally derived by using the canonical quantization, but once the path integral is defined one may use it for more general cases. The time-ordering product is easily evaluated by the path integral. The time-ordering product defined by the path integral is denoted by the $T^{\star}$ product, which agrees with the T-product in the canonical operator formalism in many cases but differs in some important cases, in particular, in the cases associated with the equations of motion.  One can define the path integral  \eqref{path integral-0} better by adding a suitable kinetic energy term and taking an appropriate limit later, but we forgo an analysis of the refinement here; the $i\epsilon$ prescription in \eqref{free propagator} is defined by this procedure. } 
\begin{eqnarray}\label{path integral-0}
\langle T^{\star}p^{k}(t)x^{l}(t^{\prime})\rangle =\int {\cal D}x^{k}{\cal D}p^{k}\{p^{k}(t)x^{l}(t^{\prime})\} \exp \left[\frac{i}{\hbar}\int dt  ~p^{k}(t)\dot{x^{k}}(t) \right].
\end{eqnarray}
A careful evaluation by adding source functions gives rise to the correlations functions 
\begin{eqnarray}\label{free propagator}
&&\int dt e^{i\omega(t-t^{\prime})}\langle T^{\star}p^{k}(t)x^{l}(t^{\prime})\rangle=\frac{\hbar\omega}{\omega^{2} +i\epsilon}\delta_{kl}, \nonumber\\
&&\langle T^{\star}x^{k}(t)x^{l}(t^{\prime})\rangle=0,\nonumber\\
&&\langle T^{\star}p^{k}(t)p^{l}(t^{\prime})\rangle=0
\end{eqnarray}
where Feynman's $i\epsilon$ prescription is generalized by $\epsilon=\epsilon_{1} + i\mu^{2}$ using two small positive parameters $\epsilon_{1}$ and $\mu^{2}$ to define the correlation function precisely.

 The BJL prescription is based on {\em two basic rules}:\\
(i) The $T^{\star}$ product is replaced by the ordinary $T$ product which is well-defined for the equal-time limit $t=t^{\prime}$ if 
\begin{eqnarray}\label{BJL-1}
\lim_{\omega\rightarrow \infty}\int dt e^{i\omega(t-t^{\prime})}\langle T^{\star}p^{k}(t)x^{l}(t^{\prime})\rangle=\lim_{\omega\rightarrow \infty}\frac{\hbar\omega}{\omega^{2} +i\epsilon}\delta_{kl}=0
\end{eqnarray}
that is an analogue of the Riemann-Lebesgue lemma, which states that the Fourier transform of a smooth function vanishes at large frequencies.\\
(ii)We define the equal-time commutation relations by the procedure
\begin{align}\label{BJL-2}
&\omega\int dt ~e^{i\omega(t-t^{\prime})}\langle T p^{k}(t)x^{l}(t^{\prime})\rangle
\nonumber\\
=&\int dt ~e^{i\omega(t-t^{\prime})}i\frac{d}{dt}\langle T p^{k}(t)x^{l}(t^{\prime})\rangle\nonumber\\
=&\int dt ~e^{i\omega(t-t^{\prime})}\left\{i\delta(t-t^{\prime})[p^{k}(t),x^{l}(t)]+\langle Ti\frac{d}{dt}p^{k}(t)x^{l}(t^{\prime})\rangle \right\}\nonumber\\
=&\omega\frac{\hbar\omega}{\omega^{2} +i\epsilon}\delta_{kl}
\end{align}
where we used $T p^{k}(t)x^{l}(t^{\prime})=\theta(t-t^{\prime}) p^{k}(t)x^{l}(t^{\prime}) + \theta(t^{\prime}-t) x^{l}(t^{\prime})p^{k}(t)$.
By considering the limit $\omega\rightarrow \infty$ in \eqref{BJL-2}, we conclude 
\begin{eqnarray}\label{BJL-3}
[p^{k}(t),x^{l}(t)] &=&-i\hbar\delta_{kl}, \nonumber\\
\int dt e^{i\omega(t-t^{\prime})}\langle Ti\frac{d}{dt}p^{k}(t)x^{l}(t^{\prime})\rangle&=&0
\end{eqnarray}
where we used the fact (i) that the Fourier transform of the ordinary 
T-product $\langle Ti\frac{d}{dt}p^{k}(t)x^{l}(t^{\prime})\rangle$ vanishes for large $\omega$.
The second expression of \eqref{BJL-3} is consistent with the equations of motion $\frac{d}{dt}p^{k}(t)=0$ given by $L_{0}$.  We thus obtain the operator relations
\begin{eqnarray}
[p^{k}(t),x^{l}(t)] = -i\hbar\delta_{kl}, \ \ \frac{d}{dt}p^{k}(t)=0.
\end{eqnarray}
From the rest of the correlation functions in \eqref{free propagator}, we obtain similarly
\begin{eqnarray}
&&[x^{k}(t), x^{l}(t)] = 0, \ \ \frac{d}{dt}x^{k}(t)=0, \nonumber\\
&&[p^{k}(t), p^{l}(t)] = 0,  \ \ \frac{d}{dt}p^{k}(t)=0.
\end{eqnarray}
We thus recover the canonical commutation relations and the quantum equations of motion implied by 
$L_{(0)}=p^{k}(t)\dot{x}^{k}(t)$.

We next illustrate the effect of the potential term using 
\begin{eqnarray}
L_{(1)}=p^{k}(t)\dot{x}^{k}(t) -V(x)
\end{eqnarray}
by treating $L^{\prime}= -V(x)$ as a perturbation. The starting correlation functions are the same as in \eqref{free propagator} except for the  correlation function which incorporates $-V(x)$ is given by 
\begin{align}
&\langle T^{\star}p^{k}(t)p^{l}(t^{\prime})\rangle
=\int {\cal D}x^{k}{\cal D}p^{k}\{p^{k}(t)p^{l}(t^{\prime})\}\exp \left\{\frac{i}{\hbar}\int dt \left[ p^{k}(t)\dot{x^{k}}(t)-V(x) \right] \right\} \nonumber\\
=&  \int dt_{1}dt_{2} \langle Tp^{k}(t)x^{l^{\prime}}(t_{1}) \rangle \left\{\frac{-i}{\hbar}\int ds\frac{\delta}{\delta x^{l^{\prime}}(t_{1})}\frac{\delta}{\delta x^{k^{\prime}}(t_{2})}V(x(s)) \right\} \langle Tx^{k^{\prime}}(t_{2})p^{l}(t^{\prime})\rangle \nonumber\\
=&\int ds\int\frac{d\omega}{2\pi}\frac{d\omega^{\prime}}{2
\pi}e^{-i\omega(t-s)}\frac{\hbar\omega}{\omega^{2} +i\epsilon}\left( \frac{-i}{\hbar} \right) \partial_{k}\partial_{l}V(x(s))e^{-i\omega^{\prime}(s-t^{\prime})}
\frac{-\hbar\omega^{\prime}}{(\omega^{\prime})^{2} +i\epsilon}.
\end{align}
Using the BJL procedure  for the Fourier transform of $\langle T^{\star}p^{k}(t)p^{l}(t^{\prime})\rangle$ by multiplying it by $\omega$ and then taking the limit $\omega \rightarrow \infty$,
one obtains (after some re-arrangements) 
\begin{eqnarray}
&&[p^{k}(t), p^{l}(t)]=0,\nonumber\\
&&\langle T[\dot{p}^{k}(t)+ \partial_{k}V(x(t))]p^{l}(t^{\prime})\rangle
=0.
\end{eqnarray}
One thus recovers the universal canonical commutation relations and the equations of motion $\dot{p}^{k}(t)+\partial_{k}V(x(t))=0$ implied by $L_{(1)}=p^{k}(t)\dot{x}^{k}(t) -V(x)$ when combined with $[x^{k}(t), x^{l}(t)] = 0, \ \ \dot{x}^{k}(t)=0$ and $[p^{k}(t),x^{l}(t)] = -i\hbar\delta_{kl}$.

The general strategy is that one uses the equations of motion when one encounters with $\dot{x}^{k}(t)$ or $\dot{p}^{k}(t)$ in the BJL prescription, and thus only the universal canonical commutation relations are obtained.
So far, the procedure is essentially a re-phrasing of the conventional canonical formulation of quantum mechanics in terms of the path integral.  Our purpose of formulating the commutation relations in the present manner is that one may be able to deal with the situation, where the equations of motion defined by the path integral are modified by an induced term at low energies but still the canonical commutation relations, which are defined in the high frequency domain, may be maintained without modifications. 

Another useful example is the particle moving in electromagnetic potentials
\begin{eqnarray}\label{Lagrangian-3}
L
&=&P_{k}(t)\dot{x^{k}}(t) - \frac{1}{2m}(P_{k} +eA_{k}(x))^{2} +  e\phi(x).
\end{eqnarray}
We then obtain the canonical commutation relations from the first term of \eqref{Lagrangian-3}
\begin{eqnarray}\label{commutator-2}
[x^{k}(t), P^{l}(t)]= i\hbar \delta_{kl}, \ \ [x^{k}(t), x^{l}(t)]= 0, \ \ [P^{k}(t), P^{l}(t)]= 0
\end{eqnarray}
which are universally valid.
If one defines  an auxiliary variable
\begin{eqnarray}\label{appendix covariant derivative}
p_{k}=P_{k}+eA_{k}(x)
\end{eqnarray}
the Lagrangian is modified to 
\begin{eqnarray}\label{Lagrangian-3-2}
L
&=&p_{k}(t)\dot{x^{k}}(t) - eA_{k}(x)\dot{x^{k}}(t)- \frac{1}{2m} p_{k}^{2} +  e\phi(x).
\end{eqnarray}
and the commutation relations  are re-written as 
\begin{eqnarray}\label{commutator-10}
&&[x^{k}(t), p^{l}(t)]= i\hbar \delta_{kl}, \ \ [x^{k}(t), x^{l}(t)]= 0,\nonumber\\
&& [p^{k}(t), p^{l}(t)]= -i\hbar e \left(\frac{\partial}{\partial x_{k}(t)}A_{l}(x(t))-\frac{\partial}{\partial x_{l}(t)}A_{k}(x(t))\right).
\end{eqnarray}
 The derivative coupling (namely, the time-derivative $\dot{x^{k}}(t)$ in $eA_{k}(x)\dot{x^{k}}(t)$), as is well-known, modifies the commutation relations written in the form \eqref{commutator-10} depending on the functional form of $A_{k}$. But the important fact is that the universally valid canonical commutation relations \eqref{commutator-2} are always valid without any modifications by higher order effects in $\hbar$; the commutators in \eqref{commutator-10} are obtained form \eqref{commutator-2} by the replacement \eqref{appendix covariant derivative}.
 
Another comment is  that the potential which generates a constant magnetic field $B$, 
\begin{eqnarray}
A^{k}(x(t))=\left(-\frac{1}{2}Bx^{2}(t), \frac{1}{2}Bx^{1}(t), 0\right),
\end{eqnarray}
 for example, contains an arbitrary large frequencies in the sense that this potential is assumed to be valid for very slow movement as well as for very fast movement of $x^{k}(t)$, by definition. In the conventional formulation of quantum mechanics, we do not write the time dependence explicitly (Schr\"{o}dinger picture) in the premise that we can generate the time dependence by the operation $e^{i\hat{H}t} A^{k}(x(0))e^{-i\hat{H}t}=A^{k}(x(t))$ (in the Heisenberg picture), which contains the frequencies covering all the spectra of $\hat{H}$ \footnote{In the BJL prescription, we use infinitely large $\omega$ to ensure that the canonical commutation relations are valid for any sensible form of the Hamiltonian.}. It may happen that a specific potential does not survive this operation 
\begin{eqnarray}
e^{i\hat{H}t} {\cal A}^{k}(x(0))e^{-i\hat{H}t}\neq {\cal A}^{k}(x(t))
\end{eqnarray}
 since an {\em assumed specific functional form} of the potential such as ${\cal A}^{k}(x(t))$ is valid only for a specific very small range of frequencies contained in $x^{k}(t)$; what we have in mind is the adiabatic Berry's phase which assumes a genuine monopole form at the limit of  adiabatic slow motion but changes its {\em functional form} at higher frequencies in non-adiabatic domain.  In such a case, the conventional canonical formulation based on the time independent variables is not reliable in general and an alternative scheme such as the BJL prescription becomes useful.

\newpage
\section{What is the monopole in Berry's phase?}

The exact characterization of the monopole appearing in Berry's phase and its possible topology change to a dipole are important in the applications of Berry's phase. This is a major issue in the classical analysis of the topological aspects of Berry's phase. We present an analysis of this issue using an exactly solvable version of Berry's original model of a spin placed in the rotating magnetic field. This exact solution reproduces the results of a generic model of Berry's phase in both adiabatic and non-adiabatic limits and provides a smooth interpolation between these two limiting cases. The topology change is visualized as a lensing effect of the monopole in the parameter space. 
\subsection{Exactly solvable model of Berry's phase}
Berry's phase is defined in association with the adiabatic theorem, and  the Dirac monopole-like configuration is found in the generic two-level crossing problem in the precise adiabatic limit.  It has been shown that the monopole disappears in the nonadiabatic limit in Section 2. It is important to understand precisely how the Dirac monopole, which appears in the adiabatic limit, disappears when one goes to the parameter domain of nonadiabatic motion.
 This problem is also related to the basic issue of the topology change of the (apparently) topological phase.  To analyze those issues in a reliable manner, we consider an exactly solvable model of Berry's phase, in which the fast moving variables are given by the spin degrees of freedom and the slow variables are given by the background C-number rotating magnetic field $\vec{B}(t)$. 
 This model is the same as the model discussed by Berry in his original paper \cite{Berry} but the time dependence of parameters which define the model are more restrictive. To be specific, we consider  the Schr\"{o}dinger equation \cite{Berry}
\begin{eqnarray}\label{starting equation}
i\hbar\partial_{t}\psi(t)=\hat{H}\psi(t)
\end{eqnarray}
for the Hamiltonian
$\hat{H}=-\mu\hbar\vec{\sigma}\cdot\vec{B}(t)$
describing the motion of a magnetic moment $\mu\hbar\vec{\sigma}$ placed in a rotating magnetic field 
\begin{eqnarray}\label{solvable model}
\vec{B}(t)=B(\sin\theta\cos\varphi(t), 
\sin\theta\sin\varphi(t),\cos\theta )
\end{eqnarray}
with $\vec{\sigma}$ standing for Pauli matrices. The naively expected level crossing takes place at the vanishing external field $B=0$. 
It has been noticed that the equation \eqref{starting equation} is exactly solved if one restricts the movement of the magnetic field to the form
$\varphi(t)=\omega t$ with constant $\omega$, and constant $B$ and $\theta$ \cite{Fujikawa1B}. It will be explained later that this particular parametrization \eqref{solvable model} maintains the essence of Berry's phase. 
The exact solution is written as
 \begin{align}\label{eq-exactamplitude1}
\psi_{\pm}(t)
&=w_{\pm}(t)\exp\left[-\frac{i}{\hbar}\int_{0}^{t}dt
w_{\pm}^{\dagger}(t)\big(\hat{H}
-i\hbar\partial_{t}\big)w_{\pm}(t)\right]\nonumber\\
&=w_{\pm}(t)\exp\left[-\frac{i}{\hbar}\int_{0}^{t}dt
w_{\pm}^{\dagger}(t)\hat{H}w_{\pm}(t)\right]\exp\left[-\frac{i}{\hbar}\int_{0}^{t}
{\cal A}_{\pm}(\vec{B})\cdot\frac{d\vec{B}}{d t}dt\right]
\end{align}
where 
\begin{eqnarray}\label{exact eigenfuntion}
w_{+}(t)&=&\left(\begin{array}{c}
            \cos\frac{1}{2}(\theta-\alpha) e^{-i\varphi(t)}\\
            \sin\frac{1}{2}(\theta-\alpha)
            \end{array}\right), \ \ \ 
w_{-}(t)=\left(\begin{array}{c}
            \sin\frac{1}{2}(\theta-\alpha) e^{-i\varphi(t)}\\
            -\cos\frac{1}{2}(\theta-\alpha)
            \end{array}\right).
\end{eqnarray}
It is important that these solutions differ from the so-called instantaneous solutions used in the adiabatic approximation, which are given by  setting $\alpha=0$; the following analysis of topology change is not feasible using the instantaneous solutions.   The parameter $\alpha(\theta,\eta)$ is defined by $\mu\hbar B\sin\alpha= (\hbar\omega/2)\sin(\theta-\alpha)$ or equivalently \cite{Fujikawa1B}
\begin{eqnarray}\label{cotangent}
\cot\alpha(\theta,\eta)=\frac{\eta+\cos\theta}{\sin\theta}
\end{eqnarray}
with $\eta=2\mu\hbar B/\hbar\omega$ for $0\leq \theta\leq\pi$, which specifies the branch of the cotangent function.  

The second term in the exponential of the exact solution \eqref{eq-exactamplitude1}
is customarily called Berry's phase and defined by a potential-like object (or connection)~\footnote{We fix the sign convention of Berry's phase by the expression \eqref{ES Berry's phase} in this section.}
\begin{eqnarray}\label{ES Berry's phase} 
{\cal A}_{\pm}(\vec{B})\equiv w_{\pm}^{\dagger}(t)\left(-i\hbar\frac{\partial}{\partial \vec{B}} \right)w_{\pm}(t).
\end{eqnarray}
This potential describes an azimuthally symmetric static magnetic monopole-like object in the present case.

The solution \eqref{eq-exactamplitude1} to the Schr\"{o}dinger equation is confirmed by evaluating
\begin{align}
i\hbar\partial_{t}\psi_{\pm}(t)
=&\{ i\hbar\partial_{t}w_{\pm}(t)+w_{\pm}(t)[w_{\pm}^{\dagger}(t)\big(\hat{H}
-i\hbar\partial_{t}\big)w_{\pm}(t)]\}\nonumber\\
&\times\exp\left[-\frac{i}{\hbar}\int_{0}^{t}dt^{\prime}
w_{\pm}^{\dagger}(t^{\prime})\big(\hat{H}
-i\hbar\partial_{t^{\prime}}\big)w_{\pm}(t^{\prime})\right]\nonumber\\
=&\{ i\hbar\partial_{t}w_{\pm}(t)+w_{\pm}(t)[w_{\pm}^{\dagger}(t)\big(\hat{H}
-i\hbar\partial_{t}\big)w_{\pm}(t)]\nonumber\\
&+w_{\mp}(t)[w_{\mp}^{\dagger}(t)\big(\hat{H}
-i\hbar\partial_{t}\big)w_{\pm}(t)]\}\nonumber\\
&\times\exp\left[-\frac{i}{\hbar}\int_{0}^{t}dt^{\prime}
w_{\pm}^{\dagger}(t^{\prime})\big(\hat{H}
-i\hbar\partial_{t^{\prime}}\big)w_{\pm}(t^{\prime})\right]\nonumber\\
=&\hat{H}\psi_{\pm}(t)
\end{align}  
where we used
$w_{\mp}^{\dagger}\big(\hat{H}-i\hbar\partial_{t}\big)w_{\pm}=0$ by noting \eqref{cotangent}, and the completeness relation $w_{+}w_{+}^{\dagger}+w_{-}w_{-}^{\dagger}=1$.

The parameter 
$\eta\geq 0$ in \eqref{cotangent} is written as 
\begin{eqnarray}\label{parameter2}
\eta=\frac{2\mu\hbar B}{\hbar\omega}=\frac{\mu BT}{\pi}
\end{eqnarray}
when one defines the period $T=2\pi/\omega$. The parameter $\eta$ is a ratio of the two different energy scales appearing in the model, namely, the static energy $2\mu\hbar B$ of the dipole moment in an external magnetic field  and the kinetic energy (rotation energy) $\hbar\omega$: $\eta\gg 1$  (for example, $T\rightarrow \infty$ for any finite $B$) corresponds to the adiabatic limit, and $\eta\ll 1$ (for example, $T\rightarrow 0$ for finite $B$) corresponds to the non-adiabatic limit. In a mathematical treatment of the adiabatic theorem, the precise adiabaticity is defined by $T\rightarrow \infty$ with fixed $B$ \cite{Simon}. 

The parameter $\alpha(\theta,\eta)$ in \eqref{cotangent} is normalized as $\alpha(0,\eta)=0$
by definition. 
 The analysis of the transition from $\eta>1$ to
$\eta<1$ through the critical value $\eta=1$ is important. 
In Fig.B.1, we show the relation between $\alpha(\theta,\eta)$ and $\theta$ with emphasis on the transition region near $\eta=1$ given by \eqref{cotangent} \cite{DF1, Fujikawa-Umetsu}.  
\begin{figure}[!htb]
 \begin{center}
    \includegraphics[width=8cm]{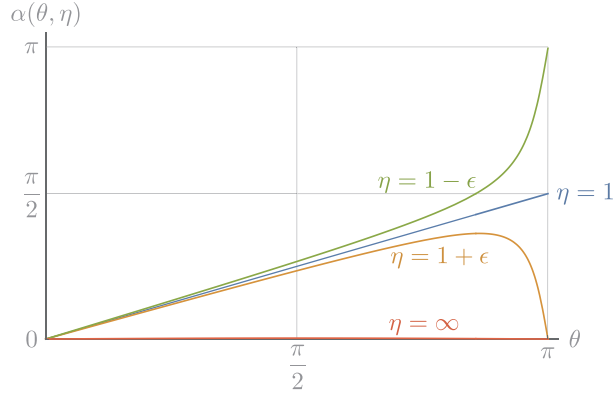} 
       \end{center}
\vspace{-3mm}      
  \caption{The topology change at the parameter value $\eta=1$ determined by Eq. \eqref{cotangent}. Note the limiting values of $\alpha(\theta,\eta)$, $\alpha(\theta,0)=\theta$ and $\alpha(\theta,\infty)=0$.}\label{Fig_02}
\end{figure}
\vspace{1mm}

For the parameters
$\eta=1\pm\epsilon$
with a small positive $\epsilon$, the value $\alpha(\theta,\eta)$ departs from the common value $\frac{1}{2}\theta$ assumed at around $\theta=0$ and splits into two branches for the values of the parameter $\theta$ close to  $\theta=\pi$. We have $\alpha(\pi,\eta)=0$ for $\eta=1+\epsilon$ and  $\alpha(\pi,\eta)=\pi$ for $\eta=1-\epsilon$, respectively,  with the slopes
\begin{eqnarray}\label{singularity}
\frac{d\alpha(\theta,\eta)}{d\theta}|_{\theta=\pi} =\mp \frac{1}{\epsilon}
\end{eqnarray}
for $\eta=1\pm \epsilon$, respectively, using \eqref{cotangent}. We thus observe the singular jump characteristic to the topology change in terms of $\alpha(\theta,\eta)$ at $\eta=1$. It will be shown  later that the topology of the monopole-like object is specified by the value 
\begin{eqnarray}
\lim_{\theta\rightarrow\pi} \alpha(\theta, \eta) = 0, \frac{1}{2}\pi, \pi,
\end{eqnarray}
for $\eta>1$, $\eta=1$ and $\eta<1$, respectively.

The extra phase factor for one period of motion is written explicitly as,
\begin{align}\label{Berry's phase1}
\exp\left[-\frac{i}{\hbar}\oint
{\cal A}_{\pm}(\vec{B})\cdot\frac{d\vec{B}}{d t}dt\right]=&\exp \left\{-i\oint \frac{-1\mp\cos(\theta-\alpha(\theta,\eta))}{2}d\varphi \right\}\nonumber\\
=&\exp \left\{-i\oint\frac{1\mp\cos(\theta-\alpha(\theta,\eta))}{2}d\varphi+2\pi i\right\}\nonumber\\
=&\exp\left\{-\frac{i}{\hbar}\Omega_{\pm} \right\},
\end{align}
with the monopole-like integrated flux \cite{DF1}
\begin{eqnarray}\label{solid-angle}
\Omega_{\pm} 
&=&\hbar\oint \frac{[1\mp\cos(\theta-\alpha(\theta,\eta))]}{2B\sin\theta}B\sin\theta d\varphi .
\end{eqnarray}
In \eqref{Berry's phase1}, we adjusted the trivial phase $2\pi i$ for the convenience of the later analysis; this is related to a  gauge transformation of Wu and Yang~\cite{Wu-Yang}.  
The corresponding energy eigenvalues are
\begin{eqnarray}\label{energy eigenvalue}
E_{\pm}= w_{\pm}^{\dagger}(t)\hat{H}w_{\pm}(t) =\mp (\mu\hbar B\cos\alpha).
\end{eqnarray}
 From now on, we concentrate on $\Omega_{+}$ associated with the lower energy eigenvalue $E_{+}$, i.e., the spin is pointing to the direction of the applied magnetic field; the monopole $\Omega_{-}$ associated with the energy eigenvalue $E_{-}$ is described by $-\Omega_{+}$ up to a gauge transformation of Wu and Yang.
 We then have an {\em azimuthally symmetric}  monopole-like potential \cite{DF1}
\begin{eqnarray}\label{new potential}
{\cal A}_{\varphi} 
= \frac{\hbar }{2B\sin\theta} [1 - \cos\Theta(\theta, \eta) ]
\end{eqnarray}
and ${\cal A}_{\theta} ={\cal A}_{B} =0$, where we defined 
\begin{eqnarray}\label{effective angle}
\Theta(\theta, \eta)=\theta-\alpha(\theta,\eta).
\end{eqnarray}
The standard Dirac monopole \cite{Dirac} is recovered when one sets $\alpha(\theta,\eta)=0$ (or in the ideal adiabatic limit $\eta=\infty$ in \eqref{cotangent}), namely, $\Theta=\theta$ in \eqref{new potential} and when $B$ is identified with the radial coordinate $r$ in the real space.
The crucial parameter $\Theta(\theta, \eta)$ is shown in Fig.B.2. 

\begin{figure}[H]
\centering
\includegraphics[width=8cm]{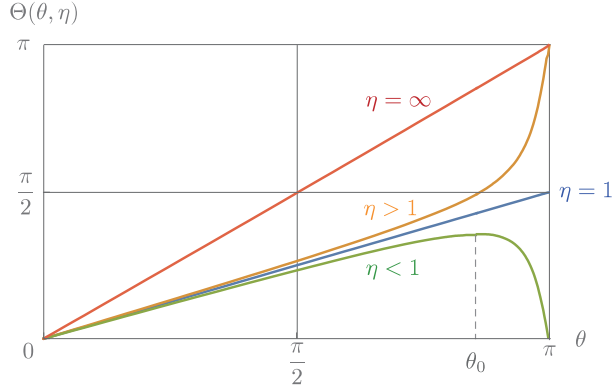}
\caption{\small The relation between $\theta$ and $\Theta(\theta, \eta)=\theta-\alpha(\theta,\eta)$ parameterized by $\eta$. We have the exact relations $\Theta(\theta, \infty)=\theta$, $\Theta(\theta, 1)=\theta/2$ and $\Theta(\theta, 0)=0$, respectively, for $\eta=\infty$, $\eta=1$ and $\eta=0$. Topologically, $\eta>1$ corresponds to a monopole, $\eta=1$ corresponds to a half-monopole, and $\eta<1$ corresponds to a dipole, respectively.  
Note that $\cos \theta_0 =-\eta$  with $\eta<1$, for which $\partial\Theta(\theta, \eta)/\partial\theta=0$.  See  \cite{DF1}.}
\end{figure}
\vspace{1mm}

The Dirac string appears at the singularity of the potential \eqref{new potential}. There exists no singularity at $\theta=0$ since $\Theta(\theta, \eta)\rightarrow 0$ for $\theta\rightarrow 0$.
The singularity does not appear at the origin $B=0$ with any fixed $T$ since $\alpha(\theta,\eta)\rightarrow \theta$ for $B\rightarrow 0$, namely, if one uses $\Theta(\theta, \eta)=\theta -\alpha(\theta,\eta) \rightarrow 0$ for $\eta=\mu BT/\pi\rightarrow 0$ in \eqref{cotangent}. In fact the potential vanishes at $B=0$ for any finite $T$; we have a useful relation in the non-adiabatic domain $\eta=\mu BT/\pi \ll 1$, namely, $\alpha(\theta)\simeq \theta -\eta \sin\theta \ {\rm for} \ 0\leq \eta \ll 1$ in \eqref{cotangent} \cite{DF1}, which implies
\begin{eqnarray}
{\cal A}_{\varphi}
&\simeq& \frac{\hbar}{4B}(\mu TB/\pi)^{2}\sin\theta
\end{eqnarray}  
that has no singularity associated with the Dirac string at $\theta=\pi$ near $B=0$ and linearly vanishes in $B$.
  Thus the Dirac string can appear only at $\theta=\pi$ and only when $\Theta(\pi, \eta)\neq 0$, namely, for $\eta\geq 1$ in Fig.B.2 or equivalently  
\begin{eqnarray}
 B\geq \frac{\pi}{\mu T}
 \end{eqnarray}
for any fixed finite $T$; the end of the Dirac string is located at $\frac{\pi}{\mu T}$ and $\theta=\pi$.  The total magnetic flux passing through a small circle C  around the Dirac string at the point $B$ and $\theta=\pi$
is given by the potential \eqref{new potential}
\begin{eqnarray}\label{Stokes}
\oint_{C} {\cal A}_{\varphi} B\sin\theta d\varphi =\frac{e_{M}}{2}(1-\cos\Theta(\pi, \eta))
\end{eqnarray}
with $e_{M}=2\pi\hbar$. See a schematic figure in Fig.B.3  \cite{DF1}.

\begin{figure}[H]
\centering
\includegraphics[width=5cm]{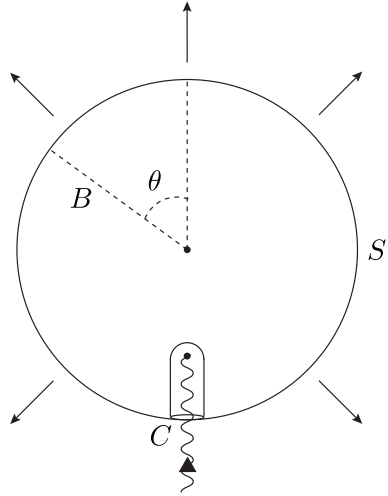}
\caption{\small A schematic picture with fixed $T$.  Geometrical center and the origin of the Dirac string are displaced by the distance $B=\pi/\mu T$. An infinitesimal circle $C$ surrounds the Dirac string indicated by a wavy line. }\label{Fig_05}
\end{figure}
\vspace{1mm}

This flux agrees with the integrated flux outgoing from a sphere with a radius $B$ covering the monopole
due to Stokes' theorem, since no singularity appears except for the Dirac string.  For $\eta>1$, one sees from Fig.B.2 that the above flux is given by $e_{M}=2\pi\hbar$ and thus Dirac's quantization condition is satisfied in the sense $\exp[-ie_{M}/\hbar]=1$. On the other hand, the flux vanishes for $\eta<1 (i.e., B<\frac{\pi}{\mu T})$ and $\Theta(\pi, \eta)=0$, and thus the object changes to a dipole \cite{DF1}.

The topology change is seen in the change of $\Theta(\pi)=\pi$ for $\eta>1$ to $\Theta(\pi)=0$ for $\eta<1$ in Fig.B.2.  But we have a well-defined potential (``half-monopole'') at the boundary $\eta=1$
\begin{eqnarray}\label{half sphere}
{\cal A}_{\varphi}=\frac{\hbar}{2B\sin\theta}\left(1-\cos\frac{1}{2}\theta \right)
\end{eqnarray}  
for $\theta\neq \pi$. 

\subsection{Fixed $T$ configurations}                                                                

We analyze the behavior of the magnetic monopole-like object \eqref{new potential} in more detail for fixed $T$ and varying B;
this is close to the description of a monopole in the real space if one identifies $B$ with the radial variable $r$ of the real space.  
The topology and topology change of Berry's phase when regarded as a magnetic monopole defined in the parameter space of $\vec{B}$ is specified by the parameter $\eta$, as is suggested  by a discrete jump of the end point $\lim_{\theta\rightarrow\pi}\Theta(\theta, \eta)$ in Fig.B.2. 

 Using the exact potential \eqref{new potential}
we have an analogue of the magnetic flux in the {\em parameter space} $\vec{B}=B(\sin\theta\cos\varphi, 
\sin\theta\sin\varphi,\cos\theta )$, 
\begin{eqnarray}\label{magnetic flux}
{\cal B}\equiv \nabla\times {\cal A}
=\frac{\hbar}{2}\frac{\frac{\partial\Theta(\theta,\eta)}{\partial\theta}\sin\Theta(\theta,\eta)}{\sin\theta}\frac{1}{B^{2}}{\bf e}_{B} - \frac{\hbar}{2}\frac{\frac{\partial\Theta(\theta,\eta)}{\partial B}\sin\Theta(\theta,\eta)}{B\sin\theta}{\bf e}_{\theta}
\end{eqnarray}
for $\theta\neq \pi$ and $B\neq 0$ with ${\bf e}_{B}=\frac{\vec{B}}{B}$, and  ${\bf e}_{\theta}$ is a unit vector in the direction $\theta$ in the spherical coordinates.
We have
\begin{eqnarray}  
\frac{\partial\Theta(\theta,\eta)}{\partial\theta} = \frac{\eta(\eta+\cos\theta)}{1+\eta^{2}+2\eta\cos\theta},
\end{eqnarray}
by noting $
\frac{\partial\alpha(\theta,\eta)}{\partial\theta}=\frac{1+\eta\cos\theta}{(\eta+\cos\theta)^{2}+\sin^{2}\theta}$ in \eqref{cotangent},
and thus 
$
\frac{\partial\Theta(\theta,\eta)}{\partial\theta}=0$ 
at $\cos\theta_{0}=-\eta$ for $\eta<1$.
The factor in the  second term in \eqref{magnetic flux} is given by recalling $\eta=\mu TB/\pi$, 
\begin{eqnarray}
\frac{\partial\Theta(\theta,\eta)}{\partial B}&=&\frac{\mu T}{\pi}\frac{\partial\Theta(\theta,\eta)}{\partial \eta}\nonumber\\
&=&\frac{\eta}{B}\frac{\sin\theta}{1+\eta^{2}+2\eta\cos\theta}
\end{eqnarray}
using \eqref{cotangent} and \eqref{effective angle}. Thus we have (by setting $e_{M}=2\pi\hbar$)
\begin{eqnarray}\label{magnetic flux2}
{\cal B}&\equiv& \nabla\times {\cal A}\nonumber\\
&=&\frac{e_{M}}{4\pi}\frac{\sin\Theta(\theta,\eta)}{\sin\theta}\frac{\eta}{B^{2}}\frac{1}{1+\eta^{2}+2\eta\cos\theta}[(\eta+\cos\theta){\bf e}_{B} -\sin\theta{\bf e}_{\theta}]
\end{eqnarray}
We also have from \eqref{cotangent},
\begin{eqnarray}\label{alpha parameter}
\cos\alpha=\frac{\eta+\cos\theta}{\sqrt{1+\eta^{2}+2\eta\cos\theta}}, \ \ \  \sin\alpha=\frac{\sin\theta} {\sqrt{1+\eta^{2}+2\eta\cos\theta}},
\end{eqnarray} 
and thus 
\begin{eqnarray}
\sin\Theta(\theta,\eta)&=&\sin(\theta-\alpha) \nonumber\\
&=&\sin\theta\cos\alpha -\cos\theta\sin\alpha\nonumber\\
&=&\frac{\eta\sin\theta}{\sqrt{1+\eta^{2}+2\eta\cos\theta}} ,\end{eqnarray}
and similarly $\cos\Theta(\theta,\eta)
=[1+\eta\cos\theta]/\sqrt{1+\eta^{2}+2\eta\cos\theta}$.

We finally have the azimuthally symmetric magnetic field from \eqref{magnetic flux2} \cite{Fujikawa-Umetsu}
\begin{eqnarray}\label{magnetic flux3}
{\cal B}
&=&\frac{e_{M}}{4\pi}\frac{\eta^{2}}{B^{2}}\frac{1}{(1+\eta^{2}+2\eta\cos\theta)^{3/2}}[(\eta+\cos\theta){\bf e}_{B} -\sin\theta{\bf e}_{\theta}].
\end{eqnarray}
We note that $B/\eta=\pi/\mu T$ and, $\theta=\pi$ define the end  point of the Dirac string in the fixed $T$ picture. The magnetic field
${\cal B}$  is not singular at $\theta=\pi$ for $\eta>1$ which shows that the Dirac string is not observable if it satisfies the Dirac quantization condition, which is the case of the present model. In the adiabatic limit $\eta\rightarrow\infty$ ($\pi/\mu T\rightarrow 0$ with fixed $B$) in \eqref{magnetic flux3}, the outgoing magnetic flux agrees with that of the Dirac monopole
\begin{eqnarray}
 {\cal B}
=\frac{e_{M}}{4\pi}\frac{1}{B^{2}}{\bf e}_{B}
\end{eqnarray}
 located at the origin (level crossing point)  in the parameter space. This is the common magnetic monopole field associated with Berry's phase in the precise adiabatic approximation. At the origin $B=0$ with {\em fixed finite} $T$, which corresponds to the nonadiabatic limit $\eta=\mu BT/\pi\rightarrow 0$, the magnetic field \eqref{magnetic flux3} approaches a constant field parallel to the z-axis 
 \begin{eqnarray}
 {\cal B}
=\frac{e_{M}}{4\pi}\left(\frac{\mu T}{\pi}\right)^{2}[\cos\theta{\bf e}_{B} -\sin\theta{\bf e}_{\theta}].
\end{eqnarray} 
A precise view of the magnetic flux generated by the monopole-like object \eqref{magnetic flux3} is shown in Fig.B.4.  Note that $ {\cal B}\rightarrow 0$ in the nonadiabatic limit $T\rightarrow 0$ with any fixed $B$.

 In passing, we comment on the notational conventions in this section: $\vec{B}(t)$ stands for the externally applied magnetic field to define the original Hamiltonian in \eqref{starting equation} and $\vec{B}$ is used to specify the parameter space to define Berry's phase, and ${\cal B}$ stands for the ``magnetic field'' generated by Berry's phase in the parameter space. The calligraphic symbols ${\cal A},\ {\cal B}$, ${\cal \nabla}$ and the bold ${\bf e}$ stand for vectors without writing arrows.

\begin{figure}[H]
\centering
\includegraphics[width=6cm]{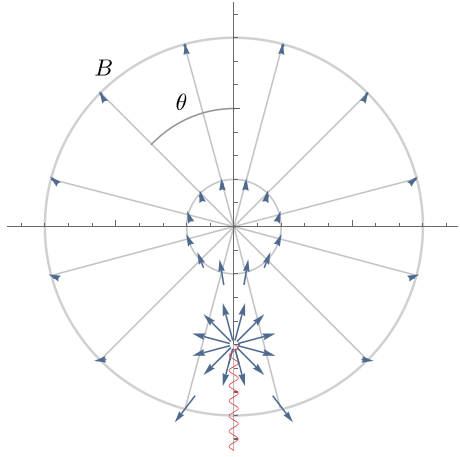}
\caption{\small Arrows indicating the direction and magnitude of the magnetic flux from the azimuthally symmetric monopole-like object associated with Berry's phase \eqref{magnetic flux3} in the fixed  $T$ picture. Two spheres with radii $B>\pi/\mu T$ (i.e., $\eta>1$) and $B<\pi/\mu T$ (i.e., $ \eta<1$) are shown. The wavy line stands for the Dirac string with the end located at $B=\pi/\mu T$ and $\theta=\pi$ from which the magnetic flux is imported. Only in the ideal adiabatic limit $T\rightarrow\infty$, the end of the Dirac string and the geometrical center of Berry's phase which is located at the origin agree.}
\end{figure}
\vspace{1mm}

\subsection{Lensing of Dirac monopole in Berry's phase}
It is  shown that the monopole associated with Berry's phase is mathematically regarded as a Dirac monopole moving away from the level crossing point of the parameter space driven by the force generated by the nonadiabatic rotating external field with finite period $T=2\pi/\omega<\infty$ in Berry's model \cite{Fujikawa-Umetsu}.  This picture corresponds to a description of the monopole-like object by changing $T$ with fixed $B$.  We consider the configuration in Fig.B.5. 

\begin{figure}[H]
\hspace{-4cm}
\includegraphics[width=16cm]{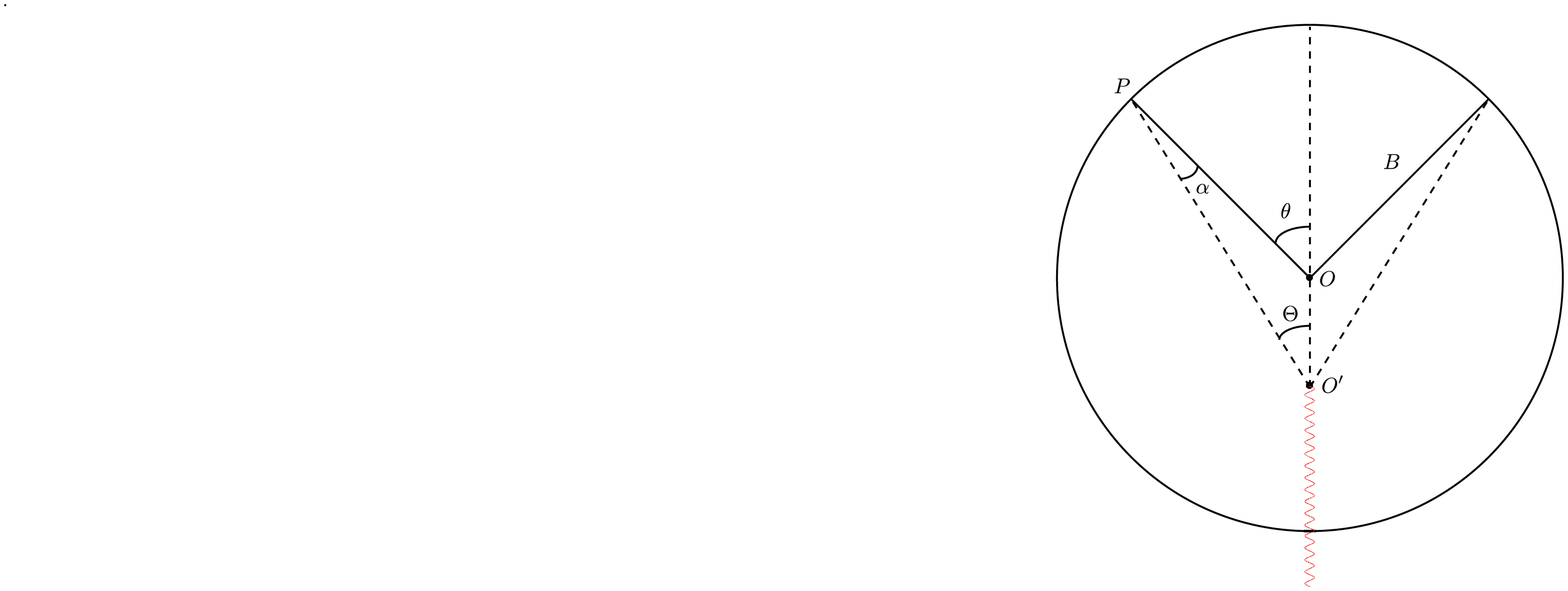}
\caption{\small A geometric picture in the 3-dimensional parameter space $\vec{B}$ with a sphere centered at $O$ and the radius $B$ by assuming azimuthal symmetry. We suppose that  a genuine azimuthally symmetric Dirac monopole is located at the point $O^{\prime}$ in the parameter space.  The distance between $O$ and $O^{\prime}$ is chosen at 
$\overline{OO^{\prime}} = B/\eta$. The three angles $\theta$, $\alpha$ and $\Theta=\theta-\alpha$ are shown. The observer is located at the point $P$. The wavy line indicates the Dirac string.}
\end{figure}
\vspace{1mm}

We then have 
\begin{align}
\overline{O^{\prime}P}^{2}
&=B^{2}+\left(\frac{B}{\eta}\right)^{2} -2B\left(\frac{B}{\eta} \right)\cos(\pi-\theta)
\nonumber\\
&=\frac{B^{2}}{\eta^{2}}\left[ 1 + \eta^{2} + 2\eta\cos\theta \right],
\end{align}
and the unit vector ${\bf e}$ in the direction of $\vec{O^{\prime}P}$ is 
\begin{align}
{\bf e}
&=\cos\alpha {\bf e}_{B} -\sin\alpha {\bf e}_{\theta}
\end{align}
with $ {\bf e}_{B}=\vec{B}/B$ and $ {\bf e}_{\theta}$ is a unit vector in the direction of $\theta$ in the spherical coordinates.
Then the magnetic flux of Dirac's monopole with a magnetic charge $e_{M}$ located at $O^{\prime}$ when observed at the point P is given by 
\begin{align}\label{Dirac monopole}
{\cal B}^{\prime} &= \frac{e_{M}}{4\pi}\frac{1}{\overline{O^{\prime}P}^{2}}{\bf e}\nonumber\\
&=\frac{e_{M}}{4\pi}\frac{\eta^{2}}{B^{2}}\frac{1}{1+\eta^{2}+2\eta\cos\theta}(\cos\alpha {\bf e}_{B} -\sin\alpha {\bf e}_{\theta}).
\end{align}
Next we fix the parameter $\alpha$. We have $
(B/\eta)^{2}=B^{2} + \overline{O^{\prime}P}^{2} -2B\overline{O^{\prime}P}\cos\alpha$
which gives 
\begin{align}
\cos\alpha&=\frac{1}{2B(B/\eta)\sqrt{1 + \eta^{2} + 2\eta\cos\theta}}\left[B^{2} + \left(\frac{B}{\eta} \right)^{2}(1 + \eta^{2} + 2\eta\cos\theta)-\left(\frac{B}{\eta} \right)^{2} \right]\nonumber\\
&=\frac{\eta+\cos\theta}{\sqrt{1 + \eta^{2} + 2\eta\cos\theta}}
\end{align}
and  from the geometrical relation $\frac{B\sin\alpha}{B\sin\theta}=\frac{B/\eta}{(B/\eta)\sqrt{1 + \eta^{2} + 2\eta\cos\theta}}$ ,
\begin{eqnarray}
\sin\alpha=\frac{\sin\theta}{\sqrt{1 + \eta^{2} + 2\eta\cos\theta}}.
\end{eqnarray}
 The parameter $\alpha$ agrees with the parameter in \eqref{alpha parameter}. 
The azimuthally symmetric flux \eqref{Dirac monopole} is thus given by \cite{Fujikawa-Umetsu}
\begin{eqnarray}\label{Dirac monopole2}
{\cal B}^{\prime}
=\frac{e_{M}}{4\pi}\frac{\eta^{2}}{B^{2}}\frac{1}{(1+\eta^{2}+2\eta\cos\theta)^{3/2}}[(\eta+\cos\theta) {\bf e}_{B} -(\sin\theta){\bf e}_{\theta}]
\end{eqnarray}
which agrees with the flux given by Berry's phase \eqref{magnetic flux3}.

This agreement of two expressions \eqref{magnetic flux3} and \eqref{Dirac monopole2}  shows that the Dirac monopole originally at the level crossing point in the parameter space  formally appears to drift away by the distance $B/\eta=\pi/\mu T$ in the parameter space when the precise adiabaticity condition $T=\infty$ \cite{Simon} is spoiled by the finite $T$.  
It is interesting  
that two dynamical parameters,  the strength of the external magnetic field and the period in Berry's model, are converted to very different geometrical parameters in Berry's phase, namely, the shape of the monopole and the distance of the deviation of the monopole from the level crossing point. 
The observed magnetic field on the sphere with a radius $B$, which is controlled by the observer, thus changes when one changes the parameter $T$ that determines the end of the Dirac string located at $\pi/\mu T$ in the parameter space.  
This geometrical picture is useful when one draws the precise magnetic flux from the monopole-like object for finite $T$ as in Fig.B.6 and it is essential when one attempts to understand the motion of a particle in the magnetic field.

\begin{figure}[H]
\centering
\includegraphics[width=14cm]{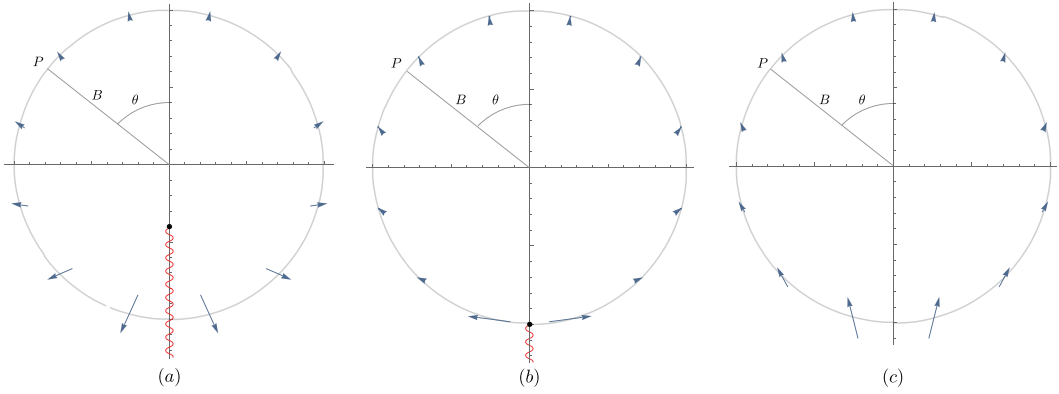}
\caption{\small Arrows indicating the direction and magnitude of the  magnetic flux observed at the point P with fixed $B$ and $\theta$ when the end of Dirac string at $\pi/\mu T$ is varied from the point $\pi/\mu T<B$ (Fig.B.6(a)) to the boundary $\pi/\mu T=B$ (Fig.B.6(b)) and then to the point $\pi/\mu T>B$ (Fig.B.6(c)), which correspond to the change of the basic parameter $\eta=\mu BT/\pi$ from the adiabatic domain $\eta=2.5>1$ to the boundary  $\eta=1$ and then to the nonadiabatic domain $\eta=0.5<1$, respectively. The wavy line stands for the Dirac string with the end at $B=\pi/\mu T$ and $\theta=\pi$. These figures after a suitable rescaling may also be interpreted as the results with the end of the Dirac string kept fixed at $\pi/\mu T$ and $\theta=\pi$ and varying the distance $B$, starting with a large $B>\pi/\mu T$  (Fig.B.6(a)) toward a small $B<\pi/\mu T$  (Fig.B.6(c)) in the parameter space, such as two spheres in Fig.B.4. See \cite{Fujikawa-Umetsu}.}
\end{figure}
\vspace{1mm}

In terms of the original physical setting of a magnetic dipole placed in a given rotating magnetic field described by the Hamiltonian \eqref{starting equation}, the cone drawn by the dipole becomes  sharper compared to the cone of the given magnetic field, which subtends the solid angle $\Omega=2\pi (1-\cos\theta)$, when the rotating speed of the external magnetic field becomes larger and the dipole moment is ``left behind'' (actually, not left behind, but rather follow the movement by making the polar angle $\theta$ smaller), namely, \cite{Fujikawa1B}
\begin{align}\label{solid angle}
\psi_{+}^{\dagger}(t)\vec{\sigma}\psi_{+}(t)&=
w_{+}^{\dagger}(t)\vec{\sigma}w_{+}(t)\nonumber\\
 &=\left( \sin\Theta\cos\varphi(t), \sin\Theta\sin\varphi(t), \cos\Theta \right)\nonumber\\
&=-\psi_{-}^{\dagger}(t)\vec{\sigma}\psi_{-}(t)
\end{align}
that subtends the solid angle 
$\Omega=2\pi (1-\cos\Theta)$ with $\Theta=\theta-\alpha$
; this sharper cone is effectively recognized as the drifting monopole in Berry's phase by an observer located at the point $P$ in Fig.B.5. 

The above agreement of the solid angle drawn by the spinor solution \eqref{solid angle} with Berry's phase is known to be generally valid for the two-component spinor \cite{Fujikawa1B}: The most general orthonormal spinor bases are parameterized as 
\begin{eqnarray}
v_{+}(t)=\left(\begin{array}{c}
            \cos\frac{1}{2}\theta(t) e^{-i\varphi(t)}\\
            \sin\frac{1}{2}\theta(t)
            \end{array}\right), \ \ \ 
v_{-}(t)=\left(\begin{array}{c}
            \sin\frac{1}{2}\theta(t)e^{-i\varphi(t)}\\
            -\cos\frac{1}{2}\theta(t)
            \end{array}\right)
\end{eqnarray}
that give the spin vector
\begin{align}
v_{+}^{\dagger}(t)\vec{\sigma}v_{+}(t)
 =\left( \sin\theta(t)\cos\varphi(t), \sin\theta(t)\sin\varphi(t), \cos\theta(t) \right)=-v_{-}^{\dagger}(t)\vec{\sigma}v_{-}(t)
\end{align}
subtending the solid angle 
$\tilde{\Omega}_{\pm}=\oint (1\mp\cos\theta(t))d\varphi(t)$ for a closed movement. On the other hand, the ``holonomy'' which is related to Berry's phase, satisfies
\begin{eqnarray}
\oint dt ~v^{\dagger}_{\pm}(t)i\partial_{t}v_{\pm}(t)=-\frac{1}{2}\oint  (1\mp \cos\theta(t))d\varphi(t) +2\pi =-\frac{1}{2}\tilde{\Omega}_{\pm} +2\pi.
\end{eqnarray}
These two quantities thus agree up to the factor $1/2$ and  up to trivial phase $2\pi$ in the case of spinor bases. The important fact is that our exact solution of the Schr\"{o}dinger equation has this structure of $v_{+}(t)$ and $v_{-}(t)$ with $\theta(t)=\theta-\alpha(\theta)$. 

Coming back to \eqref{solid angle}, one may thus prefer to understand that Fig.B.5 implies an analogue of the effect of lensing of Dirac's monopole, since the movement of the monopole in the parameter space is a mathematical one.  In the precise adiabatic limit with $T=\infty$ \cite{Simon}, the monopole is located at the level crossing point $O$, but when the effect of nonadiabatic rotation with finite $T< \infty$ is turned on, the image of the monopole is displaced to the point $O^{\prime}$ located at $\pi/\mu T$ by keeping the  topology and strength of the point-like monopole intact. In this picture, it is important that the topological monopole itself is not resolved in the nonadiabatic domain but it  disappears from  observer's view located at the point P for fixed $B$ when $\pi/\mu T = B/\eta \rightarrow {\rm large} $ with fixed $B$ (i.e., $\eta\rightarrow$ small). In the middle, the formal topology change takes place when $\pi/\mu T$ touches the sphere with the fixed radius $B$ (i.e., $\eta=1$). Even in the picture of lensing, the ``magnetic flux'' generated by Berry's phase  measured at the point in the parameter space specified by $(B,\theta)$ is the real flux.
It will be interesting to examine the possible experimental implications of these aspects of Berry's phase, which is expressed  by the magnetic field \eqref{magnetic flux3}, in the applications of Berry's phase.

It may be appropriate to mention that our exact solution agrees with the generic solution discussed in Section 2 for two limiting cases: 
At the adiabatic limit $\eta\rightarrow \infty$, the parameter $\alpha\rightarrow 0$ in Fig.B.1, and the exact solution \eqref{new potential} naturally reproduces the exact Dirac monopole solution. On the other hand, the phase in \eqref{Berry's phase1} approaches
\begin{eqnarray}
\exp\left\{-\frac{i}{\hbar}\Omega_{\pm} \right\}=\exp\left\{-i\oint\frac{1\mp\cos(\theta-\alpha(\theta,\eta))}{2}d\varphi \right\} \rightarrow 1
\end{eqnarray}
in the non-adiabatic limit $\eta\rightarrow 0$ for which
$\alpha\rightarrow \theta$ in \eqref{cotangent}. This agrees with the trivial phase in \eqref{Berry connection}  in Section 2.

\subsection{Smooth topology change}
As for the smooth transition from a monopole to a dipole, it corresponds to the process of the shrinking of the sphere with a radius $B$ covering the end of the Dirac string located at $\pi/\mu T$ to a smaller sphere for which $B<\pi/\mu T$ as in Fig.B.4. When the sphere touches the end of the Dirac string (at $\eta=1$) in the middle, one encounters a  ``half monopole'' \eqref{half sphere} with the outgoing flux which is half of the full monopole $e_{M}/2=\pi\hbar$. See Stokes' theorem \eqref{Stokes} with $\Theta(\pi, \eta=1)=\pi/2$ in Fig.B.2. At this specific point, the Dirac string becomes observable \cite{DF1}, corresponding to the Aharonov-Bohm effect \cite{Aharonov} of the electron in the magnetic flux generated by the superconducting Cooper pair \cite{Tonomura}.  As for more details, see the discussion below.
It is then natural to attach the end of the Dirac string to an infinitesimally small opening on the sphere (see Fig.B. 6b) forming a closed sphere and thus leading to the vanishing net outgoing flux, which corresponds to a dipole. The idea of the half monopole at $\eta=1$ is interesting, but it is natural to incorporate it as a part of a dipole. The monopole-like object \eqref{new potential} is always  a dipole if one counts the Dirac string as in Fig.B.4  and Stokes' theorem \eqref{Stokes} always holds.  In this sense, no real topology change takes place for the movement of $B$, from large $B$ to small $B$ with fixed $T$, except for the fact that the unobservable Dirac string becomes observable at $B=\pi/\mu T$ and triggers the topology change from a monopole to a dipole. 

A more detailed analysis of this smooth topology change goes as follows:
From a point of view of the net outgoing flux, we see the full flux with $e_{M}=2\pi\hbar$ in Fig.B.6a and the half flux with $e_{M}/2$ in Fig.B.6b and then no net flux in Fig.B.6c, corresponding to $\Theta(\pi)$ with $\pi$, $\pi/2$ and $0$, respectively, in \eqref{Stokes}. 
Thus these configurations are very distinct. 
On the other hand, Stokes' theorem \eqref{Stokes} formally shows a smooth transition among distinct topologies specified by $\Theta(\pi)$ with $\pi$, $\pi/2$ and $0$.  Our smoothness argument of topology change in Berry's phase is based on the Stokes theorem.  But the arguments of Dirac \cite{Dirac} and Wu and Yang \cite{Wu-Yang} to distinguish different configurations are important. Namely, if the Dirac string is not observable, then we ignore it physically and identify a monopole.  This unobservability critically depends on the magnetic charge of the monopole-like object and leads to the quantization of the charge in the case of the genuine Dirac monopole~\cite{Dirac, Wu-Yang}. In the present case, the magnetic charge is fixed by the formula of Berry's phase at $e_{M}=2\pi\hbar$.  Thus if the magnetic flux carried by the Dirac string satisfies the unobservability condition, we regard the monopole-like object as a physical monopole, and otherwise no physical monopole, namely, we regard only a combination of the monopole-like object and the accompanied string as a physical entity \cite{DF1}.

By keeping this criterion in mind, we start with an analysis of the adiabatic configuration with $\eta=\mu TB/\pi >1$ such as in Fig.B.6a.  The argument of Wu and Yang is to consider the singularity-free potentials in the upper and lower hemispheres        
\begin{align}
{\cal A}_{\varphi +} 
&=  \frac{e_{M}}{4\pi B\sin\theta}(1 - \cos\Theta(\theta)),\nonumber\\
{\cal A}_{\varphi -} 
&=  \frac{e_{M}}{4\pi B\sin\theta}(- 1 - \cos\Theta(\theta)),
\end{align}
using the potential in \eqref{new potential} with $e_{M}=2\pi\hbar$ by denoting the potential there as ${\cal A}_{\varphi +}$.  Note that ${\cal A}_{\varphi -}$ here is different from $\Omega_{-}$ in \eqref{solid-angle} which is associated with the energy eigenvalue $w_{-}Hw_{-}$.
These two potentials are related by a gauge transformation 
\begin{eqnarray}
{\cal A}_{\varphi -}={\cal A}_{\varphi +} - \frac{\partial\Lambda}{B\sin\theta\partial \varphi}
\end{eqnarray}
with 
\begin{eqnarray}
\Lambda=\frac{e_{M}}{2\pi} \varphi.
\end{eqnarray}
The physical condition is 
\begin{align}
\exp\left[ -\frac{i}{\hbar}\oint {\cal A}_{\varphi -}B\sin\theta d\varphi \right]&=\exp \left[-\frac{i}{\hbar}\oint{\cal A}_{\varphi +}B\sin\theta d\varphi +\frac{i}{\hbar}\oint \frac{\partial\Lambda}{B\sin\theta\partial \varphi}B\sin\theta d\varphi \right]\nonumber\\
&=\exp \left[-\frac{i}{\hbar}\oint{\cal A}_{\varphi +}B\sin\theta d\varphi \right]
\end{align}
which is in fact satisfied since the gauge term gives
$\exp[i e_{M}/\hbar]=\exp[2\pi i]=1$ and thus defines a monopole. 
Note that the physical condition in the present context is that the Schr\"{o}dinger wave function \eqref{eq-exactamplitude1} is single valued under the gauge transformation. It is confirmed that the present argument of gauge transformation is equivalent to the evaluation of the phase change induced by the Dirac string~\cite{Wu-Yang}. 
The fact that the physical condition is satisfied 
shows that the magnetic charge 
\begin{eqnarray}\label{magnetic charge}
e_{M}=2\pi \hbar
\end{eqnarray}
is properly quantized satisfying the Dirac quantization condition, although we have no analogue of an electric coupling in the present case unlike the original Dirac monopole~\cite{Dirac}.

In contrast, for the transitional domain $\eta=\mu TB/\pi =1$
such as in Fig.B.6b 
we have two potentials from \eqref{half sphere}  
\begin{align}
{\cal A}_{\varphi +}&=\frac{e_{M}}{4B\sin\theta}\left(1-\cos\frac{\theta}{2}\right)\nonumber\\
{\cal A}_{\varphi -}&=\frac{e_{M}}{4B\sin\theta}\left(-\cos\frac{\theta}{2}\right)
\end{align}
which are well-defined in the upper and lower hemispheres, respectively, and are related by the gauge transformation 
\begin{eqnarray}
{\cal A}_{\varphi -}={\cal A}_{\varphi +} - \frac{\partial\Lambda}{B\sin\theta\partial \varphi}
\end{eqnarray}
with 
\begin{eqnarray}
\Lambda=\frac{e_{M}}{4\pi} \varphi.
\end{eqnarray}
The physical condition 
\begin{align}
\exp\left[-\frac{i}{\hbar}\oint {\cal A}_{\varphi -}B\sin\theta d\varphi \right]&=\exp\left[-\frac{i}{\hbar}\oint{\cal A}_{\varphi +}B\sin\theta d\varphi +\frac{i}{\hbar}\oint \frac{\partial\Lambda}{B\sin\theta\partial \varphi}B\sin\theta d\varphi \right]\nonumber\\
&=\exp\left[-\frac{i}{\hbar}\oint{\cal A}_{\varphi +}B\sin\theta d\varphi \right]
\end{align}
is not satisfied since the gauge transformation gives
\begin{eqnarray}\label{half monopole phase}
\exp[ie_{M}/2\hbar]=\exp[i\pi ]=-1.
\end{eqnarray}
We thus conclude that the half-monopole at the transitional value of the parameter $\eta=1$ with the magnetic charge $e_{M}/2$ cannot describe a physical monopole; it is physical  as a combination of  the monopole-like object, which generates the outgoing flux, and the accompanied Dirac string  \footnote{A half-monopole with a magnetic charge $e_{M}/2$ gives a non-trivial phase \eqref{half monopole phase} and thus the Dirac string is not unobservable. In fact this phase of $\exp[i\pi]$ is the same as the Aharonov-Bohm phase of an electron in the magnetic field generated by the superconducting current of the Cooper pair in the experiment by Tonomura~\cite{Tonomura}.  In our criterion following the analysis of Wu and Yang \cite{Wu-Yang}, the Dirac string thus becomes a physical
observable just as the outgoing flux from the monopole-like object. }, although the Dirac string is actually defined only at $B=\pi/\mu T$. Topologically, it is thus the same as the dipole for $\eta<1$ in Fig.B.6c.

\subsection{ Longuet-Higgins phase change rule in an exactly solvable model}

It is instructive to see the implications of  an exactly solvable model of Berry's phase discussed in this appendix  on the analysis of Longuet-Higgins phase change rule, which is the manifestation of topological properties of Berry's phase, namely, the phase change  when the particle circles the level crossing point.
The exact solution, \eqref{new potential} with \eqref{effective angle},  gives (after adjusting the notation including $B\rightarrow r$)
\begin{eqnarray}
A^{(+)}_{\varphi} 
= \frac{\hbar }{2r\sin\frac{\pi}{2}} \left[1 - \cos\left( \frac{\pi}{2}-\alpha(\frac{\pi}{2}, \eta ) \right) \right] =\frac{\hbar }{2r} \left[1 - \sin \alpha( \frac{\pi}{2}, \eta ) \right] 
\end{eqnarray}
if one sets $\theta=\frac{\pi}{2}$ to project the monopole to the two-dimensional subspace. The phase change factor is then given by 
\begin{eqnarray}\label{phase change factor} 
\exp \left\{ -\frac{i}{\hbar}\oint \hbar A^{(+)}_{\varphi} rd\varphi \right\}
=\exp\left\{-i\pi \left[1 - \sin \alpha(\frac{\pi}{2}, \eta) \right] \right\}.
\end{eqnarray}
It is important that the quantity $\alpha(\frac{\pi}{2}, \eta)$ controls the phase change factor instead of the naively expected $\theta$. The parameter $\alpha(\frac{\pi}{2}, \eta)$ is given in Fig.B.1, and one finds $\alpha(\frac{\pi}{2}, \infty) = 0$,  \ $\alpha(\frac{\pi}{2}, 1) = \pi/4$, \ $\alpha(\frac{\pi}{2}, 0) = \pi/2$, respectively. 

In the adiabatic limit $\eta\rightarrow \infty$, we have $\alpha(\frac{\pi}{2}, \infty) = 0$ and thus the conventional phase change rule with $(-1)$ is given by \eqref{phase change factor}. For $\eta=1$, which approximately corresponds to the topology change from a monopole to a dipole in the present model,
we have the Longuet-Higgins phase change factor \eqref{phase change factor}
\begin{eqnarray}
\exp\{-i\pi[1 - 1/\sqrt{2}] \}.
\end{eqnarray}
In the nonadiabatic limit $\eta\rightarrow 0$, $\sin \alpha(\frac{\pi}{2}, \eta)\rightarrow 1$ and we have a unit phase change factor; thus the phase change rule is gone. This variation of the phase change factor, when one goes to off the precise adiabatic limit, shows how the topology is modified and eventually lost.   

A physical picture of this topology change is seen in terms of the motion of the spin placed in the rotating magnetic field in the present section. This is illustrated in Fig.B.7 for the case with fixed  $T=2\pi/\omega$ but changing $B (=r)$.\\
\\
\begin{figure}[H]
 \begin{center}
    \includegraphics[width=11cm]{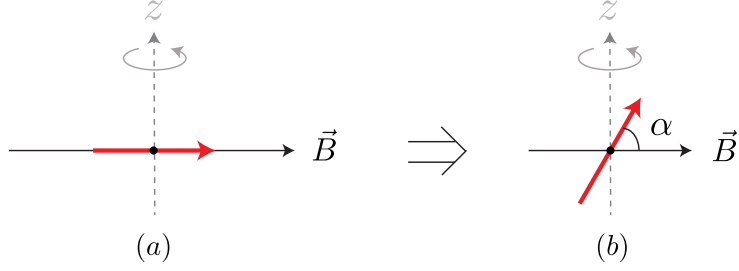} 
       \end{center}
\vspace{-3mm}      
  \caption{\small Horizontal views of the magnetic field $\vec{B}$ rotating around the origin in the horizontal plane (i.e., $x-y$-plane, and thus the vertical direction is $z$-axis) with a fixed angular velocity $\omega$ and a free spin (denoted by a bold arrow)  placed at the origin. For a slow adiabatic rotation and strong magnetic field, the spin stays on the horizontal plane and follows the rotating magnetic field, as in Fig.B.7(a) and gives the Longuet-Higgins phase change factor \eqref{phase change factor} at $(-1)$. When the magnetic field approaches the non-adiabatic configurations, which implies the weak field $\vec{B}$ in the present case (or rapid rotation), the spin starts to stand up effectively from the horizontal plane to the $z$-direction still following the rotating magnetic field with the angular velocity $\omega$ as in Fig.B.7(b); the effective polar angle becomes $\pi/2 - \alpha$.  For the angle $\alpha=\pi/2$, for which the spin becomes effectively perpendicular to the magnetic field, the Longuet-Higgins phase change factor \eqref{phase change factor} becomes unity, and no phase change rule appears. } 
\end{figure}
\vspace{1mm}

Figure B.7 shows that the spin is aligned to the direction of the magnetic field confined in the $x-y$ plane in the adiabatic limit of strong magnetic field $B$. But the spin   gradually ``stands up'' effectively to the direction of $z$ axis when the magnetic field in the $x-y$ plane becomes weaker with the fixed angular velocity $\dot{\varphi}=\omega$, although the spin follows the angular velocity $\omega$ of the magnetic field. The formula \eqref{phase change factor} shows that the Longuet-Higgins phase change factor gradually deviates from the ideal value $(-1)$. This is a picture how the phase change rule specified by Berry's phase at $\theta=\pi/2$ is gradually lost  in the non-adiabatic domain in the present model.

\subsection{Some remarks}

To understand the magnetic flux generated by the Dirac monopole in Berry's phase explicitly, we used a mathematical description of the topology change in Berry's phase in terms of a geometrical movement of Dirac's monopole caused by the nonadiabatic variations of parameters in Berry's model. It is remarkable that the monopole itself formally remains in tact without being resolved even in the nonadiabatic domain, but moves away from the level crossing point as in Fig. B.5. 
Traditionally, we are accustomed to understanding the topology change in terms of the winding and unwinding of some topological obstruction. The present geometrical description of topology change in terms of the moving monopole is a hitherto unknown mechanism. This new mechanism  partly arises from the fact that Berry's phase is not a simple monopole but rather a complex of the monopole and the level crossing point located at the origin of coordinates.  If one instead understands Berry's phase as a simple monopole, one would find a novel class of monopoles \cite{DF2, Mavromatos}.

We analyzed the monopole-like object and its topology appearing in Berry's phase by treating $\vec{B}$ as a given classical parameter.  If one adds other physical considerations, there appear some conditions on the parameters of the exactly solvable model \eqref{starting equation}. For example, the two energy levels in \eqref{energy eigenvalue} cross at $\alpha(\theta;\eta)=\pi/2$, which is related to the topology change from a monopole to a dipole in an intricate way, as is seen in Fig.B.1.


\newpage

\end{document}